\documentclass[journal=jacsat,manuscript=article]{achemso}

\usepackage[version=3]{mhchem} % Formula subscripts using \ce{}
\usepackage{braket}
\usepackage{amsmath}
\usepackage{xcolor}
\usepackage{amssymb}
\usepackage{simpler-wick}

\newcommand{\ckc}{\ket{\Phi_{oc}^{ty}}}
\newcommand{\ckd}{\ket{\Phi_{\bar{o}\bar{c}}^{\bar{t}\bar{y}}}}
\newcommand{\cke}{\ket{\Phi_{\bar{o}c}^{\bar{t}y}}}
\newcommand{\ckf}{\ket{\Phi_{o\bar{c}}^{t\bar{y}}}}
\newcommand{\ckg}{\ket{\Phi_{\bar{o}c}^{\bar{y}t}}}
\newcommand{\ckh}{\ket{\Phi_{o\bar{c}}^{y\bar{t}}}}

\author{Juan E. Arias-Martinez}
\email{juanes@berkeley.edu}
\affiliation
{{Kenneth S. Pitzer Center for Theoretical Chemistry, Department of Chemistry, University of California, Berkeley, California 94720, USA}}
\alsoaffiliation{Chemical Sciences Division, Lawrence Berkeley National Laboratory, Berkeley, California 94720, USA}
\author{Hamlin Wu}
\email{hamlin@berkeley.edu}
\affiliation
{{Kenneth S. Pitzer Center for Theoretical Chemistry, Department of Chemistry, University of California, Berkeley, California 94720, USA}}
\author{Martin Head-Gordon}
\email{mhg@cchem.berkeley.edu}
\affiliation
{{Kenneth S. Pitzer Center for Theoretical Chemistry, Department of Chemistry, University of California, Berkeley, California 94720, USA}}
\alsoaffiliation{Chemical Sciences Division, Lawrence Berkeley National Laboratory, Berkeley, California 94720, USA}

\title[An \textsf{achemso} demo]
  {Generalization of one-center non orthogonal configuration interaction 
   singles to open shell singlet reference states: Theory and application to valence-core pump-probe states in acetylacetone}

\abbreviations{IR,NMR,UV}
\keywords{American Chemical Society, \LaTeX}

\begin{document}

\begin{tocentry}
\includegraphics[width=1.00\linewidth]{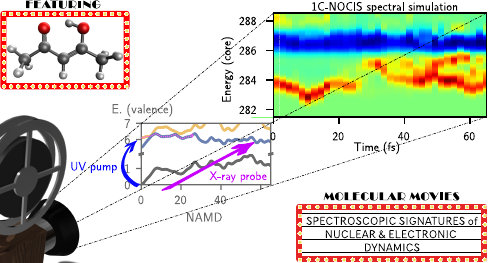}
\end{tocentry}

\begin{abstract}
We formulate a one-center non-orthogonal configuration interaction singles (1C-NOCIS) theory for the computation of core excited states of an initial singlet state with two unpaired electrons. This model, which we refer to as 1C-NOCIS two-electron open-shell (2eOS), is appropriate for computing the K-edge near-edge X-ray absorption spectra (NEXAS) of the valence excited states of closed-shell molecules relevant to pump-probe time-resolved (TR) NEXAS experiments. With inclusion of core hole relaxation effects and explicit spin adaptation, 1C-NOCIS 2eOS requires mild shifts to match experiment, is free of artifacts due to spin contamination, and can capture the high-energy region of the spectrum beyond the transitions into the singly occupied molecular orbitals (SOMO). Calculations on water and thymine illustrate the different key features of excited-state NEXAS, namely the core-to-SOMO transition as well as shifts and spin-splittings in the transitions analogous to those of the ground state. Finally, simulations of the TR-NEXAS of acetylacetone after excitation onto its $\pi \xrightarrow{} \pi^*$ singlet excited state at the carbon K-edge - an experiment carried out recently - showcases the ability of 1C-NOCIS 2eOS to efficiently simulate NEXAS based on non-adiabatic molecular dynamics simulations. 
\end{abstract}

\section{Introduction}
Time-resolved near-edge X-ray absorption (TR-NEXAS) experiments aim to track the ensuing dynamics of molecular systems after a perturbation with light by monitoring the NEXAS (equivalently referred to as NEXAFS, XANES, or often simply XAS) features of the species involved as a function of time. With advances in synchrotron slicing techniques and the advent of free-electron lasers, the time resolution of modern TR-NEXAS experiments is well into the femtosecond regime.\cite{Chergui2023, Kraus2018} Furthermore, improvements in high-harmonic generation have brought extreme UV and soft X-ray femtosecond pulses in the water window (270 - 550 eV) to table-top laser equipment.\cite{Bengtsson2017, Barreau2020} With element and site specificity, as well as strong sensitivity to the electronic environment of the species being probed, the TR-NEXAS experiments enabled by these new technologies have already provided fundamental insight into the role of dark singlet and triplet states in the electronic relaxation of organic molecules and directly tracked the nuclear motion of small molecules post strong-field ionization.\cite{Wolf2017, Bhattacherjee2017, Scutelnic2021, Green2022, Ross2022, Ridente2023} 

Before the development of ultra-fast techniques, the experimental focus of NEXAS was on characterizing stable molecules in their ground state. Accordingly, the development of electronic structure methods to aid in the assignment of NEXAS features focused predominantly on closed-shell systems as the reference state. A comprehensive review of the variety of methods available for computing NEXAS of closed shell systems is captured in recent reviews and is beyond the scope of this article,\cite{norman2018simulating, DFT-core-review_Besley2021} but we comment briefly on those most relevant to the present study. 

Within density functional theory (DFT), two main shortcomings of the associated linear response theory, namely time-dependent (TD)DFT,\cite{dreuw2005single} had to be addressed in order to extend its applicability to core excitations. The first difficulty, finding a way to target the inner roots of the effective TDDFT Hamiltonian, was addressed with techniques such as the core-valence separation (CVS) scheme and the restricted-energy window (REW) approach.\cite{Cederbaum1980, Stener2003, Zhang2012} The second difficulty, the delocalization errors and the lack of core-hole orbital relaxation inherent to TDDFT for core excitations, caused the calculated spectra to incur large shifts (on the order of 10's of eV) to align with experimental results. Specialized short-range corrected functionals tailored to core excitations laid a first solution to circumvent the latter two effects.\cite{Besley2009-SRC}. 

The recently-proposed electron-affinity (EA)-TDDFT provides a solution amenable to standard functionals by formulating a response theory that appropriately deals with the delocalization error in TDDFT, while employing core-ionized reference orbitals to address the core-hole relaxation.\cite{Carter-Fenk2022a, Carter-Fenk2022b} The latter also serves as an appropriate generalization to DFT of the earlier static-exchange (STEX) approach, which has been used to calculate NEXAS with success for decades.\cite{Hunt1969, Agren1994, Agren1997, Carravetta2001} In essence, STEX solves a single-excitation configuration interaction (CIS)\cite{foresman1992toward} in the space of core occupied and virtual orbitals, using core-ionized orbitals to account for orbital relaxation: in other words, STEX might more descriptively be called EA-CIS. EA-CIS for K-shell excitations has also been generalized to include coupling between core excitations on multiple sites via the non-orthogonal CIS (NOCIS) approach.\cite{Oosterbaan2018}

State-specific, or orbital-optimized,\cite{hait2021orbital} DFT approaches include the $\Delta$ self-consistent field (SCF) and the related restricted open-shell Kohn-Sham (ROKS) and transition-potential (TP)-SCF models provide an alternative to response theories. There are two key advantages relative to standard TDDFT that result in vastly improved accuracy: orbital relaxation is described by explicit orbital optimization in the presence of a core hole and an electron in the particle state,\cite{Filatov1999, Triguero1999, Besley2009, Ehlert2017, Michelitsch2019, Hait2020-CS, Ehlert2020, Hait2020} and avoiding the adiabatic approximation mitigates particle-hole self-interaction errors.\cite{dreuw2005single} ROKS achieves an accuracy on the order of 0.3 - 0.4 eV for the K-edge of main group elements and, with a perturbative treatment of spin-orbit coupling (SOC), L-edges of second-group elements.\cite{Hait2020-CS} In a subsequent study, it was found that accounting for scalar relativistic effects via the spin-free exact two-component (X2C) model \cite{X2C_Saue2011} extended the applicability of ROKS to the K-edge of third group elements and the first few transition metals before higher-order relativistic effects take hold.\cite{cunha2021relativistic} The main disadvantage of state-specific methods is that each of the states present in the NEXAS spectrum must be calculated separately, making them cumbersome for use in large systems. Therefore, full-spectrum ``state-universal'' methods are preferable relative to state-specific approaches when their accuracy is sufficient for the target purpose. 

Advances in wave function theory aimed at computing NEXAS have made a variety of single-reference methods based on coupled-cluster (CC) theory and algebraic diagrammatic construction (ADC) available for use on small and medium-sized systems.\cite{norman2018simulating} CC approaches are typically based on the linear response\cite{Sneskov2012} or equation of motion (EOM) approach\cite{Hitchiker-Krylov2008}, while ADC is explicitly designed for excited states from the outset.\cite{Dreuw2015} In both cases, as for TDDFT, the CVS scheme became a crucial component of EOM-CC and ADC formalisms to circumvent calculating the states between the ground state and the core-excited resonance states, and avoiding the divergences that plague the response equations.\cite{Barth1985, Wenzel2015, Coriani2015, Vidal2019, Nanda2020} An interesting approach takes inspiration from TP-SCF to provide a set of reference orbitals, optimized to a fractional occupation in the core, for a balanced treatment of both the ground state and the core excited states via the EOM-CC framework.\cite{Simons2021, simons2022transitionpotential} Finally, for the purpose of benchmarking core-excited state energies, state-specific coupled-cluster methods relying on excited-state references have recently been studied.\cite{Matthews2020, Arias-Martinez2022}

Relative to the development in electronic structure methods for computing the NEXAS of closed shell systems, the theoretical modeling of the NEXAS of open-shell radicals remains in an exploratory stage due to a collection of challenges. First, common to both closed-shell and radical systems, an electronic structure model must account for core-hole relaxation to obtain reasonable accuracy. Second, open-shell systems are often multi-configurational: aside from the case where the initial state is a high-spin open-shell, such as a doublet with a single open shell or an M$_s = \pm 1$ triplet with two open shells, multiple configurations are necessary for the proper description of a spin-pure initial state reference. Generating excited states out of an open-shell initial state compounds the challenge of ensuring spin purity. Third, electronic structure NEXAS calculations on short-lived radical species rely on molecular dynamics (MD) simulations for nuclear geometries that properly represent the evolution of the system, often involving a wide range of configurations, imposing the need for efficient generation of the spectra. 

The underdevelopment of methods for core spectroscopies on open-shell methods, coupled with the experimental advances and exciting prospects for unconvering fundamental chemical phenomena, have fueled a rapidly-growing body of work. While a variety of theories, such as 1C-NOCIS and STEX,\cite{Oosterbaan2019, Oosterbaan2020, Carravetta2022} EOM-CCSD,\cite{Faber2019} $\Delta$SCF and ROKS,\cite{hait2021orbital, Hait2022} and a number of TDDFT-based formalisms, have been extended to treat one-electron open-shell (1eOS) doublets and high-spin 2eOS triplets, we focus on the developments for 2eOS singlets and highlight relevant ideas from other open-shell cases when useful.

\begin{figure}
  \includegraphics[width=1.00\linewidth]{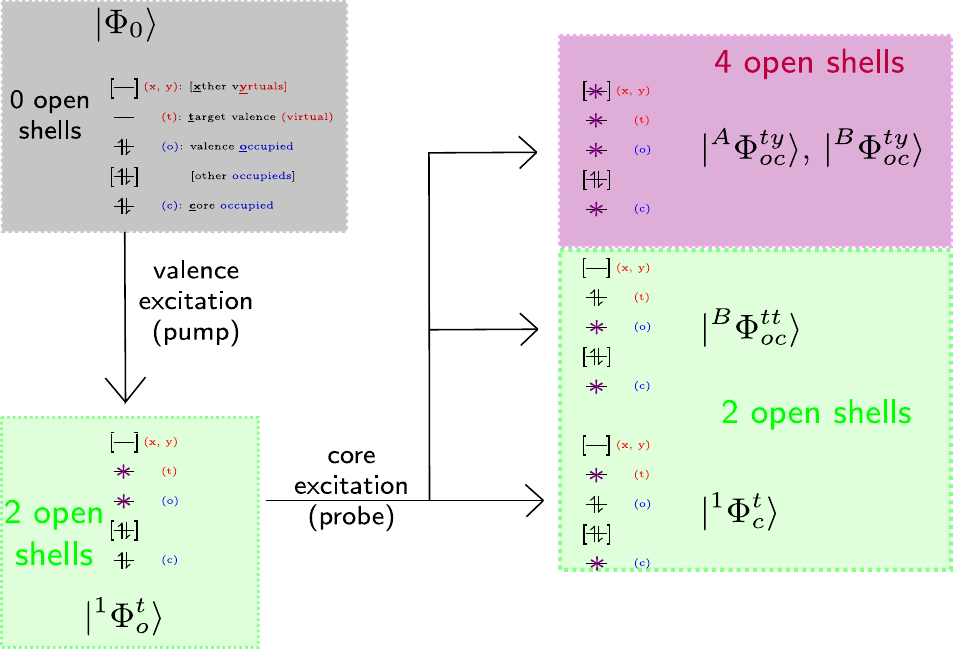}
  \caption{Electronic states resulting from valence-pump and core-probe excitations. The purple stars represent the half-filled orbitals and signify that, in general, more than one configuration is required to generate a spin-pure state. The explicit definition on the CSFs in this Figure, constructed relative to the ground state, is provided in the text (Eqs. \ref{eq:valence_excited_state} - \ref{eq:PP_S-B}). Their orbital labels are explained on the ground state panel (top-left).}
  \label{fgr:pump-probe-states}
\end{figure}

Figure \ref{fgr:pump-probe-states} provides a visual guideline to the types of excited states relevant in UV-pump X-ray-probe TR-NEXAS experiments on a molecule with a closed-shell ground state. A visible or UV pump causes one photon absorption that promotes the molecule to an optically allowed valence excited state where an electron in \textbf{o}ccupied orbital (\textbf{o}) has been promoted to a \textbf{t}arget virtual orbital (\textbf{t}): 
\begin{align}
    \ket{^{1}\Phi_{o}^{t}} &= (2)^{-1/2} \left(\ket{\Phi_{o}^{t}} + 
                                               \ket{\Phi_{\bar{o}}^{\bar{t}}} \right) \label{eq:valence_excited_state}
\end{align}
Three kinds of core excited states can be conceived out of a 2eOS singlet valence excited state.  When the \textbf{c}ore electron (\textbf{c}) re-pairs either with the now singly-occupied molecular orbital (SOMO) in the occupied space, or with the newly created particle SOMO, we can obtain two singlet spin-adapted CSFs: 
\begin{align}
    \ket{^{1}\Phi_{oc}^{to}} =\ket{^{1}\Phi_{c}^{t}} &= (2)^{-1/2} \left(\ket{\Phi_{c}^{t}} +
                                               \ket{\Phi_{\bar{c}}^{\bar{t}}} \right) \label{eq:c_to_o}\\
    \ket{^{1}\Phi_{oc}^{tt}} &= (2)^{-1/2} \left(\ket{\Phi_{\bar{o}c}^{\bar{t}t}} +
                                                 \ket{\Phi_{o\bar{c}}^{t\bar{t}}} \right) \label{eq:c_to_t}
\end{align}
The c $\xrightarrow{}$ SOMO(o) transition, associated with Eq. \ref{eq:c_to_o}, is special in that it is usually well-separated to lower energy from the NEXAS transitions in the ground state and allows for a clear detection of a valence excited state in the TR-NEXAS when bright.\cite{Bhattacherjee2017, Wolf2017} Either OO-DFT or response theories, such as TDDFT, ADC, and EOM-CC , can reliably calculate the c $\xrightarrow{}$ SOMO(o) transition as an energy difference between the valence excited state and the pump-probe core excited state, since the later can be reached from the ground state by a single excitation.\cite{Tsuru2021} On the other hand, the c $\xrightarrow{}$ SOMO(t) state associated with Eq. \ref{eq:c_to_t} is beyond traditional TDDFT because it is a double excitation out of the closed-shell reference. As it is well-established that EOM-CC requires truncation beyond doubles to properly correlate doubly-excited states, the $\ket{^{1}\Phi_{oc}^{tt}}$ excited state is beyond EOM-CCSD despite the fact that the dominant configuration exists within the associated Fock space.\cite{Loos2019-Double_excitations} For similar reasons, it is possible that the modification of ADC to describe 2eOS singlets proposed by Ruberti \textit{et al.} also fails to capture the $\ket{^{1}\Phi_{oc}^{tt}}$ excited state when truncated at second order.\cite{Ruberti2014, Leitner2022} Regardless, the former was used by Neville and coworkers to calculate the NEXAS (as well as the extended X-ray absorption fine structure - EXAFS) of a few excited states of ethylene, ammonia, carbon dioxide, and water at their FC geometries.\cite{Neville2016-JCP} Subsequently, they employed the model to predict and study the TR-NEXAS of ethylene in detail, demonstrating the sensitivity of NEXAS to nuclear motion and electronic character, even if solely judged by the c $\xrightarrow{}$ SOMO transitions.\cite{Neville2016-Faraday, Neville2018} Very recently, the aforementioned ADC methodology was also used to re-examine the TR-NEXAS of pyrazine after excitation into its B$_{2u}$ $\pi \xrightarrow[]{} \pi^*$ state at the nitrogen K-edge.\cite{Kaczun2023} Two DFT-based methods proposed in the last two years explore unusual response approaches to capture both c $\xrightarrow{}$ SOMO transitions. The first is the hole-hole Tamm-Dancoff Approximated (hh-TDA) DFT, which shows promise in its efficiency yet still suffers from a lack of orbital relaxation and the difficulties associated with converging a doubly-electron attached reference.\cite{Hohenstein2021} In contrast, multi-reference spin-flip (MR-SF)-TDDFT employs a relaxed open-shell core-excited triplet reference to access both the valence excited singlet state and the c $\xrightarrow{}$ SOMO configurations via spin-flip operations.\cite{Park2022}

Continuing with the higher energy one-electron core excitations, we arrive at those involving a core-excitation into a fully-vacant virtual orbital. These types of transitions, resulting in a four-electron open-shell (4eOS) excited state, are beyond traditional TD-DFT, ADC(2), EOM-CCSD, or hh-TDA DFT. MR-SF-TDDFT can only partially describe the 4eOS states, since some of the configurations necessary are still not accessible with the theory. There are six 4eOS M$_S$ = 0 configurations with SOMOs in o, c, t, and an arbitrary virtual orbital y.
\begin{align}
    \ckc,\; \ckd,\; \cke,\; \ckf,\; \ckg,\; \ckh
\end{align}
As demonstrated in Section 3.1 of the Supporting Information (SI), diagonalizing the S$^2$ operator in the basis of the six 4eOS, M$_S$ = 0 configurations yields two linearly independent singlet, three triplet, and one quintet CSFs.\cite{aszabo82:qchem, Pauncz1979} 
\begin{align}
    \ket{^{1_A}\Phi_{oc}^{ty}} &= (12)^{-1/2}  \left(2\ckc + 2\ckd + \cke + \ckf - \ckg - \ckh \right) \label{eq:PP_S-A} \\
    \ket{^{1_B}\Phi_{oc}^{ty}} &= (2)^{-1}   \left(\cke + \ckf + \ckg + \ckh \right) \label{eq:PP_S-B} \\
    \ket{^{3_C}\Phi_{oc}^{ty}} &= (2)^{-1/2} \left(\ckc - \ckd \right) \\
    \ket{^{3_D}\Phi_{oc}^{ty}} &= (2)^{-1/2} \left(\cke - \ckf \right) \\
    \ket{^{3_E}\Phi_{oc}^{ty}} &= (2)^{-1/2} \left(\ckg - \ckh \right) \\
    \ket{^{5_F}\Phi_{oc}^{ty}} &= (6)^{-1/2} \left(\ckc + \ckd - \cke - \ckf + \ckg + \ckh \right)
\end{align}
A common strategy to generate an excited-state NEXAS - including the 4eOS states - is to employ TDDFT or EOM-CCSD for the core excitation on top of an optimized non-Aufbau configuration representing the initial valence excited state.\cite{Bhattacherjee2017, Attar2017, Scutelnic2021, Tsuru2021} $\Delta$SCF - the aforementioned procedure to capture the valence excited states - employs a single unrestricted configuration to describe the valence excited state, rendering it severely spin-contaminated for open-shell singlets. A response theory that also disregards spin-symmetry on top of it carries over and exacerbates the spin-contamination on the excited states, artificially shifting the predicted energies and possibly predicting spurious bright excitations (Section 1 of the SI). Incorporating the developments in alternative approaches to reconcile economic standard response theories with open-shell references while addressing spin purity, such as the spin-adapted (s)-TDDFT is an exciting prospect.\cite{Li2010}

While using $\Delta$SCF in conjunction with response theories spin-contaminates the resulting pump-probe states, employing $\Delta$SCF to target both the initial valence excited states and the final pump-probe core excited states allows for spin-purification procedures, such as the approximate spin-projection (AP), to address the deficiency of employing a single configuration to describe low-spin open-shell excited states.\cite{yamaguchi1988spin} The AP is straightforward for 2eOS singlets and the spin-purification can be made rigorous by employing a single set of restricted open-shell orbitals for the AP procedure, like in the ROKS approach, instead of optimizing unrestricted configurations separately for the low-spin and high-spin orbitals. Extending the spin-recoupling schemes to 3eOS and 4eOS systems opens the door for the full NEXAS calculation beyond the c $\xrightarrow{}$ SOMO transitions of doublet radicals and 2eOS singlets using AP-$\Delta$SCF.\cite{Hait2020, hait2021orbital} In a similar vein, Zhao and co-workers have used multi-state (MS)-DFT to produce spin-pure core excited states out of open-shell radicals, with the attractive feature of addressing the correlation double-counting inherent to employing DFT orbitals in CI-like formalism.\cite{Zhao2021} Recently, a state-specific approach relying on CI employing core-relaxed orbitals to provide partially sin-complete 4eOS excited states was recently used by Garner and Neuscamman to explore the effect of spin coupling of the core excited states and their sensitivity to nuclear geometry for photochemical ring-opening of furanone.\cite{garner2023spin} The down-side to state-specific methods like AP-$\Delta$SCF, ROKS, or MS-DFT arises from their inconvenience. In the case of AP-$\Delta$SCF or MS-DFT, a number of unrestricted configurations (two, four, and eight for 2eOS singlets, 3eOS doublets, and 4eOS singlets, respectively) must be optimized independently to calculate spin-pure excited states. Difficult convergence and asserting that the configurations indeed correspond to the same set of spatial orbitals presents a challenge to automation procedures, making this approach cumbersome for systems of moderate size and for generating a large number of spectra for different nuclear configurations. While progress in this area has been accomplished for 2eOS core excited states via AP-$\Delta$SCF,\cite{Ehlert2020} no such automation procedure has been designed for 3eOS and 4eOS excited states.

Real-time (RT)-TDDFT provides an alternative to response theories for the simulation of excited state absorption; the interested reader is directed to Section 4.3 of the recent review by Li \textit{et al.}\cite{Li2020}  Leveraging said capacity, it has recently been employed to simulate TR-NEXAS.\cite{Chen2020, Moitra2023} In theory, RT-TDDFT is able to describe the whole spectrum of excited states but it suffers from practical considerations. First and foremost, and in common with standard frequency-domain TDDFT, the adiabatic approximation results in an inability to describe core-hole relaxation and causes RT-TDDFT to incur in large shifts to match experimental profiles in the X-ray regime unless using specially-tailored functionals. Furthemore, employing ``CVS-like'' schemes to decouple the core resonances from the continuum presents additional challenges.\cite{RT-Herbert2023} Second, the need to explicitly propagate the electronic density in time makes TR-NEXAS simulations in the hundred-femtosecond time-scales of nuclear motion intractable at the moment, with the ones reported at the moment extending to only a few attoseconds.\cite{Chen2020, Moitra2023}

The last two excited-state methodologies we are aware of to calculate pump-probe core excited states, including the 4eOS states absent from the majority of the theories previously described, involve explicit use of multi-reference (MR) formalism. The first is a MR configuration-interaction (CI) procedure employing DFT orbitals, namely DFT-MRCI,\cite{Grimme1999} employed by Seidu and coworkers to simulate the excited-state NEXAS of 1, 3-butadiene.\cite{Seidu2022} The second involves carrying out a complete active space (CAS)SCF calculation, or its dynamically-correlated PT2 variant, in conjunction with restricted-active space (RAS) techniques to ensure the proper occupancy of the relevant orbitals: one for the core orbital, one for the valence hole state, and one for the valence particle state. RASSCF / RASPT2 have been used to theoretically probe the relaxation pathways of malonaldehyde with XAS,\cite{List2020} and to predict the NEXAS signatures of the $\pi \xrightarrow{} \pi^*$ and N$_{LP} \xrightarrow{} \pi^*$ states of azobenzene.\cite{Carlini2023} Their own problems aside, namely electron correlation double-counting for DFT-MRCI and the computational cost plus the intruder state phenomena for RAS-SCF / RAS-PT2, these are promising approaches capable of simulating both the complex non-adiabatic dynamics relevant for relaxation after excitation, and the full NEXAS spectra of the resulting species.
 
The near-absence of electronic structure models capable of providing properly core-hole-relaxed, spin-pure pump-probe core excited states in an efficient manner motivates the present work. We present a new state-universal method which generalizes EA-CIS and the related NOCIS methods to describe K-shell core excitations from an open-shell singlet excited state of a molecule with a closed shell ground state.\cite{Oosterbaan2018, Oosterbaan2019, Oosterbaan2020} The paper begins by presenting a variety of reference orbitals with the relaxation appropriate for the description of core excited states out of a valence excited state. Relying on these orbitals, the ansatz for the spin-adapted generalization of the one-center (1C)-NOCIS is then provided. By virtue of working with configuration state functions (CSFs), the method avoids shifts in the energy due to spin-contamination and, perhaps more importantly, ensures that transitions are bright only when they ought to be. Subsequently, calculations on water and thymine as test systems are carried out to assess the general NEXAS features of valence excited states relative to each other and to the ground state. Finally the ability of 1C-NOCIS to efficiently simulate spectra is showcased by a simulation of the TR-NEXAS of acetylacetone at the carbon K-edge after excitation into its lowest $\pi \xrightarrow[]{} \pi^*$ state, based on a sequence of snapshot structures from molecular dynamics simulations.

% 1. Walk us through the theory
\section{Adaptation of 1C-NOCIS model for a 2eOS reference}
%We propose and investigate an adaptation of the 1C-NOCIS framework to 2eOS initial states to collectively address some of the challenges presented by the computation of NEXAS for excited states: electronic relaxation due to the core hole, proper treatment of the MR character of the open-shell excited state due to spin, and a one-shot generation of the whole NEXAS spectrum from a single structure to enable efficient sampling of nuclear configuration space. 

\subsection{Reference orbitals for the pump-probe excited states}
\begin{figure}
  \includegraphics[width=1.00\linewidth]{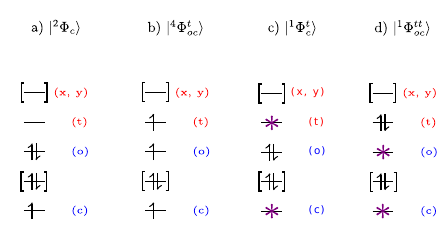}
  \caption{Visual representation of the different choice of reference orbitals considered for the construction of the pump-probe core excited states. The configurations with purple stars represent orbitals obtained from spin-pure ROKS optimizations.}
  \label{fgr:orbital_choice}
\end{figure}
We begin by introducing the reference orbitals that we will use for the description of the pump-probe core excited states, visualized in MO diagrams in Figure \ref{fgr:orbital_choice}. Appropriate core-hole orbital relaxation is key for an accurate description of core excited states, whether for closed-shell references or open-shell species. However, open-shell singlet excited states feature an additional challenge arising from the partially-filled orbitals. Ideally, the reference orbitals would also take into consideration the orbital relaxation arising from the valence excitation which, albeit smaller than the core-hole relaxation, may still be relevant. 

One attractive candidate set of orbitals comes from ROHF-optimized $M_S = \frac{1}{2}$ doublet core-ionized references, $\ket{^2\Phi_c}$, for which the presence of the core hole confer the orbitals the appropriate contraction. The deficiency of this choice lies in the poor description of valence particle states afforded by the canonical virtual orbitals of the core-ionized reference. To address the need for a well-defined target particle state \textbf{t} we rotate the doublet core ion virtual space into the natural transition orbital (NTO) basis of a regular 1C-NOCIS calculation for the ground state which provide core particle states in resemblance with valence particle states. This is similar in essence to what was done by Hait \text{et al.} to evaluate the ROKS energies from a STEX calculation employing DFT orbitals.\cite{Hait2022} The target virtual is simply chosen as the one with the highest overlap with the particle state of the valence excited state. One deficiency in this choice is that it disregards the relaxation in the orbitals due to the valence excitation. A second one, practical in nature, presents itself when the valence particle state has a fortuitously-large overlap with a core particle state of different character, or when there is simply no clear connection between the valence particle state and any core particle state. We conceive this situation taking place for high-lying valence excited states with loosely-defined particle states, such as Rydberg states in complex molecules.

An alternative choice of orbitals could come from optimizing the high-spin M$_s$ = 3 / 2 quartet core ion associated with the valence excited state, $\ket{^4\Phi_{oc}^t}$, via ROHF. The advantage of this choice comes in that the valence particle state \textbf{t} is explicitly optimized and, furthermore, the rest of the orbitals are optimized in presence of an electron in said orbital. In other words, this set of orbitals incorporate relaxation due to the valence particle state as well as the core hole. The disadvantage comes in that the spatial description of the orbitals of a singlet state may differ significantly from those optimized for a quartet reference, a phenomena that MR SF-TDDFT likely also suffers from when trying to describe singlet excited states with orbitals optimized for a triplet state.

A third choice is to construct the CSFs from the $\ket{^{1}\Phi_{c}^{t}}$ configuration optimized via ROKS. Like the $\ket{^4\Phi_{oc}^t}$ orbitals, this choice is appealing because it produces a set of orbitals relaxed in the presence of a core hole \textbf{c} and a particle state \textbf{t}, refined by the orbital optimization procedure. Unlike the $\ket{^4\Phi_{oc}^t}$ orbitals, the spatial description of the $\ket{^{1}\Phi_{c}^{t}}$ configuration is optimized within the correct multiplicity. Note that the ROKS procedure is susceptible to nonphysical mixing between the two open-shells when they possess the same spatial symmetry that must be addressed to avoid overly-intense oscillator strengths (Section S5.2 of the SI).\cite{Kowalczyk2013} A final choice of orbitals we considered come from ROKS optimization of the $\ket{^{1}\Phi_{oc}^{tt}}$ configuration. These orbitals take into consideration relaxation due to the valence hole \textbf{o} as well as the core hole \textbf{c}. We disregard this choice from here on, as the ROKS optimization of this configuration is challenging in practice. 

Finally, it is important to highlight that localization of the core orbitals (when the canonical ground-state orbitals delocalize over several atoms) prior to the SCF re-optimization in the presence of a core hole is crucial for improvements in the accuracy.\cite{Brumboiu2022} An appealing alternative choice of orbitals, unavailable to us at the moment, is evident: ROKS optimization of the 3eOS doublet core ion, where the three open-shells lie in the \textbf{c}, \textbf{o} and \textbf{t} orbitals. This configuration is perfect in that it accounts for relaxation due to all the holes and particles present in the pump-probe excited states associated with a particular valence excited state. While two linearly independent 3eOS doublet states exist (Section S3.1 of the SI), the most sensible choice for describing singlet states would be the 3eOS doublet genealogically-related with, say, the $\ket{^{1}\Phi_{c}^{t}}$ singlet.\cite{Pauncz1979}

\subsection{Ansatz for the 1C-NOCIS 2eOS wave function}
With a variety of reference orbitals available, we proceed to introduce our model for the pump-probe excited states. For convenience, we introduce a set of basis functions for each core orbital of interest in the system built out of CSFs orthogonalized against the initial 2eOS state $\ket{\Psi_i} = \ket{^1\Phi_o^t}$. The initial state can obtained by either orbital-optimization via ROKS or constructed from a CIS wave function for the state of interest rotated into the NTO basis and truncated to the dominant contributor.
\begin{align}
    \ket{^{1}\tilde{\Phi}_{c}^{t}}   &= (1 - P_i) \ket{^{1}\Phi_{c}^{t}} \\
    \ket{^{N}\tilde{\Phi}_{oc}^{tx}} &= (1 - P_i) \ket{^{N}\Phi_{oc}^{tx}} 
\end{align}
In Eqs. 11 and 12, $P_i$ represents a projector of the initial state, and in Eq. 12, the $N$ labels the linearly independent states of a specific spin (Eqs 5 - 10). A choice of orbitals from the candidates presented in the previous subsection is employed for the construction of the core-excited CSFs. The Hamiltonian in the basis of these projected-out CSFs takes the following form: 
\begin{align}
    \tilde{H}^{MN}_{x,\;y} = 
    & \bra{^{M}\Phi_{oc}^{tx}} (1 - P_i) H (1 - P_i) \ket{^{N}\Phi_{oc}^{ty}} \\ = 
    & \bra{^{M}\Phi_{oc}^{tx}} H \ket{^{N}\Phi_{oc}^{ty}} + \bra{^{M}\Phi_{oc}^{tx}} H \ket{^1\Phi_o^t} \braket{^1\Phi_o^t | ^{N}\Phi_{oc}^{ty}} + \notag\\ & 
    \braket{^{M}\Phi_{oc}^{tx} | ^1\Phi_o^t} \bra{^1\Phi_o^t} H \ket{^{N}\Phi_{oc}^{ty}} + \braket{^{M}\Phi_{oc}^{tx} | ^1\Phi_o^t} \bra{^1\Phi_o^t}H \ket{^1\Phi_o^t} \braket{^1\Phi_o^t | ^{N}\Phi_{oc}^{ty}} \\
    \mathbf{\tilde{H}^{MN}} = &\mathbf{H^{MN}} + \mathbf{H_{N.O.}^{M}} (\mathbf{S_{N.O.}^{N}})^T + \mathbf{S_{N.O.}^{M}} (\mathbf{H_{N.O.}^{N}})^T + E_i\cdot\mathbf{S_{N.O.}^{M}} (\mathbf{S_{N.O.}^{N}})^T
\end{align}
Above, the matrices $\mathbf{S_{N.O.}}$ and $\mathbf{H_{N.O.}}$ represent the overlap and Hamoltonian matrix elements between the valence excited state CSF and the final pump-probe CSFs. Because the orbital basis of the valence CSF and the pump-probe CSFs differs, these matrix elements must be built with non-orthogonal (N.O.) configuration interaction techniques. For the 4eOS configurations, we label the Hamiltonian matrix elements with only the subscripts \textbf{x} and \textbf{y}, denoting arbitrary virtual orbitals w.r.t. the closed-shell configuration,  because this is the only running index; the indexes \textbf{o}, \textbf{t}, and \textbf{c} are fixed by the valence excited state being probed and the core orbital of interest. In other words, the dimensionality of the matrix $\tilde{H}^{MN}_{x,\;y}$ is, at most, $N_{virt.} \times N_{virt.}$. 
\begin{figure}
  \includegraphics[width=1.00\linewidth]{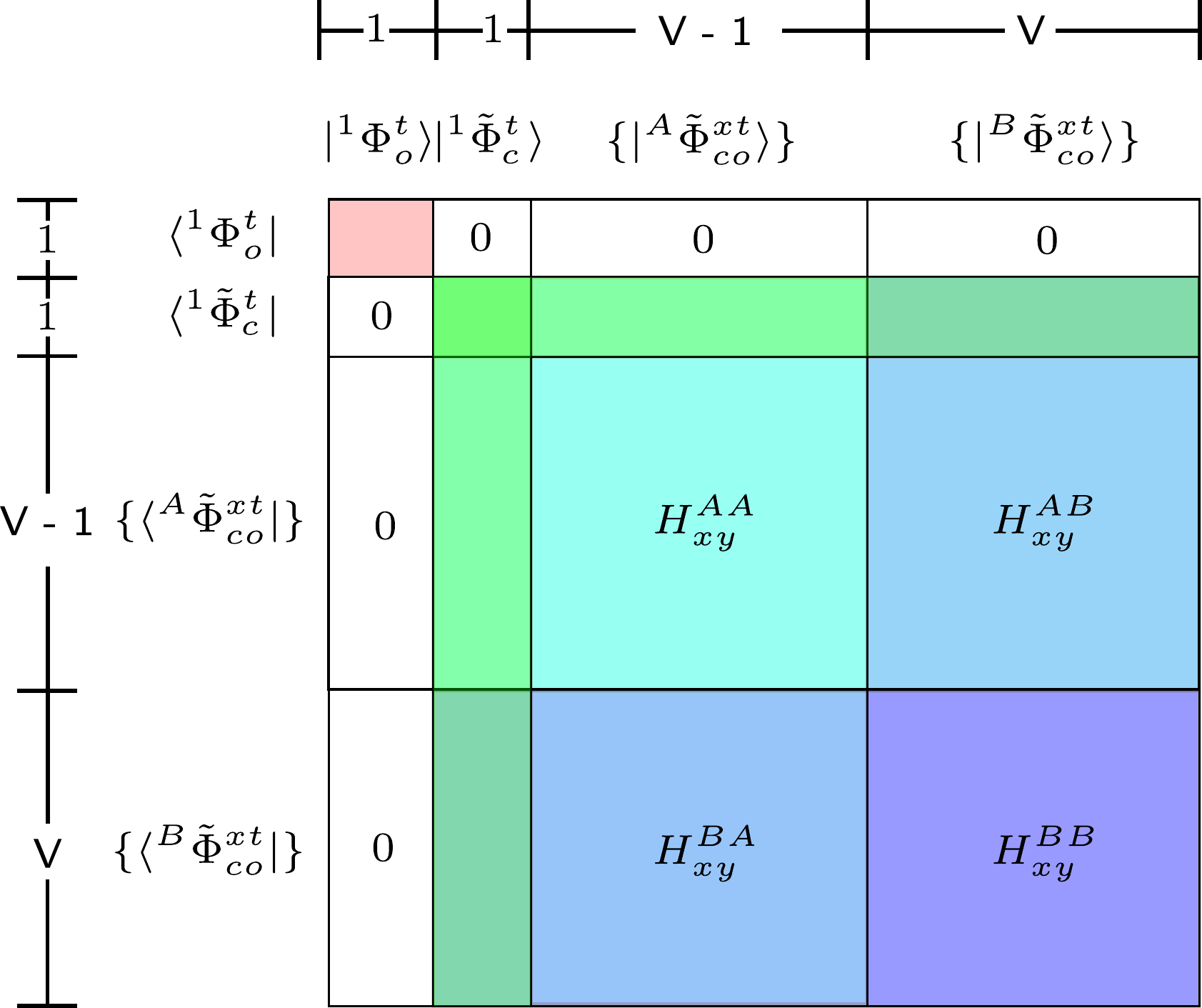}
  \caption{Visual representation of the 1C-NOCIS 2eOS Hamiltonian for singlet pump-probe excited states. The dimensionality of the specific sub-rows and sub-columns is listed to the left / above and the index labels are explained in Figure \ref{fgr:pump-probe-states}.}
  \label{fgr:H}
\end{figure}

Figure \ref{fgr:H} provides a visual representation of the 1C-NOCIS 2eOS Hamiltonian for singlet pump-probe excited states. While the 2eOS configuration associated with the c $\xrightarrow{}$ SOMO(o) transition must be explicitly included, the 2eOS configuration associated with the c $\xrightarrow{}$ SOMO(t) transition is included as a special case of the 4eOS  $\ket{^{1_B}\Phi_{oc}^{ty}}$ configuration when y = t. On the other hand, the $\ket{^{1_A}\Phi_{oc}^{ty}}$ configuration vanishes when y = t. In total, the dimensionality of the 1C-NOCIS 2eOS singlet Hamiltonian is $2N_{virt.} \times 2N_{virt.}$. Section 2 in the SI provides derivations of the standard non-relativistic electronic Hamlitonian in the basis of the relevant non-spin-adapted configurations via algebraic techniques.\cite{Harris1992} Sub-Section 3.2 provides the Hamiltonian matrix elements in the basis of CSFs by taking the appropriate linear combinations of the non-spin adapted matrix elements dictated by the eigenfunctions of the S$^2$ operator; these matrix elements correspond to the first term in Eq 16. Section 4 provides simplifications due to spin-adaptation to $\mathbf{S_{N.O.}}$ and $\mathbf{H_{N.O.}}$. Solving the eigenvalue problem
\begin{align}
    \mathbf{\tilde{H}} \mathbf{C} = &\mathbf{\tilde{S}} \mathbf{C} \mathbf{E} 
\end{align}
where the matrix 
\begin{align}
    \tilde{S}^{MN}_{x,\;y} = 
    & \bra{^{M}\Phi_{oc}^{tx}} (1 - P_i) (1 - P_i) \ket{^{N}\Phi_{oc}^{ty}} \notag\\ = 
    & \bra{^{M}\Phi_{oc}^{tx}} (1 - P_i) \ket{^{N}\Phi_{oc}^{ty}} \notag\\ =
    & \braket{^{M}\Phi_{oc}^{tx} | ^{N}\Phi_{oc}^{ty}} + \braket{^{M}\Phi_{oc}^{tx} | ^1\Phi_o^t} \braket{^1\Phi_o^t | ^{N}\Phi_{oc}^{ty}} \\
    \mathbf{\tilde{S}} = &\mathbf{I} - \mathbf{S_{N.O.}^{M}} (\mathbf{S_{N.O.}^{N}})^T 
\end{align}
accounts for the deviation from orthogonality between the CSFs due to the projection of the initial state yields Hamiltonian eigenstates
\begin{align}
    \ket{\Psi_f} &= b_c^t \ket{^{1}\tilde{\Phi}_{c}^{t}} + \sum_{M, y} b^\sigma_x \ket{^{M}\tilde{\Phi}_{oc}^{ty}}  \\
    \ket{\Psi_f} &= (1 - P_i) \left( b_c^t \ket{^{1}\Phi_{c}^{t}} + \sum_{M, x} b^M_x \ket{^{M}\Phi_{oc}^{ty}} \right)
\end{align}
with eigenvalues $E_f$. The excitation energy is simply the energy difference between the initial valence excited state and the final pump-probe states, $\Delta E_{if} = E_f - E_i$. Oscillator strengths
\begin{align*}
    f_{f \xleftarrow[]{} i} = \frac{2}{3} \Delta E_{if} |\bra{\Psi_i} \hat{\mu} \ket{\Psi_f}|^2
\end{align*}
are the final ingredient to generate a theoretical spectra. Central to this quantity is the transition dipole moment, $\bra{\Psi_i} \hat{\mu} \ket{\Psi_f}$.

% The formal scaling of the method presented herein is dictated by the O$(N^3)$ SCF optimizations required to provide the reference orbitals for the valence excited state and the pump-probe excited states. In practice, the

\section{Computational methods}
1C-NOCIS 2eOS was implemented in a development version of QChem 6.1 and will be available in the next public release.\cite{QCHEM5} Calculations were carried out with different aug-pcX-n (n = 1 - 4) basis sets on the atoms associated with the K-edge of interest, with an aug-pcseg-1 basis on the remaining atoms.\cite{Ambroise2019} Scalar relativistic effects are incorporated via the X2C model. The geometry for the preliminary calculations on water is the experimental one listed in the Computational Chemistry Comparison and Benchmark Data Base (CCCBDB) of the National Institute of Standards and Technology (NIST).\cite{FIPS1402} The geometry used for the calculations on thymine, provided in the literature,\cite{Wolf2017} was optimized to the ground state at the CCSD(T) / aug-cc-pVDZ level. Two NAMD trajectories on acetylacetone for proof-of-concept spectral simulation were carried out using the augmented fewest-switches surface-hopping (AFSSH) algorithm.\cite{Subotnik2011} They were initiated on the first $\pi \xrightarrow{} \pi^*$ state from the ground state geometry, optimized at the HF level and carried out with CIS with the aforementioned basis set combination, for consistency of the dynamics with the NEXAS calculations.

% 2. Present the spectra for the nine different excited states of water + the ground state. Speak about the features of the spectrum.
\section{Preliminary excited-state NEXAS calculations}
Before a discussion on the excited state spectra, we make an important note about the range of applicatibility of our 1C-NOCIS calculations, both for the ground state and the excited states. Without an appropriate treatment of the continuum of states beyond the K-edge ionization threshold - the region associated with the EXAFS - our 1C-NOCIS 2eOS calculations are only appropriate for computing the bound core resonances associated with the NEXAS. As is evident from Figure S3, states beyond the ionization threshold are ill-behaved and we thus set the ionization threshold as a boundary to the reliability of our theoretical spectrum. The ionisation threshold for the closed-shell ground state is simply defined as the energy difference between the ground state and the optimized core-ionized reference - namely the $\Delta$SCF ionization energy. Removing a core electron from a 2eOS valence excited state results in a 3eOS system, and thus there are three different ionization potentials for the valence excited states. These are associated with the quartet and two doublet states obtained by diagonalization of the S$^2$ operator in Section S.3.1.2 of the SI:
\begin{align*}
    \ket{^{2_G} \Phi_{oc}^{t }} = &(6)^{-1/2}\Big(2\ket{\Psi_{o\bar{c}}^{t}} - \ket{\Psi_{o\bar{c}}^{t }} - \ket{\Psi_{c\bar{o}}^t}\Big) \\
    \ket{^{2_H} \Phi_{oc}^{t}} = &(2)^{-1/2}\Big(\ket{\Psi_{o\bar{c}}^{t }} - \ket{\Psi_{c\bar{o}}^t}\Big) \\
    \ket{^{4_I} \Phi_{oc}^t} = &(3)^{-1/2}\Big(\ket{\Psi_{o\bar{c}}^{t}} + \ket{\Psi_{o\bar{c}}^{t }} +\ket{\Psi_{c\bar{o}}^t}\Big)
\end{align*}
We establish the ionization thresholds for the valence excited states as the energy difference between the valence excited state and the energy of the 3eOS CSFs evaluated with whichever orbitals are employed for the 1C-NOCIS 2eOS procedure.

% 2. Present the spectra for the nine different excited states of water + the ground state. Speak about the features of the spectrum.
\subsection{Water}
\begin{figure}
  \includegraphics[width=1.00\linewidth]{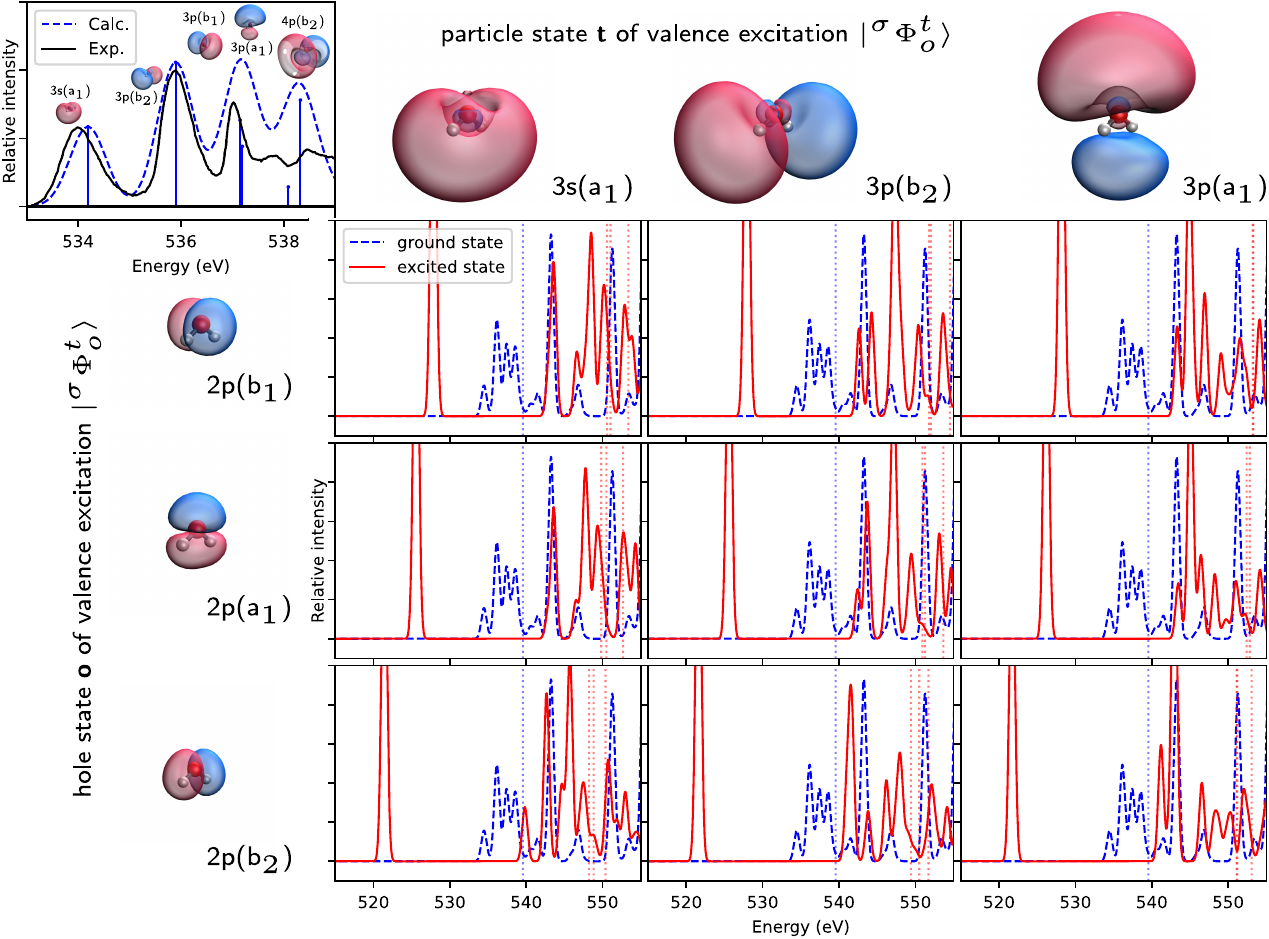}
  \caption{Top-left inset: NEXAS of the ground state of water computed by 1C-NOCIS / aug-pcX-2 (O) / aug-pcseg-1 (H), shifted by -0.36 eV, compared to the experimental spectra.\cite{Schirmer1993} Bottom-right: 1C-NOCIS / aug-pcX-2 (O) / aug-pcseg-1 (H) NEXAS for nine different singlet valence excited states of water, compared to the calculated ground state NEXAS. The columns / rows, in left-right / descending order correspond to different particle / hole states associated with the valence excitation in increasing / decreasing energy order. The ionization thresholds are plotted as vertical dotted lines.}
  \label{fgr:H2O_SX}
\end{figure}

The 1C-NOCIS NEXAS for the closed-shell ground state of water is in excellent agreement with experiment (top-left of Figure \ref{fgr:H2O_SX}), requiring only a shift of -0.36 eV to align the strongest signal. We employed the 1C-NOCIS 2eOS theory to calculate the NEXAS for nine different valence excited states of water, associated with the different combinations of 2p$_x$ (b$_1$), 2p$_z$ (a$_1$), and 2p$_y$ (b$_2$) hole states with 3s (a$_1$), 3p$_y$ (b$_2$), and 3p$_z$ (a$_1$) particle states. Before analyzing the spectra in more detail, we briefly comment here on the impact of the choice of orbitals on the excited state NEXAS, and elaborate on Section 5.3 of the SI. The spectral profile generated the ROHF-optimized $\ket{^2\Phi_{c}}$ (1C-NOCIS NTO basis) and ROKS-optimized $\ket{^{1}\Phi_{c}^{t}}$ orbitals are almost indistinguishable from each other (Figure S5). On the other hand, the spectra resulting from the ROHF-optimized $\ket{^4\Phi_{oc}^t}$ orbitals differs significantly from those generated with the other two candidate orbitals. 

The nine panels in Figure \ref{fgr:H2O_SX} plot the resulting excited-state NEXAS when employing the 1C-NOCIS NTO orbitals for the construction of the CSFs. The \textbf{o} and \textbf{t} orbitals of the reference, serving as the visual indexes for the NEXAS of the different excited states in Figure \ref{fgr:H2O_SX}, clearly resemble the particle and hole states of the valence excited states of water. The c $\xrightarrow{}$ SOMO(o) transitions dominate the NEXAS of the excited states and are red-shifted from the ground state features by several eVs. The strong intensity of the O$_{1s} \xrightarrow{}$ SOMO(o) transitions is due to the localization of both the 1s and 2p orbitals in the oxygen atom, allowing for a large transition dipole matrix element $\bra{\phi_{O_{1s}}} \hat{\mu} \ket{\phi_{O_{2p}}}$.  In contrast the overlap between the 1s orbital and the diffuse Rydberg orbitals is smaller and results in a lower transition intensity into said particle states. This region of the spectrum, containing the transitions that are analogous to the transitions out of the closed-shell ground state, is richer than the latter due to the spin-splitting in the 4eOS core excited states (Figure S6). 

A striking feature of the 4eOS states is a strong blue shift on the order of 5 - 10 eV for the high-energy features relative to the ground state transitions, with a general increase in the oscillator strength. While these states seem to be all beyond the ionization threshold of the ground state, a few of them are still within the core ionization threshold corresponding to the valence excited state itself. The two previous calculations we found on the K-edge ionization energy of the first two excited states of water at the Frank-Condon (FC) geometry differ drastically. The CVS-IP-ADC(2)-x results of Neville and coworkers report an IP of roughly 530 eV, compared to the IP of the ground state at 540 eV.\cite{Neville2016-JCP} Instead, the $\Delta$SCF / CVS-EOM-IP-CCSD results of Moitra \textit{et al.} predict the different doublet and quartet ionization potentials to lie at 544 - 549 eV.\cite{Moitra2021} The ionization thresholds computed in this work, in the range of 550 - 554 eV and visualized as vertical lines in Figure \ref{fgr:H2O_SX}, are in closer agreement to those of Moitra and co-workers. While the 1C-NOCIS 2eOS results presented here, the $\Delta$SCF / CVS-EOM-EE-CCSD calculations of Moitra \textit{et al.}, and the CVS-ADC(2)-x work of Neville \textit{et al.} agree on the strong c $\xrightarrow{}$ SOMO(o) transition red shifted from the ground state, the fate of the c $\xrightarrow{}$ SOMO(t) 2eOS and the 4eOS states differs significantly. As the authors acknowledge, and as explained in the Background section, the high-energy transitions are beyond $\Delta$SCF / CVS-EOM-EE-CCSD and are completely absent from the spectra presented in their work. CVS-ADC(2)-x, predicts these transitions to be well into the ionization continuum, beyond 560 eV. Since no experimental data exists yet for these states, the correct location in energy and relative intensities of these features remains unknown. 

% 2. move on to a more complex situation: thymine. More chemically relevant (normal excited states, so to speak) and featuring more edges to probe.
\subsection{Thymine}
\begin{figure}
  \includegraphics[width=1.00\linewidth]{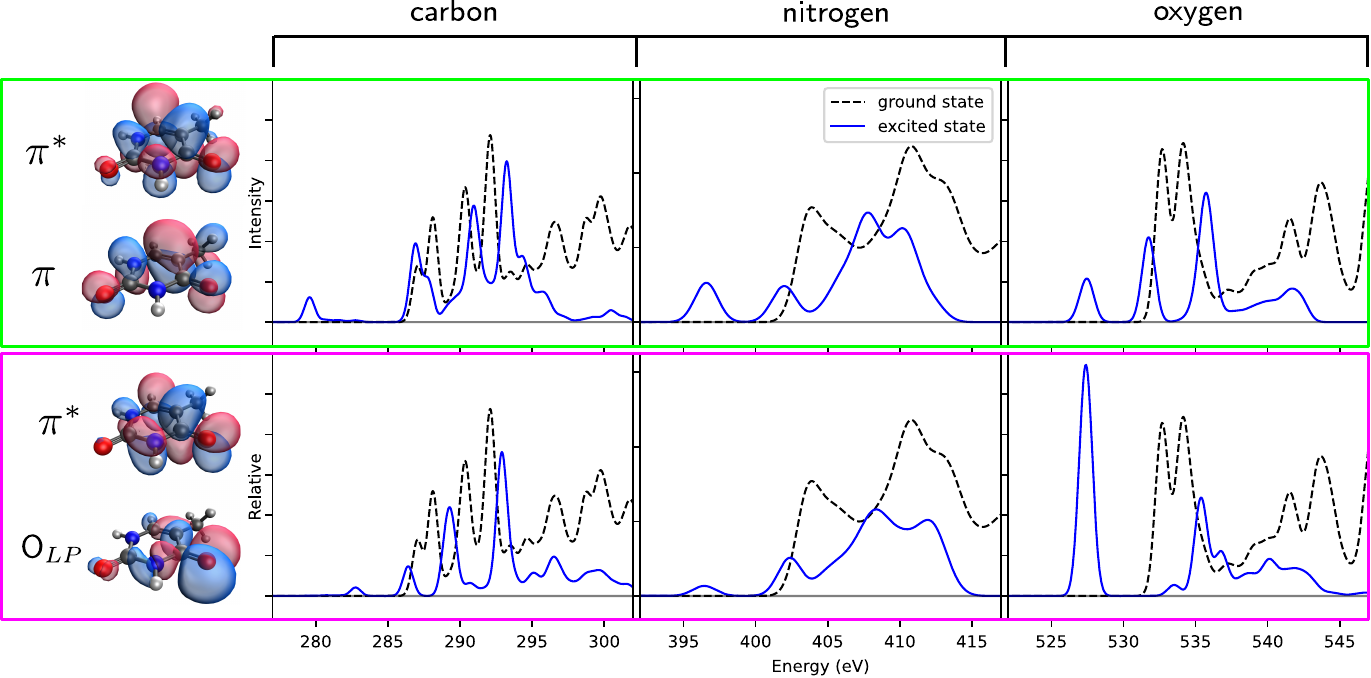}
  \caption{Theoretical NEXAS of the ground state and O$_{LP} \xrightarrow{} \pi^*$ (bottom panels) and $\pi \xrightarrow{} \pi^*$ (top pannels) excited states of thymine at the carbon (left), nitrogen (center), and oxygen (right) K-edges, calculated with 1C-NOCIS. The aug-pcX-2 basis set is used on the atoms of the edge probed, with aug-pcseg-1 on all the other atoms. A shift of -2.25, -2.35, and -1.25 eV is applied to the carbon, nitrogen, and oxygen K-edges, respectively, to align the 1C-NOCIS calculations of the ground state with experiment.\cite{Zubavichus2008}}
  \label{fgr:thymine}
\end{figure}
Thymine served as a valuable second case study because, aside from being a molecule of broad chemical and biological interest, it has a variety of heavy atoms, which allowed us to investigate the behavior of the excited-state NEXAS at different edges. Furthermore, an experimental TR-NEXAS with sub-100 fs time resolution at the oxygen K-edge was reported in the literature.\cite{Wolf2017} Figure \ref{fgr:thymine} provides the NEXAS of the O$_{LP} \xrightarrow{} \pi^*$ (bottom panels) and $\pi \xrightarrow{} \pi^*$ (top panels) excited states of thymine at the carbon, nitrogen, and oxygen K-edges, as calculated with 1C-NOCIS 2eOS. With the exception of the oxygen atom with the dominant contribution to the hole state of the (O$_{LP} \xrightarrow{} \pi^*$) state, and in stark contrast to water, the calculated ionization thresholds of the two excited states studied here lie relatively close to those of the ground state for all atoms at the C, N and the O K-edge (Section S6 of the SI). We speculate that this difference could arise from the fact that the excited states of thymine feature particle states with proper valence character, whereas those of water are all of Rydberg character at the FC geometry.\cite{Rubio2008} Excited states of Rydberg character are quasi-ionized systems, which are known to feature severe blue-shifts in their core ionization potentials.\cite{Lindblad2020, Couto2020} In regards to the strongly blue-shifted ionization potential for one of the oxygens: the (O$_{LP} \xrightarrow{} \pi^*$) excited state migrates electron density localized at said oxygen atom to a $\pi^*$ orbital delocalized over the whole molecule. In other words, it has local charge-transfer character and in a sense this scenario mimicks the effect of a Rydberg excitation or an ionization from the point of view of this oxygen atom. 

As is clearly identified in the TR-NEXAS experiment of Wolf \textit{et al.}, the c $\xrightarrow[]{}$ SOMO(o) transition at the oxygen K-edge is the most prominent, well-separated feature for identification of the O$_{LP} \xrightarrow{} \pi^*$ excited state. Direct observation of the bright $\pi \xrightarrow{} \pi^*$ state is predicted to be more challenging. One promising way could be to rely on the splitting of the O$_{1s} \xrightarrow{} \pi^*$ feature at the oxygen K-edge, which seems to be accentuated in the $\pi \xrightarrow{} \pi^*$ state. In theory, then, a positive induced absorption flanking the ground state bleach should be observed with enough time and energy resolution. In the experiment, some induced absorption is observed in this region in the ultra-fast time scales but with the resolution of the TR-NEXAS its not possible to assign it with certainty. The carbon and nitrogen K-edges of the excited states seem to overlap too much with each other and with the ground state at the FC geometry to be of use. While thymine is a relatively rigid molecule, it is possible that structural dynamics may result in shifts to the energies and intensities that disentangles the NEXAS of the different states. We relegate an investigation of the spectral consequences of nuclear motion on a more flexible molecule: acetylacetone.

% 3. Cap off with a full-fledge TR-NEXAS simulation with dynamics: acetylacetone!
\subsection{Simulation of the TR-NEXAS of acetylacetone}

No set of benchmark excited-state NEXAS exists from theory yet and only a limited number of UV-pump X-ray-probe TR-NEXAS experiments have come out in the last two decades, where the excited state NEXAS are encoded as a differential absorption from the static spectrum. The resolution of this data is compromised, in part, due to the difficulties of constructing instruments capable of the incredible time resolution required to observe these short-lived transient species and, as a result, no high-resolution data coming from experiment exists either. We chose to validate the 1C-NOCIS 2eOS model by generating a theoretical TR-NEXAS for the carbon K-edge of acetylacetone after excitation into $S_2$ - its lowest $\pi \xrightarrow{} \pi^*$ state - since the experimental TR-NEXAS data in the work of Battharchee \textit{et al.} is among the clearest ones and the excited state dynamics are well-established, providing certainty as to which excited states should be visible in the experiment.\cite{Bhattacherjee2017, Squibb2018} Specifically, the $\pi \xrightarrow{} \pi^*$ state lives on the order of 50 - 100 fs, decaying into S$_1$ (O$_{LP} \xrightarrow{} \pi^*$). In turn, S$_1$ lives for hundreds of fs before direct internal conversion into the T$_1$ ($\pi \xrightarrow{} \pi^*$) takes place; the T$_2$ (O$_{LP} \xrightarrow{} \pi^*$) state is found to play a minor role in the relaxation from S$_1$ into T$_1$. With $\sim$60 fs X-ray pump pulses, the set up employed by Battacherjee and coworkers to study these dynamics has the time resolution to unambiguously observe the $S_1$ and T$_1$ states but not the S$_2$ state.

\begin{figure}
  \includegraphics[width=1.00\linewidth]{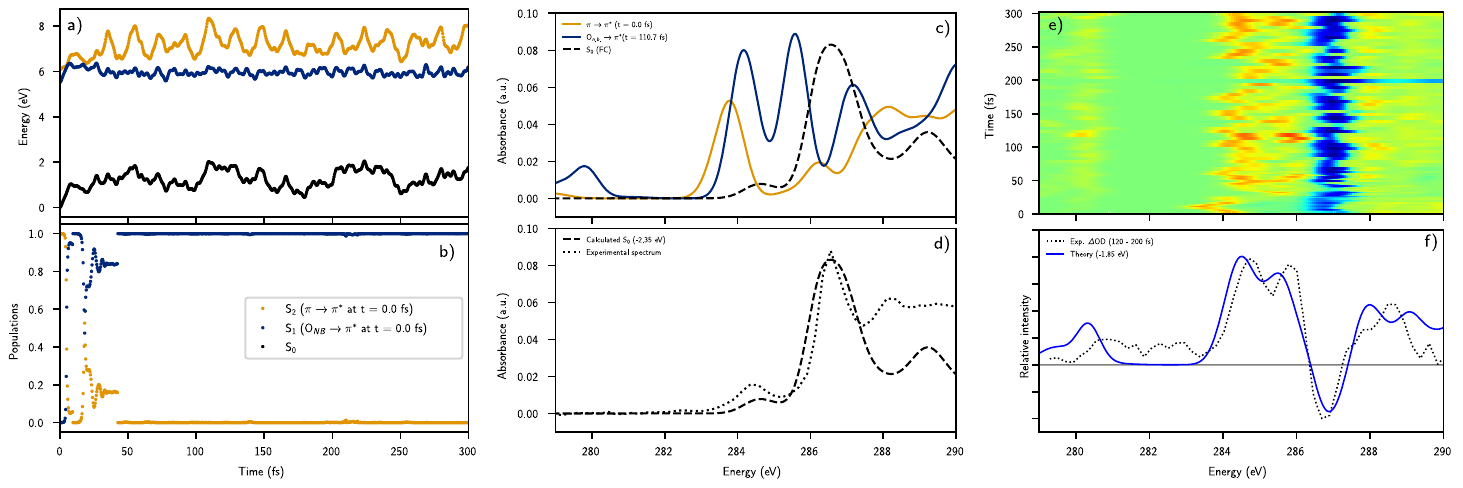}
  \caption{a) Energies of the three lowest singlet states (the ground state and the (O$_{n.b.} \xrightarrow{} \pi^*$) and ($\pi \xrightarrow{} \pi^*$) excited states) during an AFSSH NAMD simulation. b) The populations of the two excited states during the NAMD trajectory. c) Comparison of the ground state NEXAS spectrum (Wigner-broadened with 150 structures) and the excited-state NEXAS spectra at relevant times (i.e. when they are populated in the dynamics.) d) Comparison of the ground state NEXAS spectrum calculated with 1C-NOCIS / aug-pcX-1 and the experimental NEXAS of Bhattacherjee and coworkers.\cite{Bhattacherjee2017} A shift of -2.35 eV is required to match experiment and theory. e) Simulated TR-NEXAS, shifted by -1.85 eV; see text for details f) Comparison of experimental spectrum, averaged over the 120 - 200 fs time bins compared to the simulated spectrum averaged over the same time range.\cite{Bhattacherjee2017}}
  \label{fgr:acetylacetone}
\end{figure}
Figure \ref{fgr:acetylacetone}b shows the populations of the S$_1$ and S$_2$ states of acetylacetone from a sample AFSSH trajectory, initiated in the bright S$_2$ state and carried out in the adiabatic basis, with the energies of the states displayed in Figure \ref{fgr:acetylacetone}a. The system likely remains in the $\pi \xrightarrow{} \pi^*$ state in the first 20 fs of the trajectory. Within the subsequent 30 fs, the O$_{LP} \xrightarrow{} \pi^*$ state becomes populated. To prepare the construction of a theoretical TR-NEXAS based on this trajectory, the ground state spectrum is calculated with standard 1C-NOCIS and compared against the experimental static spectrum, shown in Figure \ref{fgr:acetylacetone}d. A shift of -2.35 eV, within the errors expected for 1C-NOCIS for closed-shell organic molecules, is required in the computed spectrum to match the dominant feature of the experiment.\cite{Carter-Fenk2022a} The shift required represents the remaining dynamic correlation not captured by 1C-NOCIS. Despite a complete disregard to dynamic correlation, this shift is smaller than the +3.50 and -4.25 eV shifts required by DFT-MRCI and RAS-PT2 to align the ground state spectra of butadiene and malonaldehyde, respectively, to either experiment or CVS-ADC(2)-x, and serves as a testament to the dominant relevance of orbital relaxation for a description of core excited states.\cite{Seidu2022, List2020} We proceed by calculating the excited-state NEXAS for S$_1$ and S$_2$ with 1C-NOCIS 2eOS out of structures plucked from the NAMD simulations at regular time intervals. An example of the excited state spectra at relevant times is displayed in Figure \ref{fgr:acetylacetone}c, compared against the calculated ground state spectrum. Finally, the theoretical TR-NEXAS shown in Figure \ref{fgr:acetylacetone}e is constructed by taking a linear combination of the excited state NEXAS as a function of time, with the coefficients being determined by the NAMD populations and a fraction of the ground-state NEXAS subtracted to simulate the bleach feature. For the transient spectrum, a smaller shift of -1.85 eV is applied for a better match with the experimental results, obtained from Ref \citenum{Bhattacherjee2017}.

We used a finer time-grid to select structures within the first 65 fs of the dynamics to properly capture the effect of the non-adiabatic dynamics on the spectrum, displayed in more detail in Figure S8 of the SI. A single intense absorption centered at 284 eV, corresponding to C$_{1s} \xrightarrow{}$ SOMO($\pi^*$) transitions of the $\pi \xrightarrow{} \pi^*$ state is observed in the first 20 fs of the TR-NEXAS. The effect of vibrational motion in the NEXAS absorption - oscillations in the energy of the signal on the order of 1 eV - is clearly observed. The signal changes dramatically at around 25 fs, at which point a two-peak feature - characteristic of the O$_{LP} \xrightarrow{} \pi^*$ state - emerges and oscillates in and out of the spectra for the remaining 300 fs simulated. 

The mild positive absorbance in the 281 - 283 eV regime in the experimental spectrum at 120 - 200 fs (Figure \ref{fgr:acetylacetone}f) could be attributed in part to the C$_{1s} \xrightarrow{}$ O$_{LP}$ (which would be underestimated by 1C-NOCIS, with the shift that was applied, at 280.5 eV if that is the case) or else a small fraction of triplet state already present, which absorbs in this regime. Moving up in energy we arrive at the doubled-peaked feature, clearly observed in the experiment (Figure \ref{fgr:acetylacetone}f).\cite{Bhattacherjee2017} Visualizations of the spectral contributions due to each of the carbon atoms to the O$_{LP} \xrightarrow{} \pi^*$ excited state in Figure \ref{fgr:acetylacetone}c are provided in Section S7.3 of the SI. The low-energy peak of this double-peaked signal comes from the C$_{1s} \xrightarrow{} \pi^*$ transitions out of the central atoms. The higher-energy peak arises primarily from C$_{1s} \xrightarrow{} \pi^*$ transitions out of the terminal carbon atoms, with a secondary but significant contribution from a C$_{1s} \xrightarrow{} \pi^*$ transition from the central carbon atom, where the $\pi^*$ particle state in the latter corresponds to the higher-lying, empty $\pi^*$ orbital that is not involved in the valence excitation. The absorbance into the higher-energy states beyond the SOMOs is also recovered, although the match in the profile shape begins to deteriorate, likely in part due to the onset of the ionization potentials and the resulting contributions of the continuum transitions. The results from an spectral simulations from an additional trajectory are presented in Section S7.2 of the SI.

\section{Conclusion}
% Conclusion
To conclude, we highlight the merits of 1C-NOCIS 2eOS by specifying the challenges other methods would face when attempting to simulate the TR-NEXAS of acetylacetone, as done in the previous subsection. $\Delta$SCF / TD-DFT, the theoretical model employed to calculate the excited state NEXAS in the study of Bhattacherjee \textit{et al.} requires a shift of 10.3 eV to match the calculated ground state with experiment and, as explained earlier, the TD-DFT core spectra calculated out of the $\Delta$SCF valence excited states comes with a strong degree of spin contamination. The clear double-peaked feature present in the experiment up until 1.5 ps is not observed in any of the computed spectra. In contrast, 1C-NOCIS requires a shift smaller by a factor of five to match with experiment, provides excited-state NEXAS free from spin contamination by design, and the resulting spectral profile matches well with the experimental observation.  While hh-TDA DFT may be able to efficiently capture the important c $\xrightarrow{}$ SOMO transitions without spin-contamination, it would require a large shift to align with experiment and would be incapable to produce the remaining of the spectrum. Most of the theories able to provide the full spectrum are either too cumbersome (AP-$\Delta$SCF or MS-DFT) or prohibitively expensive to simulate a TR-NEXAS for acetylacetone at the carbon K-edge for a duration on the order of hundreds of femtoseconds. 1C-NOCIS 2eOS, on the other hand, is a quasi-black-box diagonalization-based approach - where only the valence excited state and the edge of interest need to be specified - capable of sampling a range of nuclear configurations efficiently, making it amenable for the TR-NEXAS simulation presented. The only methods capable to produce comparable results, likely at an increased cost and with perhaps a larger energy shift, would be DFT-MRCI or RASPT2. 

1C-NOCIS 2eOS is far from a converged theory and there is room for development. An obvious direction for progress lies in the inclusion of dynamic correlation to further reduce shifts in the NEXAS to align with experiment and hopefully achieve sub-eV accuracy as is now plausible for closed-shell systems. Importantly, this could alleviate the possible differential shifts required for the ground state and the excited states, as was the case for our TR-NEXAS simulation of acetylacetone. A generalization of 1C-NOCIS 2eOS to DFT, as electron-affinity (EA) TD-DFT did for STEX, is an attractive candidate.\cite{Carter-Fenk2022a, Carter-Fenk2022b} Alternative options could be taking lessons from s-TDDFT, simply employing DFT orbitals in the 1C-NOCIS 2eOS model \emph{a la} DFT-MRCI, or searching for dynamic correlation within the wave function framework. Furthermore, since 1C-NOCIS 2eOS would ideally rely on NAMD simulations to simulate the excited-state NEXAS, a generalization to DFT would allow for the use of TDDFT to calculate the valence excited states themselves, and lead to NAMD trajectories and structures of a quality better than CIS. An alternative would be to simply use TDDFT NAMD as a source for the structures and populations but a mismatch in the level of theory employed for the dynamics and for the generation of the excited state NEXAS presents a book-keeping challenge, since the ordering of the valence excited states may be different in TDDFT and in CIS. Another exciting avenue for development is the inclusion of spin-orbit coupling (SOC) into the theory to calculate NEXAS beyond the K-edge. This would allow 1C-NOCIS 2eOS to serve TR-NEXAS experiments like those carried out recently at the iodine N$_{4,5}$-edge.\cite{Chang2022, Tross2023, toulson2023probing}   
\begin{acknowledgement}

JEA thanks Leonardo dos-Anjos Cunha and Kevin Carter-Fenk for stimulating discussions. This work was supported by the Director, Office of Science, Office of Basic Energy Sciences, of the U.S. Department of Energy, under Contract No. DE-AC02-05CH11231.

\end{acknowledgement}

\begin{suppinfo}

Section 1 of the SI provides the transition energies, oscillator strengths, and $\braket{S^2}$ values for the NEXAS of \ce{CO+} computed by fc-CVS EOM-CCSD and 1C-NOCIS doublets, with a comparison against the experimental results. Section 2 provides derivations of the standard non-relativistic, Born-Oppenheimer electronic Hamiltonian on the basis of the configurations (not spin-adapted) relevant to pump-probe excited states (Eqs. 2 and 4). Section 3 provides the required algebra for the explicit spin-adaptation of the 1C-NOCIS 2eOS model, including a construction of CSFs out of the aforementioned configurations by diagonalization of the $S^2$ operator (S3.1), and a construction of the Hamiltonian matrix elements in the basis of CSFs by taking the appropriate linear combination of non-spin-adapted matrix elements (S3.2). Section 4 provides simplifications due to spin-adaptation to the non-orthogonal overlap, Hamiltonian, and transition dipole matrix elements required by 1C-NOCIS 2eOS. Section 5 provides auxiliary information for the calculations on water, including the ill-behavior of states beyond the ionization threshold (S5.1), the consequences of open-shell mixing during the ROKS procedure on predicted intensities (S5.2), the impact of the choice of reference orbitals on the 1C-NOCIS 2eOS spectra (S5.4), and a protocol to visualize the contributions to the 1C-NOCIS 2eOS spectra (S5.5). Section 6 provides a more detailed view of the simulated TR-NEXAS of acetylacetone on the ultrafast timescales (S6.1), an additional simulated TR-NEXAS based on another AFFSH trajectory (S6.2), and a visualization of the different contribtuions to the 1C-NOCIS 2eOS spectra for the O$_{n.b.} \xrightarrow{} \pi^*$ excited state. All the calculations employed for this study are provided in Zenodo.

\end{suppinfo}

\bibliography{achemso-demo}

\end{document}

\title{
	Supporting Information for:\\
``Generalization of one-center non orthogonal configuration interaction 
   singles to open shell singlet reference states: Theory and application to valence-core 
   pump-probe states in acetylacetone"
}
\author{
	Juan E. Arias-Martinez,\footnote{\href{mailto:juanes@berkeley.edu}{juanes@berkeley.edu}}
 	Hamlin Wu,\footnote{\href{mailto:hamlin@berkeley.edu}{hamlin@berkeley.edu}}
	and Martin Head-Gordon\footnote{\href{mailto:mhg@cchem.berkeley.edu}{mhg@cchem.berkeley.edu}}\\
	{\em Kenneth S. Pitzer Center for Theoretical Chemistry,}\\
	{\em Department of Chemistry, University of California, Berkeley, CA 94720, USA}\\
	{\em Chemical Sciences Division, Lawrence Berkeley National Laboratory, Berkeley, CA 94720, USA}
}
\date{\today}
\maketitle
\tableofcontents
\clearpage\pagebreak

\section{Standard response theories and spin contamination in the NEXAS of open-shell species}\label{sec:spin}

The consequences of disregarding spin-symmetry via response theories may go beyond small shifts in the predicted energies of the transitions as spin-selection rules for transitions are broken and spurious bright excitations from, say, pseudo-singlets into pseudo-triplets, emerge. This is relevant because erroneously bright transitions could misguide the interpretation of experimental spectra, and while this phenomena may be mild for the c $\xrightarrow{}$ SOMO transitions calculated with the $\Delta$SCF / TD-DFT or $\Delta$SCF / EOM-CCSD approaches, it may manifest strongly for the core excitations that promote into a fully vacant orbital. 

As an illustration, we compute the oxygen K-edge of \ce{CO+}, a simple radical species, with the frozen-core core-valence separated (fc-CVS) EOM-CCSD.\cite{Vidal2019} Figure S1 shows the resulting spectra, shifted to match the SOMO transition with the experimental values (the lowest-energy peak at $\sim$528 eV);\cite{Couto2020} it predicts an intense \ce{O}$_{1s} \xrightarrow{} \pi^*$ transition that is completely absent from the experimental spectra. An analysis of the excited-state $\braket{S^2}$ reveals that the incorrectly-predicted transition actually corresponds to the quartet \ce{O}$_{1s} \xrightarrow{} \pi^*$ transition that ought to be dark, but becomes bright due to admixture of doublet character.

%The electronic excited states of open-shell species calculated with traditional response theories may come with heavy spin contamination. For example, the lowest-energy fc-CVS EOM-CCSD feature of the oxygen K-edge of \ce{CO+} aligns well with the c $\xrightarrow{}$ SOMO(o) transition - of O$_{1s} \xrightarrow{} \sigma$ character - after a shift of -1.49 eV. Said shift also alignins a strong transition into the $\pi^*$ orbitals with the experimental feature. Nonetheless, the associated quartet transition is rendered allowed by strong spin contamination, resulting in a strong bright feature that is completely abscent from experiment.

\begin{figure}[h!]
\begin{floatrow}
\capbtabbox{%
    \begin{tabular}{ccc}\toprule
    Energy (eV) & O. S. & $\braket{S^2}$ \\\midrule
    530.08 & 0.010 & 0.77 \\
    533.55 & 0.016 & 2.24 \\ 
    533.55 & 0.016 & 2.24 \\
    535.29 & 0.032 & 1.49 \\
    535.29 & 0.032 & 1.49 \\ 
    544.07 & 0.000 & 1.25 \\ 
    546.23 & 0.002 & 2.29 \\
    546.68 & 0.001 & 1.78 \\
    546.68 & 0.001 & 1.78 \\
    \bottomrule
    \end{tabular}
}{%
    \caption{Raw fc-CVS EOM-CCSD / aug-ccX-2 transition energies, oscillator strengths, and $\braket{S^2}$ values of \ce{CO+} at the oxygen K-edge.}
}
\ffigbox{%
    \includegraphics[width=1.00\linewidth]{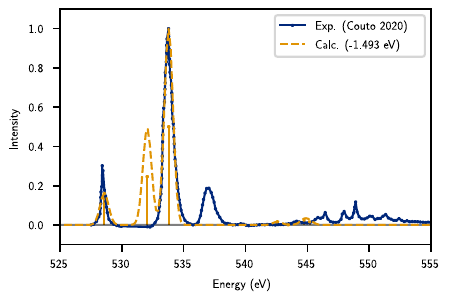}
}{%
    \caption{fc-CVS EOM-CCSD / aug-ccX-2 spectra of \ce{CO+} at the oxygen K-edge, normalized and shifted to match the experimental result.\cite{Couto2020}}
}
\end{floatrow}
\end{figure}

1C-NOCIS doublets,\cite{Oosterbaan2019} a model that includes all the configurations necessary for spin-adapted excited states for 1eOS doublets, provides a spectral profile in good quantitative agreement with the experiment (Figure S2). Like fc-CVS EOM-CCSD, 1C-NOCIS doublets also predicts a quartet \ce{O}$_{1s} \xrightarrow{} \pi^*$ transition at the same energy range but it has strictly zero oscillator strength because spin-selection rules are enforced, rendering the transition dark. A doublet radical with a SOMO in the $\sigma_{CO}$ orbital, the electronic ground state of \ce{CO+} is well described by a single configuration and the observed phenomena of erroneously-bright transitions is likely worse when employing a heavily spin contaminated reference in the first place, such as a single configuration for a singlet valence excited state obtained via $\Delta$SCF.

\begin{figure}[h!]
\begin{floatrow}
\capbtabbox{%
    \begin{tabular}{ccc}\toprule
    Energy (eV) & O. S. & $\braket{S^2}$ \\\midrule
    528.09 & 0.001 & 0.75 \\
    532.73 & 0.000 & 3.75 \\ 
    532.73 & 0.000 & 3.75 \\
    533.23 & 0.026 & 0.75 \\
    533.23 & 0.026 & 0.75 \\
    537.37 & 0.011 & 0.75 \\
    537.01 & 0.011 & 0.75 \\
    543.24 & 0.000 & 3.75 \\
    543.60 & 0.000 & 3.75 \\
    \bottomrule
    \end{tabular}
}{%
    \caption{Raw 1C-NOCIS / aug-pcX-2 transition energies, oscillator strengths, and $\braket{S^2}$ values of \ce{CO+} at the oxygen K-edge.}
}
\ffigbox{%
    \includegraphics[width=1.00\linewidth]{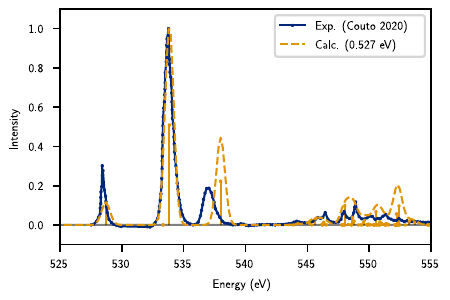}
}{%
    \caption{1C-NOCIS / aug-pcX-2 spectra of \ce{CO+} at the oxygen K-edge, normalized and shifted to match the experimental result.\cite{Couto2020}}
}
\end{floatrow}
\end{figure}
\pagebreak

\section{Derivation of orthogonal matrix elements}
We introduce some notational conventions for ease of derivations. With a closed-shell configuration $\kvac$ as a reference, we label the relevant spin orbital indices as follows:
\begin{itemize}
    \item o: \textbf{o}ccupied orbital associated with the pump valence excitation.
    \item t: \textbf{t}arget virtual associated with the pump valence excitation.
    \item c: \textbf{c}ore orbital associated with the probe excitation.
    \item x, y: arbitrary virtual orbital
\end{itemize}
The second-quantized version of the electronic Hamiltonian is written as
\begin{align}
    \mathcal{H} = f_{pq}\; p\dg q + \frac{1}{4} \bra{pq}\ket{rs}\; p\dg q\dg s r + E_0
\end{align}
where $f_{pq}$ are Fock matrix elements, $\bra{pq}\ket{rs}$ are two-electron integrals in Dirac or physicists notation, $p\dg$ and $q\dg$ are creation operators with respect to arbitrary spin orbitals, $s$ and $r$ are annihilation operators with respect to arbitrary spin orbitals, and $E_0$ is the energy of the reference. Note that we also have fully-unrestricted sums that are implicit here. The span on configurations relevant for the description of M$_s$ = 0 pump-probe excited states is 
\begin{align}
    \Big\{ \cka,\; \ckb,\; \ckc,\; \ckd,\; \cke,\; \ckf,\; \ckg,\; \ckh
           \Big\}
\end{align}
In their second-quantized form, and introducing an alphabetic label for convenience, the configurations are
\begin{align}
    &\Big\{\tka \kvac, \;
           \tkb \kvac, \;
           \tkc \kvac, \;
           \tkd \kvac, \;
           \tke \kvac, \;
           \tkf \kvac, \;
           \tkg \kvac, \;
           \tkh \kvac
           \Big\} \\
    &\Big\{ \ket{A}, \ket{B}, \ket{C}, \ket{D}, \ket{E}, \ket{F}, \ket{G}, \ket{H} \Big\}
\end{align}
We now derive the matrix elements for the Hamiltonian in the span, $\mathbf{H}$.
\begin{align}
    \mathbf{H} = 
    \begin{bmatrix}
     H_{AA} & H_{AB} & H_{AC} & H_{AD} & H_{AE} & H_{AF} & H_{AG} & H_{AH} \\
     H_{BA} & H_{BB} & H_{BC} & H_{BD} & H_{BE} & H_{BF} & H_{BG} & H_{BH} \\ 
     H_{CA} & H_{CB} & H_{CC} & H_{CD} & H_{CE} & H_{CF} & H_{CG} & H_{CH} \\ 
     H_{DA} & H_{DB} & H_{DC} & H_{DD} & H_{DE} & H_{DF} & H_{DG} & H_{DH} \\ 
     H_{EA} & H_{EB} & H_{EC} & H_{ED} & H_{EE} & H_{EF} & H_{EG} & H_{EH} \\ 
     H_{FA} & H_{FB} & H_{FC} & H_{FD} & H_{FE} & H_{FF} & H_{FG} & H_{FH} \\ 
     H_{GA} & H_{GB} & H_{GC} & H_{GD} & H_{GE} & H_{GF} & H_{GG} & H_{GH} \\ 
     H_{HA} & H_{HB} & H_{HC} & H_{HD} & H_{HE} & H_{HF} & H_{HG} & H_{HH} \\ 
    \end{bmatrix}
\end{align}
As a Hermitian operator, only the 36 upper-triangular matrix elements need to be calculated explicitly. for the specific configurations needed for a description of pump-probe excited states. We break up the Hamiltonian matrix elements into its one- and two-electron contributions.
\begin{align}
    \mathbf{H} = \mathbf{f} + \frac{1}{4} \mathbf{g}
\end{align}

\subsection{One-electron contributions: 4eOS Case}
A one-electron operator cannot couple configurations that differ by more than one orbital. Aside of the diagonal terms of $\mathbf{f}$, only a few off-diagonal terms are non-zero. We use the derivation of the f$_{CC}$  matrix elements to establish conventions to keep the notation as clean and simple as possible while retaining its completeness. We begin with the expression of the matrix element itself and make the relevant contractions
\begin{align}\label{eq:example_contractions}
    f_{CC} = \; &
f_{pq} \bvac \wick{
(o\dg c\dg \c2 x \c1 t) \cdot 
(\c1p\dg \c1q)          \cdot 
(\c1 t\dg \c2 y\dg c o)
} \kvac + \notag\\ &
f_{pq} \bvac \wick{
(o\dg c\dg \c1 x \c2 t) \cdot 
(\c1p\dg \c1q)          \cdot 
(\c1 t\dg \c2 y\dg c o)
} \kvac + \notag\\ &
f_{pq} \bvac \wick{
(o\dg c\dg \c2 x \c1 t) \cdot 
(\c1p\dg \c1q)          \cdot 
(\c2 t\dg \c1 y\dg c o)
} \kvac + \notag\\ &
f_{pq} \bvac \wick{
(o\dg c\dg \c1 x t)     \cdot 
(\c1p\dg \c1q)          \cdot 
(t\dg \c1 y\dg c o)
} \kvac + \notag\\ &
f_{pq} \bvac \wick{
(o\dg \c2c\dg \c3 x t)  \cdot 
(\c1p\dg \c2q)          \cdot 
(t\dg \c3y\dg \c1 c o)
} \kvac + \notag\\ &
f_{pq} \bvac \wick{
(o\dg \c2c\dg \c4 x \c3 t) \cdot 
(\c1p\dg \c2q)             \cdot 
(\c4 t\dg \c3y\dg \c1 c o)
} \kvac + \notag\\ &
f_{pq} \bvac \wick{
(\c2 o\dg c\dg \c3 x t) \cdot 
(\c1p\dg \c2q)          \cdot 
(t\dg \c3y\dg c \c1 o)
} \kvac + \notag\\ &
f_{pq} \bvac \wick{
(\c2 o\dg c\dg \c4 x \c3 t) \cdot 
(\c1p\dg \c2q)          \cdot 
(\c4 t\dg \c3y\dg c \c1 o)
} \kvac
\end{align}
We avoid explicitly showing the trivial contractions. For example, in the seventh term, the only operator with which $c\dg$ can contract with is $c$. The sign of the seventh term is already clear from the non-trivial contractions shown: this should carry a negative sign due to the odd number (three) of line crossings, and explicitly adding the remaining contractions - $\wick{\c c \; \c c\dg}$ and $\wick{\c {t} \; \c {t}\dg}$ - will only add an even number of line crossings and thus not affect the sign of the term. We do explicitly show the contraction $\wick{\c x \; \c y\dg}$ when it occurs because the ensuing $\delta_{xy}$ will not be used to kill the arbitrary summations associated with the operators - $\sum_{pq}$ for $f_{pq}$ and $\sum_{pqrs}$ for $\bra{pq} \ket{rs}$, it will survive and it will imply that those terms should only be added to the diagonal of the Hamiltonian. Eq. \ref{eq:example_contractions} simplifies to 
\begin{align}\label{eq:example_intermediate}
    f_{CC}= \;
&+ f_{pq} \; (\delta_{pt} \; \delta_{qt}  \; \delta_{xy}) \; \notag\\
&- f_{pq} \; (\delta_{px} \; \delta_{qt}  \; \delta_{ty}) \; \notag\\
&- f_{pq} \; (\delta_{pt} \; \delta_{qy}  \; \delta_{xt}) \; \notag\\
&+ f_{pq} \; (\delta_{px} \; \delta_{qy}) \; \notag\\
&- f_{pq} \; (\delta_{pc} \; \delta_{qc}  \; \delta_{xy}) \; \notag\\
&+ f_{pq} \; (\delta_{pc} \; \delta_{qc}  \; \delta_{xt}\delta_{ty}) \; \notag\\
&- f_{pq} \; (\delta_{po} \; \delta_{qo}  \; \delta_{xy}) \; \notag\\
&+ f_{pq} \; (\delta_{po} \; \delta_{qo}  \; \delta_{xt}\delta_{ty}) 
\end{align}
which, after using the $\delta$'s associated with p and q to kill the summations, and retaining the $\delta_{xy}$, further simplifies to 
\begin{align}\label{eq:final_fock}
    f_{CC} = \;
&f_{xy} - \delta_{ty} f_{xt} - \delta_{xt} f_{ty} + \delta_{xy} (f_{tt} - f_{cc} - f_{oo}) + \delta_{xt}\delta_{ty} (f_{cc} + f_{oo})
\end{align}
The final step is to integrate spin out. In the case of $f_{CC}$ this is trivial since all orbitals involved have the same spin and the resulting expression is identical. By virtue of corresponding to the matrix element of the spin compliment of $\ket{C} = \tkc \kvac$, the matrix element $f_{DD}$ will be identical after repeating the above procedure. We repeat the procedure of carrying out the contractions, resolving the $\delta$'s, and integrating spin out for $f_{EE}$
\begin{align}
    f_{EE} = \; &
f_{pq} \bvac \wick{
(\bar{o}\dg c\dg \c2 x \c1 {\bar{t}}) \cdot 
(\c1p\dg \c1q)                        \cdot 
(\c1 {\bar{t}\dg} \c2 y\dg c \bar{o})
} \kvac + \notag\\ &
f_{pq} \bvac \wick{
(\bar{o}\dg c\dg \c1 x {\bar{t}}) \cdot 
(\c1p\dg \c1q)                        \cdot 
({\bar{t}\dg} \c1 y\dg c \bar{o})
} \kvac + \notag\\ &
f_{pq} \bvac \wick{
({\bar{o}\dg} \c2 c\dg \c3 x {\bar{t}}) \cdot 
(\c1p\dg \c2q)                        \cdot 
({\bar{t}\dg} \c3 y\dg \c1 c \bar{o})
} \kvac + \notag\\ &
f_{pq} \bvac \wick{
(\c2 {\bar{o}\dg} c\dg \c3 x {\bar{t}}) \cdot 
(\c1p\dg \c2q)                        \cdot 
({\bar{t}\dg} \c3 y\dg c \c1 {\bar{o}})
} \kvac \notag\\ = &
f_{xy} + \delta_{xy} (f_{tt} - f_{cc} - f_{oo})
\end{align}
This time we avoid contractions that are necessarily zero due to orthogonality of spin orbitals in different spin spaces, such as $\wick{\c {\bar{t}} \; \c {y}\dg}$. The terms $f_{FF}$, $f_{GG}$, $f_{HH}$ give the exact same result as $f_{EE}$
\begin{align}
    f_{EE} = & f_{FF} = f_{GG} = f_{HH} \notag\\ = & 
               f_{xy} + \delta_{xy} (f_{tt} - f_{cc} - f_{oo})
\end{align}
The simpler terms $f_{AA}$ and $f_{BB}$, equivalent by virtue of corresponding to spin compliments, are 
\begin{align}
    f_{AA}, f_{BB} = &
    f_{pq} \bvac \wick{
    (c\dg \c1 t)   \cdot 
    (\c1p\dg \c1q) \cdot 
    (\c1 t\dg c )
    } \kvac + \notag\\ & 
    f_{pq} \bvac \wick{
    (\c2 c\dg  t)   \cdot 
    (\c1 p\dg \c2 q) \cdot 
    (t\dg \c1 c )
    } \kvac \\ = &
    f_{tt} - f_{cc}
\end{align}
Now, working out the non-zero non-diagonal matrix elements. First, the ones coupling 2eOS configurations with 4eOS configurations.
\begin{align}
    f_{AC} = \; &
f_{pq} \bvac \wick{
(\c4 c\dg \c3 t) \cdot 
(\c2 p\dg \c1 q)                        \cdot 
(\c1 t\dg \c3 y\dg \c4 c \c2 o)
} \kvac + \notag\\ &
f_{pq} \bvac \wick{
(\c4 c\dg \c3 t) \cdot 
(\c2 p\dg \c1 q)                        \cdot 
(\c3 t\dg \c1 y\dg \c4 c \c2 o)
} \kvac \notag\\ = &
\delta_{ty} f_{ot} - f_{oy} \\
    f_{AE} = \; &
f_{pq} \bvac \wick{
(\c4 c\dg \c3 t) \cdot 
(\c2 p\dg \c1 q)                        \cdot 
(\c1 {\bar{t}\dg} \c3 y\dg \c4 c \c2 {\bar{o}})
} \kvac + \notag\\ = &
\delta_{ty} f_{ot} \\
    f_{AG} = \; &
f_{pq} \bvac \wick{
(\c4 c\dg \c3 t) \cdot 
(\c2 p\dg \c1 q)                        \cdot 
(\c1 {\bar{y}\dg} \c3 t\dg \c4 c \c2 {\bar{o}})
} \kvac + \notag\\ = &
f_{oy}
\end{align}
Second, the ones corresponding to those of a 4eOS configuration with its ``anti-spin compliment''. For example:
\begin{align}
    f_{EG} = \; &
f_{pq} \bvac \wick{
(\bar{o}\dg c\dg \c2 x \c1 {\bar{t}}) \cdot 
(\c1p\dg \c1q)                        \cdot 
(\c1 {\bar{y}\dg} \c2 t\dg c \bar{o})
} \kvac + \notag\\ &
f_{pq} \bvac \wick{
(\bar{o}\dg c\dg \c1 x \c2 {\bar{t}}) \cdot 
(\c1p\dg \c1q)                        \cdot 
(\c2 {\bar{y}\dg} \c1 t\dg c \bar{o})
} \kvac + \notag\\ &
f_{pq} \bvac \wick{
(\bar{o}\dg \c4 c\dg \c3 x \c2 {\bar{t}}) \cdot 
(\c1p\dg \c4q)                        \cdot 
(\c2 {\bar{y}\dg} \c3 t\dg \c1 c \bar{o})
} \kvac + \notag\\ = &
f_{pq} \bvac \wick{
(\c4 {\bar{o}\dg} c\dg \c3 x \c2 {\bar{t}}) \cdot 
(\c1p\dg \c4q)                        \cdot 
(\c2 {\bar{y}\dg} \c3 t\dg c \c1 {\bar{o}})
} \kvac \notag\\ = &
\delta_{xt} f_{ty} + \delta_{ty} f_{xt} - \delta_{xt}\delta_{ty} (f_{cc} + f_{oo})
\end{align}
All of these cases will be the same and accounting for the spin-compliment matrix elements is trivial.
\begin{align}
    f_{EG} = f_{GE} = f_{FH} = F_{HF}
\end{align}
We collect all the non-zero results below.
\begin{align*}
    f_{AA} = & f_{tt} - f_{cc}\\
    f_{AC} = & \delta_{ty} f_{ot} - f_{oy}\\
    f_{AE} = & \delta_{ty} f_{ot} \\
    f_{AG} = & f_{oy}
    f_{BB} = & f_{tt} - f_{cc}\\
    f_{CC} = & f_{xy} - \delta_{ty} f_{xt} - \delta_{xt} f_{ty} + \delta_{xy} (f_{tt} - f_{cc} - f_{oo}) + \delta_{xt}\delta_{ty} (f_{cc} + f_{oo})\\
    f_{DD} = & f_{xy} - \delta_{ty} f_{xt} - \delta_{xt} f_{ty} + \delta_{xy} (f_{tt} - f_{cc} - f_{oo}) + \delta_{xt}\delta_{ty} (f_{cc} + f_{oo})\\
    f_{EE} = & f_{xy} + \delta_{xy} (f_{tt} - f_{cc} - f_{oo})\\
    f_{EG} = & \delta_{xt} f_{ty} + \delta_{ty} f_{xt} - \delta_{xt}\delta_{ty} (f_{cc} + f_{oo})\\
    f_{FF} = & f_{xy} + \delta_{xy} (f_{tt} - f_{cc} - f_{oo})\\
    f_{FH} = & \delta_{xt} f_{ty} + \delta_{ty} f_{xt}\\
    f_{GE} = & \delta_{xt} f_{ty} + \delta_{ty} f_{xt} - \delta_{xt}\delta_{ty} (f_{cc} + f_{oo})\\
    f_{GG} = & f_{xy} + \delta_{xy} (f_{tt} - f_{cc} - f_{oo})\\
    f_{HF} = & \delta_{xt} f_{ty} + \delta_{ty} f_{xt} - \delta_{xt}\delta_{ty} (f_{cc} + f_{oo})\\
    f_{HH} = & f_{xy} + \delta_{xy} (f_{tt} - f_{cc} - f_{oo})\\
\end{align*}
\subsection{Two-electron contributions: 4eOS Case}
The two-electron contributions become more complicated because a two-electron operator can couple configurations that differ by two orbitals. We make use of the derivation conventions employed in the one-electron contribution sub-section. For ease of visualization, we write out the matrix $\mathbf{g}$ with the explicit configurations. \\
%\resizebox{\textwidth}{!}{%
\begin{equation}
\begin{blockarray}{ccccccccc}
& A & B & C & D & E & F & G & H \\
\begin{block}{c[cccccccc]}
     A & \cba g \cka & \cba g \ckb & \cba g \ckc & \cba g \ckd & \cba g \cke & \cba g \ckf & \cba g \ckg & \cba g \ckh \\
     B & & \cbb g \ckb & \cbb g \ckc & \cbb g \ckd & \cbb g \cke & \cbb g \ckf & \cbb g \ckg & \cbb g \ckh \\
     C & & & \cbc g \ckc & \cbc g \ckd & \cbc g \cke & \cbc g \ckf & \cbc g \ckg & \cbc g \ckh \\
     D & & & & \cbd g \ckd & \cbd g \cke & \cbd g \ckf & \cbd g \ckg & \cbd g \ckh \\
     E & & & & & \cbe g \cke & \cbe g \ckf & \cbe g \ckg & \cbe g \ckh \\
     F & & & & & & \cbf g \ckf & \cbf g \ckg & \cbf g \ckh \\
     G & & & & & & & \cbg g \ckg & \cbg g \ckh \\
     H & & & & & & & & \cbh g \ckh \notag\\
\end{block}
\end{blockarray} 
\end{equation} 
%} \\

First, we make full use of the equivalence of matrix elements arising from spin-compliments, which are
\begin{align*}
    \ket{A} = \cka &\longleftrightarrow \ket{B} = \ckb \\
    \ket{C} = \ckc &\longleftrightarrow \ket{D} = \ckd \\
    \ket{E} = \cke &\longleftrightarrow \ket{F} = \ckf \\
    \ket{G} = \ckg &\longleftrightarrow \ket{H} = \ckh 
\end{align*}
The following matrix elements are equivalent after integrating spin out.
\begin{align*}
    g_{AA} & = g_{BB} & g_{CC} & = g_{DD} & g_{EE} & = g_{FF} & g_{GG} & = g_{HH} \\
    g_{AC} & = g_{BD} & g_{CE} & = g_{DF} & g_{EG} & = g_{FH} & & \\
    g_{AD} & = g_{BC} & g_{CF} & = g_{DE} & g_{EH} & = g_{FG} & & \\
    g_{AE} & = g_{BF} & g_{CG} & = g_{DH} & & & & \\
    g_{AF} & = g_{BE} & g_{CH} & = g_{DG} & & & & \\
    g_{AG} & = g_{BH} & & & & & & \\
    g_{AH} & = g_{BG} & & & & & & 
\end{align*}
Now, we make use of the fact that a two-electron operator cannot couple configurations that differ by three orbitals or more.
\begin{align*}
    g_{AD} & = 0  & g_{CD} & = 0 
\end{align*}

The remaining unique contributions that must be derived explicitly are the following

%\resizebox{\textwidth}{!}{%
\begin{equation}
\begin{blockarray}{ccccccccc}
& A & B & C & D & E & F & G & H \\
\begin{block}{c[cccccccc]}
     A & \cba g \cka & \cba g \ckb & \cba g \ckc & & \cba g \cke & \cba g \ckf & \cba g \ckg & \cba g \ckh \\
     B & & & & & & & & \\
     C & & & \cbc g \ckc & & \cbc g \cke & \cbc g \ckf & \cbc g \ckg & \cbc g \ckh \\
     D & & & & & & & & \\
     E & & & & & \cbe g \cke & \cbe g \ckf & \cbe g \ckg & \cbe g \ckh \\
     F & & & & & & & & \\
     G & & & & & & & \cbg g \ckg & \cbg g \ckh \\
     H & & & & & & & & \notag\\
\end{block}
\end{blockarray} 
\end{equation} 
%} \\
We employ the same procedure used for the one-electron integrals 
\begin{itemize}
    \item Carry out contraction, resolve the $\delta$'s, and integrate spin out in one step
    \item Avoid writing trivial contractions such as $\wick{\c o \; \c o\dg}$, $\wick{\c c \; \c c\dg}$ and $\wick{\c {t} \; \c {t}\dg}$ \textbf{when it is abundantly clear that they        make an even contribution to the number of lines-crossings.}
    \item Avoid making un-necessary contraction that are zero due to connecting operators of different spin, such as $\wick{\c {\bar{t}} \; \c {y}\dg}$.
\end{itemize}
As an additional derivation speed-up tools, we employ the symmetry of the two-electron  integrals to only write one of four equivalent contributions and introduce the classification of contractions into unique ``Families'' for book-keeping. These two tools will be explained through deriving the contribution g$_{EE}$ in detail. We begin with Family 1: both $p\dg$ and $q\dg$ contract to the left, and $s$ and $r$ contract to the right. The family will be denoted with a super-script.
\begin{align}
    g_{EE}^{F1} = \;
    &\bra{pq} \ket{rs} \; \bvac \wick{
({\bar{o}}\dg c\dg \c2 x \c1 {\bar{t}}) \cdot 
(\c1 p\dg \c2 q\dg \c2 s \c1 r)         \cdot 
(\c1 {\bar{t}}\dg \c2 y\dg c {\bar{o}})
} \kvac + \notag\\[12pt]
    &\bra{pq} \ket{rs} \; \bvac \wick{
({\bar{o}}\dg c\dg \c1 x \c2 {\bar{t}}) \cdot 
(\c1 p\dg \c2 q\dg \c2 s \c1 r)         \cdot 
(\c1 {\bar{t}}\dg \c2 y\dg c {\bar{o}})
} \kvac + \notag\\[12pt]
    &\bra{pq} \ket{rs} \; \bvac \wick{
({\bar{o}}\dg c\dg \c2 x \c1 {\bar{t}}) \cdot 
(\c1 p\dg \c2 q\dg \c2 s \c1 r)         \cdot 
(\c2 {\bar{t}}\dg \c1 y\dg c {\bar{o}})
} \kvac + \notag\\[12pt]
    &\bra{pq} \ket{rs} \; \bvac \wick{
({\bar{o}}\dg c\dg \c1 x \c2 {\bar{t}}) \cdot 
(\c1 p\dg \c2 q\dg \c2 s \c1 r)         \cdot 
(\c2 {\bar{t}}\dg \c1 y\dg c {\bar{o}})
} \kvac
\end{align}
This results in 
\begin{align*}
    g_{EE}^{F1} = \; &
\bra{\bar{t} x} \ket{\bar{t} y} - \bra{x \bar{t}} \ket{\bar{t} y} - 
\bra{\bar{t} x} \ket{y \bar{t}} + \bra{x \bar{t}} \ket{y \bar{t}} 
\end{align*}
These four terms are identical (with the same sign) due to the nature of the anti-symmetrized two-electron integrals. Adding the four contributions and integrating spin out,
\begin{align*}
    g_{EE}^{F1} = \; &
    4 \cdot \bra{\bar{t} x} \ket{\bar{t} y} \\
    = \; & 
    4 \cdot \braket{t x | t y}
\end{align*}
Family 2 corresponds to both $p \dg$ and $q \dg$ contracting to the left, and $s$ and $r$ contracting to the right. Implying the factor of four previously explained, and just showing the unique non-zero contraction contraction:
\begin{align}
    g_{EE}^{F2} = \; 
    & \bra{pq} \ket{rs} \; \bvac \wick{
(\c4 {\bar{o}}\dg \c3 c\dg \c5 x {\bar{t}}) \cdot 
(\c2 p\dg \c1 q\dg \c3 s \c4 r)         \cdot 
({\bar{t}}\dg \c5 y\dg \c1 c \c2 {\bar{o}})
} \kvac \notag\\
= \; &\bra{\bar{o}c} \ket{\bar{o}c} \; \delta_{xy} \notag\\
= \; &\braket{oc|oc} \; \delta_{xy} 
\end{align}
For Family 3, we allow for either $p\dg$ or $q\dg$ to contract either to the right or to the left, but not to the same side. The same condition holds for $s$ and $r$. This one is the broadest, in that it will result in four distinct kinds of two electron integrals (each four-fold degenerate)
\begin{align}
    g_{EE}^{F3} = \;
    &\bra{pq} \ket{rs} \; \bvac \wick{
({\bar{o}}\dg \c2 c\dg \c4 x \c1 {\bar{t}}) \cdot 
(\c1 p\dg \c3 q\dg \c2 s \c1 r)             \cdot 
(\c1 {\bar{t}}\dg \c4 y\dg \c3 c {\bar{o}})
} \kvac + \notag\\
    &\bra{pq} \ket{rs} \; \bvac \wick{
({\bar{o}}\dg \c2 c\dg \c1 x  {\bar{t}}) \cdot 
(\c1 p\dg \c3 q\dg \c2 s \c1 r)             \cdot 
({\bar{t}}\dg \c1 y\dg \c3 c {\bar{o}})
} \kvac + \notag\\
    &\bra{pq} \ket{rs} \; \bvac \wick{
(\c2 {\bar{o}}\dg c\dg \c1 x  {\bar{t}}) \cdot 
(\c1 p\dg \c3 q\dg \c2 s \c1 r)             \cdot 
({\bar{t}}\dg \c1 y\dg  c \c3 {\bar{o}})
} \kvac + \notag\\
    &\bra{pq} \ket{rs} \; \bvac \wick{
(\c2 {\bar{o}}\dg c\dg \c4 x \c1 {\bar{t}}) \cdot 
(\c1 p\dg \c3 q\dg \c2 s \c1 r)             \cdot 
(\c1 {\bar{t}}\dg \c4 y\dg  c \c3 {\bar{o}})
} \kvac \notag\\
= \; &
- \bra{\bar{t}c}\ket{\bar{t}c} \; \delta_{xy} -
  \bra{xc}\ket{yc} - 
  \bra{x\bar{o}} \ket{y\bar{o}} -
  \bra{\bar{t}\bar{o}} \ket{\bar{t}\bar{o}} \; \delta_{xy} \notag\\
= \; &
- \braket{tc | tc} \; \delta_{xy} -
  \braket{xc | yc} + 
  \braket{xc | cy} - 
  \braket{xo | yo} -
  \braket{to | to} \; \delta_{xy} + 
  \braket{to | ot} \; \delta_{xy}\notag\\
\end{align}
The whole term $g_{EE}$, after collecting the contributions from the three families and accounting for the four-fold degeneracy of each integral is
\begin{align*}
    \frac{1}{4} g_{EE} = 
    \braket{t x | t y} - \braket{xc | yc} + \braket{xc | cy} - \braket{xo | yo} + \\ 
    \delta_{xy} \big( 
    \braket{o c | o c} - \braket{tc | tc} - \braket{to | to} + \braket{to | ot}
    \big)
\end{align*}
There are eight distinct terms in total. Four of them carry a $\delta_{xy}$ with them. The two terms associated with only virtuals or only occupieds, $\braket{t x | t y}$ $\braket{o c | o c}$ - carry a positive sign. Interestingly, for two contributions, $\bra{xc}\ket{yc}$ and $\bra{\bar{t}\bar{o}} \ket{\bar{t}\bar{o}}$ both the Coulomb and exchange terms survive after integrating spin out. We now derive the rest of the terms.

\subsubsection{$g_{AA} = \cba g \cka$}
\begin{align*}
    g_{AA} = \;
&\bra{pq} \ket{rs} \; \bvac \wick{
(\c3 c\dg \c1 t) \cdot 
(\c1 p\dg \c2 q\dg \c3 s \c1 r)         \cdot 
(\c1 t\dg \c2 c)
} \kvac  \\ = & - \bra{tc} \ket{tc} 
          \\ = & \braket{tc | ct} - \braket{tc | tc} 
\end{align*}

\subsubsection{$g_{AB} = \cba g \ckb$}
\begin{align*}
    g_{AB} = \;
&\bra{pq} \ket{rs} \; \bvac \wick{
(\c3 c\dg \c1 t) \cdot 
(\c1 p\dg \c2 q\dg \c3 s \c1 r)         \cdot 
(\c1 {\bar{t}\dg} \c2 {\bar{c}})
} \kvac  \\ = & - \bra{t\bar{c}} \ket{\bar{t}c}
          \\ = & \braket{tc | ct}
\end{align*}

\subsubsection{$g_{AC} = \cba g \ckc$}
\begin{align*}
    g_{AC} = \; &
\bra{pq} \ket{rs} \; \bvac \wick{
(\c4 c\dg \c1 t) \cdot 
(\c1 p\dg \c3 q\dg \c2 s \c1 r)         \cdot 
(\c1 t\dg \c2 y\dg \c4 c \c3 o)
} \kvac + \\ &
\bra{pq} \ket{rs} \; \bvac \wick{
(\c1 c\dg \c4 t) \cdot 
(\c3 p\dg \c2 q\dg \c1 s \c1 r)         \cdot 
(\c1 t\dg \c4 y\dg \c2 c \c3 o)
} \kvac + \\ &
\bra{pq} \ket{rs} \; \bvac \wick{
(\c1 c\dg \c4 t) \cdot 
(\c3 p\dg \c2 q\dg \c1 s \c1 r)         \cdot 
(\c4 t\dg \c1 y\dg \c2 c \c3 o)
} \kvac \\ = & 
- \bra{to} \ket{ty} - \delta_{ty} \bra{oc} \ket{tc} + \bra{oc} \ket{yc} \\ = &
\braket{oc | yc} - \braket{oc | cy} 
- \braket{to | ty} + \braket{to | yt} + \delta_{yt} \big(
   \braket{oc | ct} -
   \braket{oc | tc} \big)
\end{align*}

\subsubsection{$g_{AD} = \cba g \ckd$}
\begin{align*}
    g_{AD} = \; 0
\end{align*}
\subsubsection{$g_{AE} = \cba g \cke$}
\begin{align*}
    g_{AE} = \; &
\bra{pq} \ket{rs} \; \bvac \wick{
(\c4 c\dg \c1 t) \cdot 
(\c1 p\dg \c3 q\dg \c2 s \c1 r)         \cdot 
(\c1 {\bar{t}}\dg \c2 y\dg \c4 c \c3 {\bar{o}})
} \kvac + \\ & 
\bra{pq} \ket{rs} \; \bvac \wick{
(\c1 c\dg \c4 t) \cdot 
(\c3 p\dg \c2 q\dg \c1 s \c1 r)         \cdot 
(\c1 {\bar{t}}\dg \c4 y\dg \c2 c \c3 {\bar{o}})
} \kvac \\ = & 
- \bra{t\bar{o}} \ket{\bar{t}y} - \delta_{ty} \bra{\bar{o}c} \ket{\bar{t}c} \\ = &
\braket{to | yt} - \delta_{ty} \braket{oc | tc}
\end{align*}

\subsubsection{$g_{AF} = \cba g \ckf$}
\begin{align*}
    g_{AF} = \; &
\bra{pq} \ket{rs} \; \bvac \wick{
(\c1 c\dg \c4 t) \cdot 
(\c3 p\dg \c2 q\dg \c1 s \c1 r)         \cdot 
(\c4 t\dg \c1 {\bar{y}}\dg \c2 {\bar{c}} \c3 o)
} \kvac \\ = & 
\bra{o\bar{c}} \ket{\bar{y}c} \\ = & 
 - \braket{oc | cy}
\end{align*}

\subsubsection{$g_{AG} = \cba g \ckg$}
\begin{align*}
    g_{AG} = \; &
\bra{pq} \ket{rs} \; \bvac \wick{
(\c4 c\dg \c1 t) \cdot 
(\c1 p\dg \c3 q\dg \c2 s \c1 r)         \cdot 
(\c1 {\bar{y}}\dg \c2 t\dg \c4 c \c3 {\bar{o}})
} \kvac + \\ & 
\bra{pq} \ket{rs} \; \bvac \wick{
(\c1 c\dg \c4 t) \cdot 
(\c3 p\dg \c2 q\dg \c1 s \c1 r)         \cdot 
(\c1 {\bar{y}}\dg \c4 t\dg \c2 c \c3 {\bar{o}})
} \kvac \\ = & 
- \bra{t\bar{o}} \ket{\bar{y}t} - \bra{\bar{o}c} \ket{\bar{y}c} \\ = &
\braket{to | ty} - \braket{oc | yc}
\end{align*}

\subsubsection{$g_{AH} = \cba g \ckh$}
\begin{align*}
    g_{AH} = \; &
\bra{pq} \ket{rs} \; \bvac \wick{
(\c1 c\dg \c4 t) \cdot 
(\c3 p\dg \c2 q\dg \c1 s \c1 r)         \cdot 
(\c4 y\dg \c1 {\bar{t}}\dg \c2 {\bar{c}} \c3 o)
} \kvac \\ = & 
\delta_{ty} \bra{o\bar{c}} \ket{\bar{t}c} \\ = &
 - \delta_{ty} \braket{oc | ct}
\end{align*}

\subsubsection{$g_{CC} = \cbc g \ckc$}
\begin{align*}
    g_{CC}^{F1} = 
&\bra{pq} \ket{rs} \; \bvac \wick{
(o\dg c\dg \c2 x \c1 t) \cdot 
(\c1 p\dg \c2 q\dg \c2 s \c1 r)         \cdot 
(\c1 t\dg \c2 y\dg c o)
} \kvac \\
             = \; & \bra{t x} \ket{t y} \\[12 pt]
    g_{CC}^{F2} = 
&\bra{pq} \ket{rs} \; \bvac \wick{
(\c4 o\dg \c3 c\dg \c5 x t) \cdot 
(\c2 p\dg \c1 q\dg \c3 s \c4 r)         \cdot 
(t\dg \c5 y\dg \c1 c \c2 o)
} \kvac + \\
&\bra{pq} \ket{rs} \; \bvac \wick{
(\c4 o\dg \c3 c\dg \c5 x \c6 t) \cdot 
(\c2 p\dg \c1 q\dg \c3 s \c4 r)         \cdot 
(\c5 t\dg \c6 y\dg \c1 c \c2 o)
} \kvac \\
             = \; & \bra{o c} \ket{o c} \; \delta_{xy} - 
                    \bra{o c} \ket{o c} \; \delta_{xt} \delta_{yt}\\[12 pt]
    g_{CC}^{F3} = \;
&\bra{pq} \ket{rs} \; \bvac \wick{
(o\dg \c3 c\dg \c4 x \c1 t) \cdot 
(\c1 p\dg \c2 q\dg \c3 s \c1 r)         \cdot 
(\c1 t\dg \c4 y\dg \c2 c o)
} \kvac + \\[12pt]
&\bra{pq} \ket{rs} \; \bvac \wick{
(o\dg \c3 c\dg \c4 x \c1 t) \cdot 
(\c1 p\dg \c2 q\dg \c3 s \c1 r)         \cdot 
(\c4 t\dg \c1 y\dg \c2 c o)
} \kvac + \\[12pt]
&\bra{pq} \ket{rs} \; \bvac \wick{
(o\dg \c3 c\dg \c1 x t) \cdot 
(\c1 p\dg \c2 q\dg \c3 s \c1 r)         \cdot 
(t\dg \c1 y\dg \c2 c o)
} \kvac + \\[12pt]
&\bra{pq} \ket{rs} \; \bvac \wick{
(o\dg \c3 c\dg \c1 x \c4 t) \cdot 
(\c1 p\dg \c2 q\dg \c3 s \c1 r)         \cdot 
(\c1 t\dg \c4 y\dg \c2 c o)
} \kvac + \\[12pt]
&\bra{pq} \ket{rs} \; \bvac \wick{
(\c3 o\dg c\dg \c1 x t) \cdot 
(\c1 p\dg \c2 q\dg \c3 s \c1 r)         \cdot 
(t\dg \c1 y\dg c \c2 o)
} \kvac + \\[12pt]
&\bra{pq} \ket{rs} \; \bvac \wick{
(\c3 o\dg c\dg \c1 x \c4 t) \cdot 
(\c1 p\dg \c2 q\dg \c3 s \c1 r)         \cdot 
(\c1 t\dg \c4 y\dg c \c2 o)
} \kvac + \\[12pt]
&\bra{pq} \ket{rs} \; \bvac \wick{
(\c3 o\dg c\dg \c4 x \c1 t) \cdot 
(\c1 p\dg \c2 q\dg \c3 s \c1 r)         \cdot 
(\c1 t\dg \c4 y\dg c \c2 o)
} \kvac + \\[12 pt]
&\bra{pq} \ket{rs} \; \bvac \wick{
(\c3 o\dg c\dg \c4 x \c1 t) \cdot 
(\c1 p\dg \c2 q\dg \c3 s \c1 r)         \cdot 
(\c4 t\dg \c1 y\dg c \c2 o)
} \kvac \\
=
&- \bra{t c} \ket{t c} \; \delta_{xy} +
   \bra{t c} \ket{y c} \; \delta_{xt} -
   \bra{x c} \ket{y c} + 
   \bra{x c} \ket{t c} \; \delta_{yt}  \\
&-  \bra{x o} \ket{y o} + 
   \bra{x o} \ket{t o} \; \delta_{yt} -
   \bra{t o} \ket{t o} \; \delta_{xy} +
   \bra{t o} \ket{y o} \; \delta_{xt}
\end{align*}
\begin{align*}
\frac{1}{4} g_{CC} = & \; g_{CC}^{F1} + g_{CC}^{F2} + g_{CC}^{F3} \\
             = & \; 
\bra{tx}\ket{ty} - \bra{xc}\ket{yc} - \bra{xo}\ket{yo} + \\ & 
\delta_{xt} \big( 
\bra{tc}\ket{yc} + \bra{to}\ket{yo} \big) \\ & 
\delta_{yt} \big(
\bra{xc}\ket{tc} + \bra{xo}\ket{to} \big) + \\ & 
\delta_{xy} \big( 
\bra{oc}\ket{oc} - \bra{tc}\ket{tc} - \bra{to} \ket{to} \big) - \\ &
\delta_{xt} \delta_{yt} \bra{oc}\ket{oc}
\end{align*}

\subsubsection{$g_{CE} = \cbc g \cke$}
\begin{align*}
    g_{CE}^{F1} = \; &0 \\
    g_{CE}^{F2} = \; &0 \\
    g_{CE}^{F3} = \;
&\bra{pq} \ket{rs} \; \bvac \wick{
(\c3 o\dg c\dg \c4 x \c1 t) \cdot 
(\c1 p\dg \c2 q\dg \c3 s \c1 r)         \cdot 
(\c1 {\bar{t}}\dg \c4 y\dg c \c2 {\bar{o}})
} \kvac + \\
&\bra{pq} \ket{rs} \; \bvac \wick{
(\c3 o\dg c\dg \c1 x \c4 t) \cdot 
(\c1 p\dg \c2 q\dg \c3 s \c1 r)         \cdot 
(\c1 {\bar{t}}\dg \c4 y\dg c \c2 {\bar{o}})
} \kvac \\
             = \; & - \bra{t \bar{o}} \ket{\bar{t} o} \; \delta_{xy} + 
                      \bra{x \bar{o}} \ket{\bar{t} o} \; \delta_{yt} \\
g_{CE} = \; & g_{CE}^{F1} + g_{CE}^{F2} + g_{CE}^{F3} \\
             = \; & \delta_{xy} \braket{to | ot} - 
                    \delta_{yt} \braket{xo | ot}
\end{align*}

\subsubsection{$g_{CF} = \cbc g \ckf$}
\begin{align*}
    g_{CF}^{F1} = \; & 0 \\
    g_{CF}^{F2} = \; & 0 \\
    g_{CF}^{F3} = \;
&\bra{pq} \ket{rs} \; \bvac \wick{
(o\dg \c3 c\dg \c1 x t) \cdot 
(\c1 p\dg \c2 q\dg \c3 s \c1 r) \cdot 
(t \dg \c1 {\bar{y}}\dg \c2 {\bar{c}} o)
} \kvac + \\
&\bra{pq} \ket{rs} \; \bvac \wick{
(o\dg \c3 c\dg \c4 x \c1 t) \cdot 
(\c1 p\dg \c2 q\dg \c3 s \c1 r) \cdot 
(\c4 t \dg \c1 {\bar{y}}\dg \c2 {\bar{c}} o)
} \kvac \\
             = \; & - \bra{x \bar{c}} \ket{\bar{y} c} 
                    + \bra{t \bar{c}} \ket{\bar{y} c} \delta_{xt} \\
g_{CF} = \; & g_{CF}^{F1} + g_{CF}^{F2} + g_{CF}^{F3}    \\
             = \; & \braket{xc | cy} - \delta_{xt}
                    \braket{tc | cy} 
\end{align*}

\subsubsection{$g_{CG} = \cbc g \ckg$}
\begin{align*}
    g_{CG}^{F1} = \; & 0 \\
    g_{CG}^{F2} = \; & 0 \\
    g_{CG}^{F3} = \;
&\bra{pq} \ket{rs} \; \bvac \wick{
(\c3 o\dg c\dg \c1 x \c4 t) \cdot 
(\c1 p\dg \c2 q\dg \c3 s \c1 r)         \cdot 
(\c1 {\bar{y}}\dg \c4 t\dg c \c2 {\bar{o}})
} \kvac + \\
&\bra{pq} \ket{rs} \; \bvac \wick{
(\c3 o\dg c\dg \c4 x \c1 t) \cdot 
(\c1 p\dg \c2 q\dg \c3 s \c1 r)         \cdot 
(\c1 {\bar{y}}\dg \c4 t\dg c \c2 {\bar{o}})
} \kvac \\
             = \; & \bra{x \bar{o}} \ket{\bar{y} o} - 
                    \bra{t \bar{o}} \ket{\bar{y} o} \delta_{xt} \\
g_{CG} = \; & g_{CG}^{F1} + g_{CG}^{F2} + g_{CG}^{F3}    \\
             = \; & - \braket{xo | oy} 
                    + \delta_{xt} \braket{to | oy}
\end{align*}

\subsubsection{$g_{CH} = \cbc g \ckh$}
\begin{align*}
    g_{CH}^{F1} = \; &0 \\
    g_{CH}^{F2} = \; &0 \\
    g_{CH}^{F3} = \;
&\bra{pq} \ket{rs} \; \bvac \wick{
(o\dg \c3 c\dg \c4 x \c1 t) \cdot 
(\c1 p\dg \c2 q\dg \c3 s \c1 r)         \cdot 
(\c4 y \dg \c1 {\bar{t}}\dg \c2 {\bar{c}} o)
} \kvac + \\
&\bra{pq} \ket{rs} \; \bvac \wick{
(o\dg \c3 c\dg \c1 x \c4 t) \cdot 
(\c1 p\dg \c2 q\dg \c3 s \c1 r)         \cdot 
(\c4 y \dg \c1 {\bar{t}}\dg \c2 {\bar{c}} o)
} \kvac \\[12pt]
             = \; & \bra{t \bar{c}} \ket{\bar{t} c} \; \delta_{xy} -
                    \bra{x \bar{c}} \ket{\bar{t} c} \; \delta_{yt}\\
g_{CH} = \; & g_{CH}^{F1} + g_{CH}^{F2} + g_{CH}^{F3}                 \\
             = \; & - \delta_{xy} \braket{tc | ct} 
                    + \delta_{yt} \braket{xc | ct}
\end{align*}
\subsubsection{$g_{EE} = \cbe g \cke$}
Done as an example at the beginning of the section. 
\subsubsection{$g_{EF} = \cbe g \ckf$}
\begin{align*}
     g_{EF}^{F1} = \; &0 \\
     g_{EF}^{F2} = \; & 
\bra{pq} \ket{rs} \; \bvac \wick{
(\c4 {\bar{o}}\dg \c3 c\dg \c5 x \c6 {\bar{t}}) \cdot 
(\c2 p\dg \c1 q\dg \c3 s \c4 r)         \cdot 
(\c5 t\dg \c6 {\bar{y}}\dg \c2 {\bar{c}} \c1 o)
} \kvac \\
             = \; & \bra{\bar{c} o} \ket{\bar{o} c} \; \delta_{xt} \delta_{\bar{y}\bar{t}} \\
     g_{EF}^{F3} = \; &0 \\
g_{EF} = \; & g_{EF}^{F1} + g_{EF}^{F2} + g_{EF}^{F3} \\
 = \; & \delta_{xt} \delta_{yt} \braket{co | oc}
\end{align*}

\subsubsection{$g_{EG} = \cbe g \ckg$}
\begin{align*}
     g_{EG}^{F1} = \; & 
\bra{pq} \ket{rs} \; \bvac \wick{
({\bar{o}}\dg c\dg \c2 x \c1 {\bar{t}}) \cdot 
(\c1 p\dg \c2 q\dg \c2 s \c1 r)         \cdot 
(\c1 {\bar{y}}\dg \c2 t\dg c {\bar{o}})
} \kvac \\
              = \; & \bra{\bar{t}x} \ket{\bar{y}t} \\
     g_{EG}^{F2} = \; & 
\bra{pq} \ket{rs} \; \bvac \wick{
(\c4 {\bar{o}}\dg \c3 c\dg \c6 x \c5 {\bar{t}}) \cdot 
(\c2 p\dg \c1 q\dg \c3 s \c4 r)         \cdot 
(\c5 {\bar{y}}\dg \c6 t\dg \c1 c \c2 {\bar{o}})
} \kvac \\
              = \; & \bra{\bar{o}c} \ket{\bar{o}c} \delta_{xt} \delta_{\bar{t}\bar{y}}\\
     g_{EG}^{F3} = \; & 
\bra{pq} \ket{rs} \; \bvac \wick{
({\bar{o}}\dg \c3 c\dg \c4 x \c1 {\bar{t}}) \cdot 
(\c1 p\dg \c2 q\dg \c3 s \c1 r)         \cdot 
(\c1 {\bar{y}}\dg \c4 t\dg \c2 c {\bar{o}})
} \kvac + \\ &
\bra{pq} \ket{rs} \; \bvac \wick{
({\bar{o}}\dg \c3 c\dg \c1 x \c4 {\bar{t}}) \cdot 
(\c1 p\dg \c2 q\dg \c3 s \c1 r)         \cdot 
(\c4 {\bar{y}}\dg \c1 t\dg \c2 c {\bar{o}})
} \kvac + \\ &
\bra{pq} \ket{rs} \; \bvac \wick{
(\c3 {\bar{o}}\dg c\dg \c1 x \c4 {\bar{t}}) \cdot 
(\c1 p\dg \c2 q\dg \c3 s \c1 r)         \cdot 
(\c4 {\bar{y}}\dg \c1 t\dg  c \c2 {\bar{o}})
} \kvac + \\ &
\bra{pq} \ket{rs} \; \bvac \wick{
(\c3 {\bar{o}}\dg c\dg \c4 x \c1 {\bar{t}}) \cdot 
(\c1 p\dg \c2 q\dg \c3 s \c1 r)         \cdot 
(\c1 {\bar{y}}\dg \c4 t\dg  c \c2 {\bar{o}})
} \kvac \\
              = \; & - \bra{\bar{t} c} \ket{\bar{y} c} \delta_{xt} 
                     - \bra{x c} \ket{t c} \delta_{\bar{t} \bar{y}} 
                     - \bra{x \bar{o}} \ket{t \bar{o}} \delta_{\bar{t} \bar{y}} 
                     - \bra{\bar{t} \bar{o}} \ket{\bar{y} \bar{o}} \delta_{xt} \\
g_{EG} = \; & g_{EG}^{F1} + g_{EG}^{F2} + g_{EG}^{F3} \\
         = \; & \braket{tx | yt} 
                  + \delta_{xt} (
                    \braket{to | oy}
                  - \braket{to | yo}  
                  - \braket{tc | yc} )
                  + \delta_{ty} (
                    \braket{xc | ct}
                  - \braket{xc | tc}  
                  - \braket{xo | to} ) 
                  + \delta_{xt} \delta_{ty} 
                    \braket{oc | oc}
\end{align*}

\subsubsection{$g_{EH} = \cbe g \ckh$}
\begin{align*}
    g_{EH}^{F1} = \; & 0 \\
    g_{EH}^{F2} = \; &
\bra{pq} \ket{rs} \; \bvac \wick{
(\c4 {\bar{o}}\dg \c3 c\dg \c5 x \c6 {\bar{t}}) \cdot 
(\c2 p\dg \c1 q\dg \c3 s \c4 r)         \cdot 
(\c5 y\dg \c6 {\bar{t}}\dg \c1 {\bar{c}} \c2 o)
} \kvac \\
             = &\; - \bra{o\bar{c}} \ket{\bar{o}c} \; \delta_{xy} \\
    g_{EH}^{F3} = \; & 0 \\ 
g_{EH} = \; & g_{EH}^{F1} + g_{EH}^{F2} + g_{EH}^{F3} \\
          = \; & \delta_{xy} \; \braket{oc | co}
\end{align*}

\subsubsection{$g_{GG} = \cbg g \ckg$}

\begin{align*}
    g_{GG}^{F1} = 
&\bra{pq} \ket{rs} \; \bvac \wick{
({\bar{o}}\dg c\dg \c2 t \c1 {\bar{x}}) \cdot 
(\c1 p\dg \c2 q\dg \c2 s \c1 r)         \cdot 
(\c1 {\bar{y}}\dg \c2 t\dg c {\bar{o}})
} \kvac \\
=&\bra{\bar{x}t} \ket{\bar{y}t} \notag\\[12 pt]
    g_{GG}^{F2} = 
&\bra{pq} \ket{rs} \; \bvac \wick{
(\c4{\bar{o}}\dg \c3 c\dg t \c5 {\bar{x}}) \cdot 
(\c2 p\dg \c1 q\dg \c3 s \c4 r)         \cdot 
(\c5 {\bar{y}}\dg  t\dg \c1 c \c2 {\bar{o}})
} \kvac \\
=&\bra{\bar{o}c} \ket{\bar{o}c} \delta_{\bar{x}\bar{y}} \notag\\[12 pt] 
    g_{GG}^{F3} = 
&\bra{pq} \ket{rs} \; \bvac \wick{
({\bar{o}}\dg \c3 c\dg t \c1 {\bar{x}}) \cdot 
(\c1 p\dg \c2 q\dg \c3 s \c1 r)         \cdot 
(\c1 {\bar{y}}\dg  t\dg \c2 c {\bar{o}})
} \kvac + \\
&\bra{pq} \ket{rs} \; \bvac \wick{
({\bar{o}}\dg \c3 c\dg \c1 t \c4 {\bar{x}}) \cdot 
(\c1 p\dg \c2 q\dg \c3 s \c1 r)         \cdot 
(\c4 {\bar{y}}\dg \c1 t\dg \c2 c {\bar{o}})
} \kvac + \\
&\bra{pq} \ket{rs} \; \bvac \wick{
(\c3 {\bar{o}}\dg c\dg \c1 t \c4 {\bar{x}}) \cdot 
(\c1 p\dg \c2 q\dg \c3 s \c1 r)         \cdot 
(\c4 {\bar{y}}\dg \c1 t\dg c \c2 {\bar{o}})
} \kvac + \\
&\bra{pq} \ket{rs} \; \bvac \wick{
(\c3 {\bar{o}}\dg c\dg t \c1 {\bar{x}}) \cdot 
(\c1 p\dg \c2 q\dg \c3 s \c1 r)         \cdot 
(\c1 {\bar{y}}\dg t\dg c \c2 {\bar{o}})
} \kvac \\
=& - \bra{\bar{x}c} \ket{\bar{y}c} 
   - \bra{tc} \ket{tc} \delta_{\bar{x}\bar{y}}
   - \bra{t\bar{o}} \ket{t\bar{o}} \delta_{\bar{x}\bar{y}}
   - \bra{\bar{x}\bar{o}} \ket{\bar{y}\bar{o}} \\
g_{GG} = \; & g_{GG}^{F1} + g_{GG}^{F2} + g_{GG}^{F3} \\
          = \; &
               \braket{xt | yt} - 
               \braket{xc | yc} - 
               \braket{xo | yo} + 
               \braket{xo | oy}\\ 
               & \delta_{xy} (
               \braket{oc | oc} - 
               \braket{to | to} -
               \braket{tc | tc} +
               \braket{tc | ct}
                ) \\  
\end{align*}

\subsubsection{$g_{GH} = \cbg g \ckh$}
\begin{align*}
    g_{GH}^{F1} = \; & 0 \\
    g_{GH}^{F2} = 
&\bra{pq} \ket{rs} \; \bvac \wick{
(\c4 {\bar{o}}\dg \c3 c\dg \c6 t \c5 {\bar{x}}) \cdot 
(\c2 p\dg \c1 q\dg \c3 s \c4 r)         \cdot 
(\c6 y \c5 {\bar{t}}\dg \c1 {\bar{c}} \c2 o)
} \kvac \\
= &- \bra{o\bar{c}} \ket{\bar{o}c} \delta_{\bar{x}\bar{t}} \delta_{ty} \\
    g_{GH}^{F3} = \; & 0 \\
    g_{GH} = \; & g_{GH}^{F1} + g_{GH}^{F2} + g_{GH}^{F3} \\
           = & \delta_{xt} \delta_{yt} \braket{oc | co} 
\end{align*}

\subsection{One-electron contributions: 3eOS Case}
We begin by considering the three $m_s=\frac{1}{2}$ configurations possible for the case of three electrons in three orbitals.
\begin{align*}
    \ket{I} = \ket{\Psi_{\bar{o}\bar{c}}^{\bar{t}}}\\
    \ket{J} = \ket{\Psi_{o\bar{c}}^{t}}\\
    \ket{K} = \ket{\Psi_{c\bar{o}}^{t}}
\end{align*}
With the determinants represented by their canonical ordering, the operator string associated with the $\ket{I}$ determinant requires a negative sign to represent the correct ordering. The one-electron contributions to the matrix elements are evaluated below.

$F_{II} =\bra{\Psi_{\bar{o}\bar{c}}^{\bar{t}}}f\ket{\Psi_{\bar{o}\bar{c}}^{\bar{t}}}$
\begin{align*}
    F_{II} =& f_{pq}\bvac\wick{ \c2{\bar{c}} \dg \bar{o} \dg \bar{t} \cdot(\c1 p\dg \c2 q)\cdot \bar{t}\dg \bar{o} \c1 {\bar{c}}}\kvac + \\
     &f_{pq}\bvac\wick{ \bar{c}\dg \c2 {\bar{o}}\dg \bar{t} \cdot(\c1 p \dg \c2 q)\cdot \bar{t}\dg  \c1 {\bar{o}} \bar{c} }\kvac  + \\
     &f_{pq}\bvac\wick{\bar{c}\dg \bar{o}\dg \c2{\bar{t}} \cdot(\c1 p \dg \c2 q)\cdot \c1 {\bar{t}}\dg \bar{o} \bar{c}}\kvac + \\
    =&f_{tt} - f_{cc} - f_{oo}
\end{align*}
$F_{JJ} = \bra{\Psi_{o\bar{c}}^{t}}f\ket{\Psi_{o\bar{c}}^{t}}$
\begin{align*}
    F_{JJ} =& f_{pq}\bvac\wick{\c2{\bar{c}}\dg o\dg  t \cdot(\c1 p\dg \c2 q)\cdot t\dg o \c1{\bar{c}}}\kvac + \\
    & f_{pq}\bvac\wick{{\bar{c}}\dg \c2 o\dg t \cdot(\c1 p\dg \c2 q)\cdot t\dg \c1 o {\bar{c}}} \kvac + \\
    &f_{pq}\bvac\wick{{\bar{c}}\dg o\dg \c1 t \cdot(\c1 p\dg \c1 q)\cdot \c1 t\dg  o {\bar{c}}} \kvac\\
    =& f_{tt} - f_{cc} - f_{oo}
\end{align*}
$F_{KK} = \bra{\Psi_{c\bar{o}}^{t}}f\ket{\Psi_{\bar{o}c}^{t}}$
\begin{align*}
    F_{KK} =& f_{pq}\bvac\wick{\c2 c\dg \bar{o} \dg t \cdot (\c1 p \dg \c2 q)\cdot t\dg \bar{o} \c1 c} \kvac + \\
    &f_{pq}\bvac\wick{ c\dg \c2{\bar{o}} \dg t \cdot (\c1 p \dg \c2 q)\cdot t\dg \c1{\bar{o}} c} \kvac + \\
    &f_{pq}\bvac\wick{ c\dg \bar{o} \dg \c1 t \cdot (\c1 p \dg \c1 q)\cdot \c1 t\dg \bar{o} c} \kvac \\
    &= f_{tt}- f_{cc} - f_{oo}
\end{align*}

$F_{IJ} = \bra{\Psi_{\bar{o}\bar{c}}^{\bar{t}}}f\ket{\Psi_{o\bar{c}}^{t}}$
\begin{align*}
    F_{IJ} = 0
\end{align*}
$F_{IK} = \bra{\Psi_{\bar{o}\bar{c}}^{\bar{t}}}f\ket{\Psi_{c\bar{o}}^{t}}$
\begin{align*}
    F_{IK} = 0
\end{align*}
$F_{JK} = \bra{\Psi_{o\bar{c}}^{t}}f\ket{\Psi_{c\bar{o}}^{t}}$
\begin{align*}
    F_{JK} = 0
\end{align*}

\subsection{Two-electron contributions: 3eOS Case}
The two-electron contributions for the three $m_s=\frac{1}{2}$ configurations are evaluated below. All elements are scaled by a factor of $\frac{1}{4}$ to yield the necessary matrix element. In each case, the 12 relevant contractions are evaluated to find the overall matrix element.

$g_{II} = \bra{\Psi_{\bar{o}\bar{c}}^{\bar{t}}}g\ket{\Psi_{\bar{o}\bar{c}}^{\bar{t}}}$
\allowdisplaybreaks
\begin{align*}
    \frac{1}{4}\cdot g_{II} =& \frac{1}{4}\cdot\Bigl[\bra{pq}\ket{rs}\bvac\wick{\bar{c}\dg \c3{\bar{o}} \dg \c2{\bar{t}} \cdot (\c2 p\dg \c2 q \dg \c1 s \c3 r)\cdot \c1{\bar{t}}\dg \c2{\bar{o}} \bar{c}} \kvac + \\
    &\bra{pq}\ket{rs}\bvac\wick{ {\bar{c}}\dg \c3{\bar{o}} \dg \c2{\bar{t}} \cdot ( \c2 p\dg \c2 q \dg \c3 s  \c1 r)\cdot \c1{\bar{t}}\dg \c2{\bar{o}} {\bar{c}}} \kvac + \\
    &\bra{pq}\ket{rs}\bvac\wick{\c3{\bar{c}}\dg {\bar{o}} \dg  \c1{\bar{t}} \cdot ( \c1 p\dg  \c2 q \dg  \c1 s \c3 r)\cdot \c1{\bar{t}}\dg  {\bar{o}} \c2{\bar{c}}} \kvac + \\   
    &\bra{pq}\ket{rs}\bvac\wick{\c2{\bar{c}}\dg {\bar{o}} \dg  \c1{\bar{t}} \cdot ( \c1 p\dg \c3 q \dg \c2 s \c1 r)\cdot \c1{\bar{t}}\dg {\bar{o}} \c3{\bar{c}}} \kvac + \\
    &\bra{pq}\ket{rs}\bvac\wick{{\bar{c}}\dg \c3{\bar{o}} \dg \c1{\bar{t}} \cdot (\c2  p\dg \c1 q \dg  \c1 s  \c3 r)\cdot \c1{\bar{t}}\dg \c2{\bar{o}} {\bar{c}}} \kvac + \\
    &\bra{pq}\ket{rs}\bvac\wick{{\bar{c}}\dg \c3{\bar{o}} \dg \c1{\bar{t}} \cdot (\c2  p\dg \c1 q \dg  \c3 s  \c1 r)\cdot \c1{\bar{t}}\dg \c2{\bar{o}} {\bar{c}}} \kvac + \\
    &\bra{pq}\ket{rs}\bvac\wick{\c3{\bar{c}} \dg \c4{\bar{o}} \dg {\bar{t}} \cdot (\c1  p\dg \c2 q \dg \c3 s \c4 r)\cdot {\bar{t}}\dg \c1{\bar{o}} \c2{\bar{c}}} \kvac + \\
    &\bra{pq}\ket{rs}\bvac\wick{\c4{\bar{c}}\dg \c3{\bar{o}} \dg  {\bar{t}} \cdot (\c1  p\dg \c2 q \dg \c3 s \c4 r)\cdot {\bar{t}}\dg \c1{\bar{o}} \c2{\bar{c}}} \kvac + \\
    &\bra{pq}\ket{rs}\bvac\wick{\c3{\bar{c}}\dg \c4{\bar{o}} \dg  {\bar{t}} \cdot (\c1  p\dg \c2 q \dg \c3 s \c4 r)\cdot {\bar{t}}\dg \c2{\bar{o}} \c1{\bar{c}}} \kvac + \\
    &\bra{pq}\ket{rs}\bvac\wick{\c4{\bar{c}}\dg \c3{\bar{o}} \dg  {\bar{t}} \cdot ( \c2 p\dg \c1 q \dg \c3 s \c4 r)\cdot {\bar{t}}\dg \c1{\bar{o}} \c2{\bar{c}}} \kvac + \\
    &\bra{pq}\ket{rs}\bvac\wick{\c3{\bar{c}}\dg {\bar{o}} \dg  \c1{\bar{t}} \cdot (\c2  p\dg \c1 q \dg \c3 s \c1 r)\cdot \c1{\bar{t}}\dg {\bar{o}} \c2{\bar{c}}} \kvac + \\
    &\bra{pq}\ket{rs}\bvac\wick{\c3{\bar{c}}\dg {\bar{o}} \dg  \c1{\bar{t}} \cdot (\c2  p\dg  \c1 q \dg \c1 s \c3 r)\cdot \c1{\bar{t}}\dg {\bar{o}} \c2{\bar{c}}} \kvac\Bigr] \\
    =& \frac{1}{4}\cdot\Bigl[\bra{\bar{t}\bar{o}}\ket{\bar{o}\bar{t}} - \bra{\bar{t}\bar{o}}\ket{\bar{t}\bar{o}} + \bra{\bar{t}\bar{c}}\ket{\bar{c}\bar{t}} - \bra{\bar{t}\bar{c}}\ket{\bar{t}\bar{c}} -\bra{\bar{o}\bar{t}}\ket{\bar{o}\bar{t}} + \bra{\bar{o}\bar{t}}\ket{\bar{t}\bar{o}} + \\
    &\bra{\bar{o}\bar{c}}\ket{\bar{o}\bar{c}} - \bra{\bar{o}\bar{c}}\ket{\bar{c}\bar{o}} - \bra{\bar{c}\bar{o}}\ket{\bar{o}\bar{c}} + \bra{\bar{c}\bar{o}}\ket{\bar{c}\bar{o}} + \bra{\bar{c}\bar{t}}\ket{\bar{t}\bar{c}} - \bra{\bar{c}\bar{t}}\ket{\bar{c}\bar{t}} \Bigr]\\
    =&\braket{to|ot} - \braket{to|to} + \braket{tc|ct} - \braket{tc|tc} + \braket{oc|oc} - \braket{oc|co}
\end{align*}

$g_{JJ} = \bra{\Psi_{o\bar{c}}^{t}}f\ket{\Psi_{o\bar{c}}^{t}}$
\begin{align*}
    \frac{1}{4}\cdot g_{JJ} =& \frac{1}
        {4}\cdot\Bigl[ \bra{pq}\ket{rs}\bvac \wick{{\bar{c}} \dg \c3 o \dg  \c1 t \cdot(
        \c1 p \dg \c2 q \dg \c1 s \c3 r ) \cdot \c1 t \dg \c2 o {\bar{c}}} \kvac + \\
        &\bra{pq}\ket{rs}\bvac \wick{{\bar{c}} \dg \c3 o \dg \c2 t \cdot(\c2 p \dg \c2 q \dg \c3 s \c1 r ) \cdot \c1 t \dg \c2 o {\bar{c}}} \kvac + \\
        &\bra{pq}\ket{rs}\bvac \wick{\c3{\bar{c}} \dg o \dg  \c2 t \cdot(\c2 p \dg \c2 q \dg \c1 s \c3 r ) \cdot \c1 t \dg  o \c2{\bar{c}}} \kvac + \\
        &\bra{pq}\ket{rs}\bvac \wick{\c3{\bar{c}} \dg o \dg  \c2 t \cdot(\c2 p \dg \c2 q \dg \c3 s \c1 r ) \cdot \c1 t \dg o \c2{\bar{c}}} \kvac + \\
        &\bra{pq}\ket{rs}\bvac \wick{{\bar{c}} \dg \c3 o \dg \c1 t \cdot( \c2 p \dg\c1 q \dg \c1 s \c3 r ) \cdot\c1 t \dg \c2 o {\bar{c}}} \kvac + \\
        &\bra{pq}\ket{rs}\bvac \wick{ {\bar{c}} \dg \c3 o \dg \c1 t \cdot(\c2 p \dg\c1 q \dg \c3 s  \c1 r ) \cdot\c1 t \dg \c2 o {\bar{c}}} \kvac + \\
        &\bra{pq}\ket{rs}\bvac \wick{\c3{\bar{c}} \dg \c4 o \dg t \cdot(\c1 p \dg\c2 q \dg \c3 s \c4 r ) \cdot t \dg \c1 o \c2{\bar{c}}} \kvac + \\
        &\bra{pq}\ket{rs}\bvac \wick{\c4{\bar{c}} \dg \c3 o \dg t \cdot(\c1 p \dg\c2 q \dg \c3 s \c4 r ) \cdot t \dg \c1 o \c2{\bar{c}}} \kvac + \\
        &\bra{pq}\ket{rs}\bvac \wick{\c3{\bar{c}} \dg \c4 o \dg t \cdot(\c1 p \dg\c2 q \dg \c3 s \c4 r ) \cdot t \dg \c1 o \c2{\bar{c}}} \kvac + \\
        &\bra{pq}\ket{rs}\bvac \wick{\c4{\bar{c}} \dg \c3  o \dg t \cdot(\c2 p \dg\c1 q \dg \c3 s \c4 r ) \cdot t \dg \c1 o \c2{\bar{c}}} \kvac + \\
        &\bra{pq}\ket{rs}\bvac \wick{\c4{\bar{c}} \dg o \dg \c3 t \cdot(\c2 p \dg\c3 q \dg \c4 s \c1 r ) \cdot\c1 t \dg o \c2{\bar{c}}} \kvac + \\
        &\bra{pq}\ket{rs}\bvac \wick{\c4{\bar{c}} \dg o \dg \c3 t \cdot(\c2 p \dg\c3 q \dg \c1 s \c4 r ) \cdot\c1 t \dg o \c2{\bar{c}}} \kvac\\
        =& \frac{1}{4}\cdot\Bigl[\bra{to}\ket{ot} - \bra{to}\ket{to} + \bra{t\bar{c}}\ket{\bar{c}t} - \bra{t\bar{c}}\ket{t\bar{c}} - \bra{ot}\ket{ot} + \bra{ot}\ket{to} + \\
        &\bra{o\bar{c}}\ket{o\bar{c}} - \bra{o\bar{c}}\ket{\bar{c}o} - \bra{\bar{c}o}\ket{o\bar{c}} + \bra{\bar{c}o}\ket{\bar{c}o} + \bra{\bar{c}t}\ket{t\bar{c}} - \bra{\bar{c}t}\ket{\bar{c}t}\Bigr]\\
        =& \braket{to|ot} - \braket{to|to} - \braket{tc|tc} + \braket{oc|oc}
\end{align*}

$g_{KK} = \bra{\Psi_{c\bar{o}}^{t}}g\ket{\Psi_{c\bar{o}}^{t}}$
\begin{align*}
    \frac{1}{4}\cdot g_{KK} =& \frac{1}{4}\cdot \Bigl[ \bra{pq}\ket{rs} \bvac \wick{c \dg \c3{\bar{o}} \dg \c2 t \cdot (\c2 p \dg \c2 q \dg \c1 s \c3 r )\cdot \c1 t \dg \c2{\bar{o}} c}\kvac + \\
    &\bra{pq}\ket{rs} \bvac \wick{c \dg \c3{\bar{o}} \dg\c2 t \cdot (\c2 p \dg\c2 q \dg\c3 s\c1 r )\cdot \c1t \dg \c2{\bar{o}} c}\kvac + \\
    &\bra{pq}\ket{rs} \bvac \wick{\c3 c \dg {\bar{o}} \dg\c2 t \cdot (\c2 p \dg\c2 q \dg\c1 s\c3 r )\cdot \c1 t \dg {\bar{o}}\c2 c}\kvac + \\
    &\bra{pq}\ket{rs} \bvac \wick{\c3 c \dg {\bar{o}} \dg\c2 t \cdot (\c2 p \dg\c2 q \dg \c3 s \c1 r )\cdot\c1 t \dg {\bar{o}} \c2 c}\kvac + \\
    &\bra{pq}\ket{rs} \bvac \wick{c \dg \c4{\bar{o}} \dg\c3 t \cdot (\c2 p \dg\c3 q \dg\c1 s\c4 r )\cdot \c1 t \dg \c2{\bar{o}} c}\kvac + \\
    &\bra{pq}\ket{rs} \bvac \wick{c \dg \c4{\bar{o}} \dg\c3 t \cdot (\c2 p \dg\c3 q \dg\c4 s \c1 r )\cdot \c1 t \dg \c2{\bar{o}} c}\kvac + \\
    &\bra{pq}\ket{rs} \bvac \wick{\c3 c \dg \c4{\bar{o}} \dg t \cdot (\c1 p \dg\c2 q \dg\c3 s\c4 r )\cdot t \dg \c1{\bar{o}}\c2 c}\kvac + \\
    &\bra{pq}\ket{rs} \bvac \wick{\c4 c \dg \c3{\bar{o}} \dg t \cdot (\c2 p \dg\c1  q \dg\c3 s\c4 r )\cdot t \dg \c2{\bar{o}} \c1 c}\kvac + \\
    &\bra{pq}\ket{rs} \bvac \wick{\c3 c \dg \c4{\bar{o}} \dg t \cdot (\c2 p \dg\c1 q \dg\c3 s\c4 r )\cdot t \dg \c1{\bar{o}}\c2 c}\kvac + \\
    &\bra{pq}\ket{rs} \bvac \wick{\c4 c \dg \c3{\bar{o}} \dg t \cdot (\c2 p \dg\c1  q \dg\c3 s\c4 r )\cdot t \dg \c1{\bar{o}}\c2 c}\kvac + \\
    &\bra{pq}\ket{rs} \bvac \wick{\c3 c \dg {\bar{o}} \dg\c1 t \cdot (\c2 p \dg\c1  q \dg\c3 s \c1 r )\cdot \c1 t \dg {\bar{o}} \c2 c}\kvac + \\
    &\bra{pq}\ket{rs} \bvac \wick{\c3 c \dg {\bar{o}} \dg \c1 t \cdot (\c2 p \dg\c1  q \dg\c1  s \c3 r )\cdot\c1 t \dg {\bar{o}}\c2 c}\kvac \Bigr] \\
    =& \frac{1}{4}\cdot\Bigl[\bra{t\bar{o}}\ket{\bar{o}t} - \bra{t\bar{o}}\ket{t\bar{o}} + \bra{tc}\ket{ct} - \bra{tc}\ket{tc} - \bra{\bar{o}t}\ket{\bar{o}t} + \bra{\bar{o}t}\ket{t\bar{o}} + \\
    &\bra{\bar{o}c}\ket{\bar{o}c} - \bra{\bar{o}c}\ket{c\bar{o}} - \bra{c\bar{o}}\ket{\bar{o}c} + \bra{c\bar{o}}\ket{c\bar{o}} + \bra{ct}\ket{tc} - \bra{ct}\ket{ct}\Bigr]\\
    =& \braket{tc|ct} - \braket{tc|tc} - \braket{to|to} + \braket{oc|oc}
\end{align*}

\noindent For the the sake of conciseness, the contractions involving 3 orbitals of same spin (and 1 opposite spin), are omitted due to spin orthogonality making their contribution 0 to the overall element. As a result, only 4 terms contribute to each of the following elements.

$g_{IJ} = \bra{\Psi_{\bar{o}\bar{c}}^{\bar{t}}}g\ket{\Psi_{o\bar{c}}^{t}}$

\begin{align*}
    \frac{1}{4}\cdot g_{IJ} =& -\frac{1}{4} \cdot \Bigl[
    \bra{pq}\ket{rs}\bvac\wick{{\bar{c}} \dg \c3{\bar{o}} \dg \c2{\bar{t}} \cdot (\c2 p \dg\c2 q \dg\c1 s\c3 r ) \cdot\c1 t \dg \c2 o {\bar{c}}}\kvac + \\
    &\bra{pq}\ket{rs}\bvac\wick{{\bar{c}} \dg \c3{\bar{o}} \dg \c2{\bar{t}} \cdot (\c2 p \dg\c2 q \dg\c3 s \c1 r ) \cdot\c1 t \dg\c2 o {\bar{c}}}\kvac + \\
    &\bra{pq}\ket{rs}\bvac\wick{{\bar{c}} \dg \c3{\bar{o}} \dg \c1{\bar{t}} \cdot (\c2 p \dg \c1 q \dg \c1 s \c3 r ) \cdot\c1 t \dg \c2 o {\bar{c}}}\kvac + \\
    &\bra{pq}\ket{rs}\bvac\wick{{\bar{c}} \dg \c3{\bar{o}} \dg \c1{\bar{t}} \cdot (\c2 p \dg\c1 q \dg \c3 s \c1 r ) \cdot\c1  t \dg \c2 o {\bar{c}}}\kvac
    \Bigr] \\
    =& -\frac{1}{4}\cdot\Bigl[\bra{\bar{t}o}\ket{\bar{o}t} - \bra{\bar{t}o}\ket{t\bar{o}} - \bra{o\bar{t}}\ket{\bar{o}t} + \bra{o\bar{t}}\ket{t\bar{o}}
    \Bigr] \\
    =& -\braket{to|ot}
\end{align*}

$g_{IK} = \bra{\Psi_{\bar{o}\bar{c}}^{\bar{t}}}f\ket{\Psi_{c\bar{o}}^{t}}$

\begin{align*}
    \frac{1}{4}\cdot g_{IK} =& -\frac{1}{4}\cdot\Bigl[ \bra{pq}\ket{rs}\bvac\wick{\c3{ \bar{c}}\dg {\bar{o}}\dg  \c2{\bar{t}} \cdot (\c2 p \dg\c2  q \dg \c1 s \c3 r ) \cdot \c1 t \dg {\bar{o}}\c2 c}\kvac + \\
    &\bra{pq}\ket{rs}\bvac\wick{\c3{\bar{c}}\dg {\bar{o}}\dg \c2{\bar{t}} \cdot (\c2 p \dg\c2 q \dg \c3 s \c1 r ) \cdot \c1 t \dg {\bar{o}} \c2 c)}\kvac + \\
    &\bra{pq}\ket{rs}\bvac\wick{\c3{\bar{c}}\dg {\bar{o}}\dg  \c1{\bar{t}} \cdot (\c2 p \dg\c1 q \dg \c3 s \c1 r ) \cdot\c1  t \dg {\bar{o}}\c2  c)}\kvac + \\
    &\bra{pq}\ket{rs}\bvac\wick{\c3{\bar{c}}\dg {\bar{o}}\dg \c1{\bar{t}} \cdot (\c2 p \dg \c1 q \dg \c1 s \c3 r ) \cdot\c1 t \dg {\bar{o}}\c2 c)}\kvac
    \Bigr]\\
    =& -\frac{1}{4}\cdot\Bigl[ \bra{\bar{t}c}\ket{\bar{c}t} - \bra{\bar{t}c}\ket{t\bar{c}} + \bra{c\bar{t}}\ket{t\bar{c}} - \bra{c\bar{t}}\ket{\bar{c}t}\Bigr] \\
    =& -\braket{tc|ct}
\end{align*}

$g_{JK} = \bra{\Psi_{o\bar{c}}^{t}}f\ket{\Psi_{c\bar{o}}^{t}}$
\begin{align*}
    \frac{1}{4}\cdot g_{JK} =& \frac{1}{4} \cdot \Bigl[
    \bra{pq}\ket{rs}\bvac\wick{\c3{\bar{c}} \dg \c4 o \dg t \cdot (\c2 p \dg\c1 q \dg\c3 s\c4 r) \cdot t \dg\c2 {\bar{o}}\c1 c }\kvac + \\
    &\bra{pq}\ket{rs}\bvac\wick{ \c4{\bar{c}} \dg \c3 o \dg t \cdot (\c2 p \dg\c1 q \dg\c3 s\c4 r) \cdot t \dg \c2{\bar{o}}\c1 c }\kvac + \\
    &\bra{pq}\ket{rs}\bvac\wick{\c3{\bar{c}} \dg \c4 o \dg t \cdot (\c2 p \dg\c1 q \dg\c3 s\c4 r) \cdot t \dg \c1{\bar{o}}\c2 c }\kvac + \\
    &\bra{pq}\ket{rs}\bvac\wick{\c4{\bar{c}} \dg \c3 o \dg t \cdot (\c2 p \dg\c1 q \dg\c3 s\c4 r) \cdot t \dg \c1{\bar{o}}\c2 c }\kvac 
    \Bigr]\\
    =& \frac{1}{4}\cdot\Bigl[ \bra{\bar{o}c}\ket{o\bar{c}} - \bra{\bar{o}c}\ket{\bar{c}o} - \bra{c\bar{o}}\ket{o\bar{c}} + \bra{c\bar{o}}\ket{\bar{c}o}\Bigr]\\
    =& -\braket{oc|co}
\end{align*}
\subsection{Collection of matrix elements}
\begin{align*}
    H_{AA} = & \cba H \cka \\
           = & f_{tt} - f_{cc} + \braket{tc | ct} - \braket{tc | tc} \\
    H_{AB} = & \cba H \ckb \\
           = & \braket{tc | ct} \\
    H_{AC} = & \cba H \ckc \\
             & \delta_{ty} f_{ot} - f_{oy} + \\
             & \braket{oc | yc} - 
               \braket{oc | cy} -
               \braket{to | ty} +
               \braket{to | yt} +  
               \delta_{yt} \big(
                 \braket{oc | ct} -
                 \braket{oc | tc} \big) \\
    H_{AD} = & \cba H \ckd \\ 
           = & 0 \\ 
    H_{AE} = & \cba H \cke \\
           = & \delta_{ty} f_{ot} + \braket{to | yt} - \delta_{yt} \braket{oc | tc} \\ 
    H_{AF} = & \cba H \ckf \\
           = & - \braket{oc | cy} \\
    H_{AG} = & \cba H \ckg \\
           = & f_{oy} +\braket{to | ty} - \braket{oc | yc} \\
    H_{AH} = & \cba H \ckg \\
           = & - \delta_{ty} \braket{oc | ct}
\end{align*}
\begin{align*}
    H_{CC} = & \cbc H \ckc  = H_{DD} = \cbd H \ckd \\
           = & f_{xy} - \delta_{ty} f_{xt} - \delta_{xt} f_{ty} + \delta_{xy} (f_{tt} - f_{cc} - f_{oo}) + \delta_{xt}\delta_{ty} (f_{cc} + f_{oo}) + \\
             & \braket{tx | ty} - 
               \braket{tx | yt} -
               \braket{xc | yc} +
               \braket{xc | cy} - 
               \braket{xo | yo} +
               \braket{xo | oy} + \\ & 
               \delta_{xt} \big( 
                 \braket{tc | yc} -
                 \braket{tc | cy} + 
                 \braket{to | yo} - 
                 \braket{to | oy} \big) \\ & 
               \delta_{yt} \big(
                 \braket{xc | tc} -
                 \braket{xc | ct} + 
                 \braket{xo | to} - 
                 \braket{xo | ot} \big) + \\ & 
               \delta_{xy} \big( 
                 \braket{oc | oc} - 
                 \braket{oc | co} -
                 \braket{tc | tc} +
                 \braket{tc | ct} - 
                 \braket{to | to} + 
                 \braket{to | ot} \big) - \\ &
               \delta_{xt} \delta_{yt} \big(
                 \braket{oc | oc} - 
                 \braket{oc | co} \big) \\
    H_{CD} = & \cbc H \ckd \\ 
           = & 0 \\ 
    H_{CE} = & \cbc H \cke = H_{DF} = \cbd H \ckf \\
           = & \delta_{xy} \braket{to | ot} - \delta_{yt} \braket{xo | ot} \\ 
    H_{CF} = & \cbc H \ckf = H_{DE} = \cbd H \cke \\
           = & \braket{xc | cy} - \delta_{xt} \braket{tc | cy} \\
    H_{CG} = & \cbc H \ckg = H_{DH} = \cbc H \ckh \\
           = & \delta_{xt} \braket{to | oy} - \braket{xo | oy} \\
    H_{CH} = & \cbc H \ckg = H_{DG} = \cbd H \ckg \\
           = & \delta_{yt} \braket{xc | ct} - \delta_{xy} \braket{tc | ct} \\
\end{align*}
\begin{align*}
    H_{EC} = & \cbe H \ckc \\
           = & \delta_{xy} \braket{to | ot} - \delta_{xt} \braket{to | oy} \\
    H_{ED} = & \cbe H \ckd \\
           = & \braket{yc | cx} - \delta_{yt} \braket{tc | cx} \\
    H_{EE} = & \cbe H \cke  = H_{FF} = \cbf H \ckf \\
           = & f_{xy} + \delta_{xy} (f_{tt} - f_{cc} - f_{oo}) + \\
             & \braket{t x | t y} - \braket{xc | yc} + \braket{xc | cy} - \braket{xo | yo} + \\ &
             \delta_{xy} \big( 
             \braket{o c | o c} - \braket{tc | tc} - \braket{to | to} + \braket{to | ot}
             \big) \\
    H_{EF} = & \cbe H \ckf \\ 
           = & \delta_{xt} \delta_{yt} \braket{co | oc} \\ 
    H_{EG} = & \cbe H \ckg = H_{FH} = \cbf H \ckh \\
           = & \delta_{xt} f_{ty} + \delta_{ty} f_{xt} - \delta_{xt}\delta_{ty} (f_{cc} + f_{oo}) + \\
             & \braket{tx | yt} 
                  + \delta_{xt} \big ( 
                  \braket{to | oy} -
                  \braket{to | yo} -
                  \braket{tc | yc}  \big)
                  + \delta_{ty} \big(
                  \braket{xc | ct} -
                  \braket{xc | tc} -
                  \braket{xo | to}  \big)  
                  + \delta_{xt} \delta_{ty} 
                  \braket{oc | oc} \\ 
    H_{EH} = & \cbe H \ckh = H_{FG} = \cbf H \ckg \\
           = & \delta_{xy} \; \braket{oc | co} \\
\end{align*}
\begin{align*} 
    H_{GC} = & \cbg H \ckc \\
           = & \delta_{yt} \braket{to | ox} - \braket{yo | ox} \\
    H_{GD} = & \cbg H \ckd \\
           = & \delta_{xt} \braket{yc | ct} - \delta_{xy} \braket{tc | ct} \\
    H_{GE} = & \cbg H \cke \\
           = & \delta_{xt} f_{ty} + \delta_{ty} f_{xt} - \delta_{xt}\delta_{ty} (f_{cc} + f_{oo}) + \\
             & \braket{ty | xt} 
                  + \delta_{ty} \big ( 
                  \braket{to | ox} -
                  \braket{to | xo} -
                  \braket{tc | xc} \big)
                  + \delta_{xt} \big(
                  \braket{yc | ct} -
                  \braket{yc | tc} -
                  \braket{yo | to} \big) 
                  + \delta_{xt} \delta_{ty} 
                  \braket{oc | oc}\\
    H_{GF} = & \cbg H \ckf \\
           = & \delta_{xy} \; \braket{oc | co} \\
    H_{GG} = & \cbg H \ckg  = H_{HH} = \cbh H \ckh \\
           = & f_{xy} + \delta_{xy} (f_{tt} - f_{cc} - f_{oo}) +  \\
             & \braket{xt | yt} - 
               \braket{xc | yc} - 
               \braket{xo | yo} + 
               \braket{xo | oy} \\ 
               %There may be a missing + here%
               & \delta_{xy} \big(
               \braket{oc | oc} - 
               \braket{to | to} -
               \braket{tc | tc} +
               \braket{tc | ct} \big) \\
    H_{GH} = & \cbg H \ckh \\ 
           = & \delta_{xt} \delta_{yt} \braket{oc | co}
\end{align*}

\begin{align*}
    H_{II} = &f_{tt} - f_{cc} - f_{oo} + \braket{to|ot} - \braket{to|to} + \braket{tc|ct} - \braket{tc|tc} + \braket{oc|oc} - \braket{oc|co} \\
    H_{JJ} = &f_{tt} - f_{cc} - f_{oo} + \braket{to|ot} - \braket{to|to} - \braket{tc|tc} + \braket{oc|oc} \\
    H_{KK} = & f_{tt} - f_{cc} - f_{oo} + \braket{tc|ct} - \braket{tc|tc} - \braket{to|to} + \braket{oc|oc} \\
    %Sign on these..?
    H_{IJ} = & -\braket{to|ot} \\
    H_{IK} = & -\braket{tc|ct} \\
    H_{JK} = & -\braket{oc|co}
\end{align*}

\section{Spin-adaptation of orthogonal matrix elements}
\subsection{Construction of configuration-state functions}
\subsubsection{Two electrons in two orbitals}
The possible M$_s$ = 0 states for a two-electron open-shell (2eOS) state are well known and are given without derivation. The 2eOS states relevant in pump-probe experiments correspond to pairing of the core electrons with the singly-occupied molecular orbitals (SOMOs). In particular, the \textbf{c} $\xrightarrow[]{}$ SOMO(\textbf{o}) singlet state is 
\begin{align}
    \ket{^{1}\Phi_{oc}^{to}} =\ket{^{1}\Phi_{c}^{t}} &= (2)^{-1/2} \left(\ket{\Phi_{c}^{t}} +
                                               \ket{\Phi_{\bar{c}}^{\bar{t}}} \right)
\end{align}
and the triplet state is 
\begin{align}
    \ket{^{3}\Phi_{oc}^{to}} =\ket{^{3}\Phi_{c}^{t}} &= (2)^{-1/2} \left(\ket{\Phi_{c}^{t}} -
                                               \ket{\Phi_{\bar{c}}^{\bar{t}}} \right)
\end{align}
Note that they can be represented as single excitations out of the closed-shell reference. There is also one singlet and one triplet associated with the \textbf{c} $\xrightarrow[]{}$ SOMO(\textbf{t}) transitions
\begin{align}
    \ket{^{1}\Phi_{oc}^{tt}} &= (2)^{-1/2} \left(\ket{\Phi_{\bar{o}c}^{\bar{t}t}} +
                                                 \ket{\Phi_{o\bar{c}}^{t\bar{t}}} \right) \\
    \ket{^{3}\Phi_{oc}^{tt}} &= (2)^{-1/2} \left(\ket{\Phi_{\bar{o}c}^{\bar{t}t}} -
                                                 \ket{\Phi_{o\bar{c}}^{t\bar{t}}} \right) 
\end{align}
that also appear as special cases of the 4eOS CSFs.

\subsubsection{Three electrons in three orbitals}
The possible $M_s = \frac{1}{2}$ states for a 3eOS system are obtained by diagonalizing the $S^2$ matrix in the basis of $M_s = \frac{1}{2}$ 3eOS determinants, resulting in two doublets and one quartet. The off-diagonal terms are evaluated using the rules explained in the following section regarding determinants differing by two orbitals. The diagonal terms are determined below by considering the action of the $S^2$ operator on a 3x3 determinant.

Using the 3x3 determinant of the form

\begin{align*}
    \frac{1}{\sqrt{6}}\cdot \big( c^{\alpha}_1(t^{\beta}_2 o^{\alpha}_3 - o^{\alpha}_2 t^{\beta}_3) - 
    t^{\beta}_1(c^{\alpha}_2 o^{\alpha}_3 - o^{\alpha}_2 c^{\alpha}_3) + 
    o^{\alpha}_1(c^{\alpha}_2 t^{\beta}_3 - t^{\beta}_2 c^{\alpha}_3)\big)
\end{align*}

the diagonal terms may be calculated 
\begin{align*}
    \bra{\Psi_{\bar{o}\bar{c}}^{\bar{t}}}S^2\ket{\Psi_{\bar{o}\bar{c}}^{\bar{t}}} = & \frac{1}{6}\cdot\Bigl[(
    \bra{c^{\alpha}_1t^{\beta}_2 o^{\alpha}_3}-\bra{c^{\alpha}_1 o^{\alpha}_2 t^{\beta}_3} - \bra{t^{\beta}_1 c^{\alpha}_2 o^{\alpha}_3} + \bra{t^{\beta}_1 o^{\alpha}_2 c^{\alpha}_2} + \bra{o^{\alpha}_1 c^{\alpha}_2 t^{\beta}_3} - \bra{o^{\alpha}_1 t^{\beta}_2 c^{\alpha}_3})\times \\
    & \Big((s^-_1 + s^-_2 + s^-_3)\cdot(s^+_1+s^+_2+s^+_3) + s_z + s_z^2\Big)\times \\
    &(
    \ket{c^{\alpha}_1t^{\beta}_2 o^{\alpha}_3}-\ket{c^{\alpha}_1 o^{\alpha}_2 t^{\beta}_3} - \ket{t^{\beta}_1 c^{\alpha}_2 o^{\alpha}_3} + \ket{t^{\beta}_1 o^{\alpha}_2 c^{\alpha}_2} + \ket{o^{\alpha}_1 c^{\alpha}_2 t^{\beta}_3} - \ket{o^{\alpha}_1 t^{\beta}_2 c^{\alpha}_3})
    \Bigr]\\
    =& \frac{1}{6}\cdot \Bigl[
    (
    \bra{c^{\alpha}_1t^{\beta}_2 o^{\alpha}_3}-\bra{c^{\alpha}_1 o^{\alpha}_2 t^{\beta}_3} - \bra{t^{\beta}_1 c^{\alpha}_2 o^{\alpha}_3} + \bra{t^{\beta}_1 o^{\alpha}_2 c^{\alpha}_2} + \bra{o^{\alpha}_1 c^{\alpha}_2 t^{\beta}_3} - \bra{o^{\alpha}_1 t^{\beta}_2 c^{\alpha}_3})\times \\
    & (s^-_1 + s^-_2 + s^-_3) \times \\
    &\Big(-\ket{c^{\alpha}_1 o^{\alpha}_2 t^{\alpha}_3} + \ket{o^{\alpha}_1 c^{\alpha}_2 t^{\alpha}_3} + \ket{c^{\alpha}_1 t^{\alpha}_2 o^{\alpha}_3} - \ket{o^{\alpha}_1 t^{\alpha}_2 c^{\alpha}_3} - \ket{t^{\alpha}_1 c^{\alpha}_2 o^{\alpha}_3} + \ket{t^{\alpha}_1 o^{\alpha}_2 c^{\alpha}_3} + \\
    &\frac{1}{2}\ket{c^{\alpha}_1t^{\beta}_2 o^{\alpha}_3}-\frac{1}{2}\ket{c^{\alpha}_1 o^{\alpha}_2 t^{\beta}_3} - \frac{1}{2}\ket{t^{\beta}_1 c^{\alpha}_2 o^{\alpha}_3} + \frac{1}{2}\ket{t^{\beta}_1 o^{\alpha}_2 c^{\alpha}_2} + \frac{1}{2}\ket{o^{\alpha}_1 c^{\alpha}_2 t^{\beta}_3} - \frac{1}{2}\ket{o^{\alpha}_1 t^{\beta}_2 c^{\alpha}_3} + \\
    &\frac{1}{4}\ket{c^{\alpha}_1t^{\beta}_2 o^{\alpha}_3}-\frac{1}{4}\ket{c^{\alpha}_1 o^{\alpha}_2 t^{\beta}_3} - \frac{1}{4}\ket{t^{\beta}_1 c^{\alpha}_2 o^{\alpha}_3} + \frac{1}{4}\ket{t^{\beta}_1 o^{\alpha}_2 c^{\alpha}_2} + \frac{1}{4}\ket{o^{\alpha}_1 c^{\alpha}_2 t^{\beta}_3} - \frac{1}{4}\ket{o^{\alpha}_1 t^{\beta}_2 c^{\alpha}_3}
    \Big)
    \Bigr] +  \\
    =&\frac{1}{6}\cdot \Bigl[
    \big(
    \bra{c^{\alpha}_1t^{\beta}_2 o^{\alpha}_3}-\bra{c^{\alpha}_1 o^{\alpha}_2 t^{\beta}_3} - \bra{t^{\beta}_1 c^{\alpha}_2 o^{\alpha}_3} + \bra{t^{\beta}_1 o^{\alpha}_2 c^{\alpha}_2} + \bra{o^{\alpha}_1 c^{\alpha}_2 t^{\beta}_3} - \bra{o^{\alpha}_1 t^{\beta}_2 c^{\alpha}_3}\big)\times \\
    &\big(-\ket{c^{\alpha}_1 o^{\alpha}_2 t^{\beta}_3} + \ket{o^{\alpha}_1 c^{\alpha}_2 t^{\beta}_3} + \ket{c^{\alpha}_1 t^{\alpha}_2 o^{\beta}_3} - \ket{o^{\alpha}_1 t^{\alpha}_2 c^{\beta}_3} - \ket{t^{\alpha} c^{\alpha}_2 o^{\beta}_3} + \ket{t^{\alpha}_1 o^{\alpha}_2 c^{\beta}_3} - \\
    &\ket{c^{\alpha}_1 o^{\beta}_2 t^{\alpha}_3} + \ket{o^{\alpha}_1 c^{\beta}_2 t^{\alpha}_3} + \ket{c^{\alpha}_1 t^{\beta}_2 o^{\alpha}_3} - \ket{o^{\alpha}_1 t^{\beta}_2 c^{\alpha}_3} - \ket{t^{\alpha}_1 c^{\beta}_2 o^{\alpha}_3} + \ket{t^{\alpha}_1 o^{\beta}_2 c^{\alpha}_3} - \\
    &\ket{c^{\beta}_1 o^{\alpha}_2 t^{\alpha}_3} + \ket{o^{\beta}_1 c^{\alpha}_2 t^{\alpha}_3} + \ket{c^{\beta}_1 t^{\alpha}_2 o^{\alpha}_3} - \ket{o^{\beta}_1 t^{\alpha}_2 o^{\alpha}_3} - \ket{t^{\beta}_1 c^{\alpha}_2 o^{\alpha}_3} + \ket{t^{\beta}_1 o^{\alpha}_2 c^{\alpha}_3} 
    + \\
    &\frac{1}{2}\ket{c^{\alpha}_1t^{\beta}_2 o^{\alpha}_3}-\frac{1}{2}\ket{c^{\alpha}_1 o^{\alpha}_2 t^{\beta}_3} - \frac{1}{2}\ket{t^{\beta}_1 c^{\alpha}_2 o^{\alpha}_3} + \frac{1}{2}\ket{t^{\beta}_1 o^{\alpha}_2 c^{\alpha}_2} + \frac{1}{2}\ket{o^{\alpha}_1 c^{\alpha}_2 t^{\beta}_3} - \frac{1}{2}\ket{o^{\alpha}_1 t^{\beta}_2 c^{\alpha}_3} + \\
    &\frac{1}{4}\ket{c^{\alpha}_1t^{\beta}_2 o^{\alpha}_3}-\frac{1}{4}\ket{c^{\alpha}_1 o^{\alpha}_2 t^{\beta}_3} - \frac{1}{4}\ket{t^{\beta}_1 c^{\alpha}_2 o^{\alpha}_3} + \frac{1}{4}\ket{t^{\beta}_1 o^{\alpha}_2 c^{\alpha}_2} + \frac{1}{4}\ket{o^{\alpha}_1 c^{\alpha}_2 t^{\beta}_3} - \frac{1}{4}\ket{o^{\alpha}_1 t^{\beta}_2 c^{\alpha}_3}
    \big)\Bigr] \\
    =& \frac{1}{6} \cdot (6 + 6/4 + 6/2)\\
    =& \frac{7}{4}
\end{align*}

\noindent The $S^2$ matrix is then constructed in the basis of $M_s = \frac{1}{2}$ determinants to be

\renewcommand{\kbldelim}{(}
\renewcommand{\kbrdelim}{)}
\[
    \kbordermatrix{
    & \ket{\Psi_{\bar{o}\bar{c}}^{\bar{t}}} & \ket{\Psi_{o\bar{c}}^{t}} & \ket{\Psi_{\bar{o}c}^{t}} \\
    \ket{\Psi_{\bar{o}\bar{c}}^{\bar{t}}}& \frac{7}{4} & 1 & 1  \\
    \ket{\Psi_{o\bar{c}}^{t}} & 1 & \frac{7}{4} & 1 \\
    \ket{\Psi_{\bar{o}c}^{t}} & 1 & 1 & \frac{7}{4}    }
\]

\noindent Diagonalization yields the eigenvalue matrix

\[
\renewcommand{\arraystretch}{0.75}
    \begin{blockarray}{ccc}
    \begin{block}{(ccc)}
     \frac{3}{4} & 0 & 0 \\
     0 & \frac{3}{4} & 0 \\
     0 & 0 & \frac{15}{4} \\
     \end{block}
    \end{blockarray}
\]
\noindent representing the 2 linearly independent doublets and one quartet, and the eigenvector matrix

\renewcommand{\kbldelim}{(}
\renewcommand{\kbrdelim}{)}
\[
    \kbordermatrix{
    & \ket{\Psi_{o\bar{c}}^{t}} & \ket{\Psi_{o\bar{c}}^{t}} & \ket{\Psi_{\bar{o}c}^{t}} \\
    \ket{\Psi_{o\bar{c}}^{t}}& -1 & -1 & 1  \\
    \ket{\Psi_{o\bar{c}}^{t}} & 0 & 1&1 \\
    \ket{\Psi_{\bar{o}c}^{t}} & 1 &0 & 1 }
\]
\vspace*{0.3 cm}

\noindent Taking appropriate linear combinations when necessary, the 2 linearly independent doublets and  quartet $M_s = \frac{1}{2}$ configuration state functions are constructed below, with the proper normalization constant.

\begin{align*}
    \ket{^{2_G} \Phi_{oc}^{t }} = &(6)^{-1/2}\Big(2\ket{\Psi_{o\bar{c}}^{t}} - \ket{\Psi_{o\bar{c}}^{t }} - \ket{\Psi_{c\bar{o}}^t}\Big) \\
    \ket{^{2_H} \Phi_{oc}^{t}} = &(2)^{-1/2}\Big(\ket{\Psi_{o\bar{c}}^{t }} - \ket{\Psi_{c\bar{o}}^t}\Big) \\
    \ket{^{4_I} \Phi_{oc}^t} = &(3)^{-1/2}\Big(\ket{\Psi_{o\bar{c}}^{t}} + \ket{\Psi_{o\bar{c}}^{t }} +\ket{\Psi_{c\bar{o}}^t}\Big)
\end{align*}

\subsubsection{Four electrons in four orbitals}
%Hamlin, this one is yours.
The possible M$_s$ = 0 states for a 4eOS system may be obtained by building, and diagonalizing, the S$^2$ matrix in the basis of M$_s$ = 0 determinants. We may begin by considering the off-diagonal terms, represented by the following calculation.
\[
    \bra{\Psi_{\bar{a}}^{\bar{s}}}S^2\ket{\Psi_a^s} = \frac{1}{2}\cdot\Bigl[\big(\bra{a^{\alpha}_1s^{\beta}_2} - \bra{s^{\beta}_1a^{\alpha}_2}\big)\times(s_1^- + s_2^-)(s^+_1 + s^+_2)\times \big(\ket{s^\alpha_1a^\beta_2} - \ket{a^\beta_1s^\alpha_2}\big)\Bigr] 
\]
\begin{align*}
    \bra{\Psi_{\bar{a}}^{\bar{s}}}S^2\ket{\Psi_a^s} =& \frac{1}{2}\cdot\Bigl[\big(\bra{a^{\alpha}_1s^{\beta}_2} - \bra{s^{\beta}_1a^{\alpha}_2}\big)\times\big(s_1^- + s_2^-\big)\times\big(\ket{s_1^{\alpha}a^{\alpha}_2} - \ket{a^{\alpha}_1s^{\alpha}_2}\big)\Bigr] \\
    =&\frac{1}{2}\cdot\Bigl[\big(\bra{a^{\alpha}_1s^{\beta}_2} - \bra{s^{\beta}_1a^{\alpha}_2}\big)\times\big(\ket{s^{\beta}_1a^{\alpha}_2} - \ket{a_1^{\beta}s_2^{\alpha}} + \ket{s_1^{\alpha}a_2^{\beta}} - \ket{a_1^{\alpha}s_2^{\beta}}\big)\Bigr]\\
    =&\frac{1}{2}\cdot\Bigl[0-0+0-1-1+0-0+0\Bigr]\\
    =&-1
\end{align*}

The diagonal terms are represented by the matrix element of a 4x4 determinant corresponding to the open-shell segment. Using the 4x4 determinant of the form
\begin{align*}
    \frac{1}{\sqrt{24}}\cdot\Big(&t_1^{\alpha}(o_2^{\beta}(y_3^{\alpha}c_4^{\beta}-c_3^{\beta}y_4^{\alpha}) - y_2^{\alpha}(o_3^{\beta}c_4^{\beta}-c_3^{\beta}o_4^{\beta}) + c_2^{\beta}(o_3^{\beta}y_4^{\alpha}-y_3^{\alpha}o_4^{\beta})) - \\
    &t_2^{\alpha}(o_1^{\beta}(y_3^{\alpha}c_4^{\beta} - c_3^{\beta}y_4^{\alpha}) - y_1^{\alpha}(o_3^{\beta}c_4^{\beta} - c_3^{\beta}o_4^{\beta}) + c_1^{\beta}(o_3^{\beta}y_4^{\alpha}-y_3^{\alpha}o_4^{\beta})) + \\
    &t_3^{\alpha}(o_1^{\beta}(y_2^{\alpha}c_4^{\beta} - c_2^{\beta}y_4^{\alpha}) - y_1^{\alpha}(o_2^{\beta}c_4^{\beta}-c_2^{\beta}o_4^{\beta}) + c_1^{\beta}(o_2^{\beta}y_4^{\alpha} - y_2^{\alpha}o_4^{\beta})) - \\
    &t_4^{\alpha}(o_1^{\beta}(y_2^{\alpha}c_3^{\beta}-c_2^{\beta}y_3^{\alpha}) - y_1^{\alpha}(o_2^{\beta}c_3^{\beta}-c_2^{\beta}o_3^{\beta}) + c_1^{\beta}(o_2^{\beta}y_3^{\alpha} - y_2^{\alpha}o_3^{\beta}))\Big)
\end{align*}

the diagonal terms can be represented by
\begin{align*}
    \bra{\Psi_{oc}^{ty}}S^2\ket{\Psi_{oc}^{ty}} =& \frac{1}{24}\cdot\Bigl[\big(\bra{t_1^{\alpha}o_2^{\beta}y_3^{\alpha}c_4^{\beta}} - \bra{t_1^{\alpha}o_2^{\beta}c_3^{\beta}y_4^{\alpha}} - \bra{t_1^{\alpha}y_2^{\alpha}o_3^{\beta}c_4^{\beta}} + \bra{t_1^{\alpha}y_2^{\alpha}c_3^{\beta}o_4^{\beta}} + \bra{t_1^{\alpha}c_2^{\beta}o_3^{\beta}y_4^{\alpha}} - \bra{t_1^{\alpha}c_2^{\beta}y_3^{\alpha}o_4^{\beta}} - \\
    &\bra{o_1^{\beta}t_2^{\alpha}y_3^{\alpha}c_4^{\beta}} + \bra{o_1^{\beta}t_2^{\alpha}c_3^{\beta}y_4^{\alpha}} + \bra{y_1^{\alpha}t_2^{\alpha}o_3^{\beta}c_4^{\beta}} - \bra{y_1^{\alpha}t_2^{\alpha}c_3^{\beta}o_4^{\beta}} - \bra{c_1^{\beta}t_2^{\alpha}o_3^{\beta}y_4^{\alpha}} + \bra{c_1^{\beta}t_2^{\alpha}y_3^{\alpha}o_4^{\beta}} + \\
    &\bra{o_1^{\beta}y_2^{\alpha}t_3^{\alpha}c_4^{\beta}} - \bra{o_1^{\beta}c_2^{\beta}t_3^{\alpha}y_4^{\alpha}} - \bra{y_1^{\alpha}o_2^{\beta}t_3^{\alpha}c_4^{\beta}} + \bra{y_1^{\alpha}c_2^{\beta}t_3^{\alpha}o_4^{\beta}} + \bra{c_1^{\beta}o_2^{\beta}t_3^{\alpha}y_4^{\alpha}} - \bra{c_1^{\beta}y_2^{\alpha}t_3^{\alpha}o_4^{\beta}} - \\
    &\bra{o_1^{\beta}y_2^{\alpha}c_3^{\beta}t_4^{\alpha}} + \bra{o_1^{\beta}c_2^{\beta}y_3^{\alpha}t_4^{\alpha}} + \bra{y_1^{\alpha}o_2^{\beta}c_3^{\beta}t_4^{\alpha}} - \bra{y_1^{\alpha}c_2^{\beta}o_3^{\beta}t_4^{\alpha}} - \bra{c_1^{\beta}o_2^{\beta}y_3^{\alpha}t_4^{\alpha}} + \bra{c_1^{\beta}y_2^{\alpha}o_3^{\beta}t_4^{\alpha}}\big)\times\\
    &(s_1^-+s_2^-+s_3^-+s_4^-)(s_1^++s_2^++s_3^++s_4^+)\times\\
    &\big(\ket{t_1^{\alpha}o_2^{\beta}y_3^{\alpha}c_4^{\beta}} - \ket{t_1^{\alpha}o_2^{\beta}c_3^{\beta}y_4^{\alpha}} - \ket{t_1^{\alpha}y_2^{\alpha}o_3^{\beta}c_4^{\beta}} + \ket{t_1^{\alpha}y_2^{\alpha}c_3^{\beta}o_4^{\beta}} + \ket{t_1^{\alpha}c_2^{\beta}o_3^{\beta}y_4^{\alpha}} - \ket{t_1^{\alpha}c_2^{\beta}y_3^{\alpha}o_4^{\beta}} - \\
    &\ket{o_1^{\beta}t_2^{\alpha}y_3^{\alpha}c_4^{\beta}} + \ket{o_1^{\beta}t_2^{\alpha}c_3^{\beta}y_4^{\alpha}} + \ket{y_1^{\alpha}t_2^{\alpha}o_3^{\beta}c_4^{\beta}} - \ket{y_1^{\alpha}t_2^{\alpha}c_3^{\beta}o_4^{\beta}} - \ket{c_1^{\beta}t_2^{\alpha}o_3^{\beta}y_4^{\alpha}} + \ket{c_1^{\beta}t_2^{\alpha}y_3^{\alpha}o_4^{\beta}} + \\
    &\ket{o_1^{\beta}y_2^{\alpha}t_3^{\alpha}c_4^{\beta}} - \ket{o_1^{\beta}c_2^{\beta}t_3^{\alpha}y_4^{\alpha}} - \ket{y_1^{\alpha}o_2^{\beta}t_3^{\alpha}c_4^{\beta}} + \ket{y_1^{\alpha}c_2^{\beta}t_3^{\alpha}o_4^{\beta}} + \ket{c_1^{\beta}o_2^{\beta}t_3^{\alpha}y_4^{\alpha}} - \ket{c_1^{\beta}y_2^{\alpha}t_3^{\alpha}o_4^{\beta}} - \\
    &\ket{o_1^{\beta}y_2^{\alpha}c_3^{\beta}t_4^{\alpha}} + \ket{o_1^{\beta}c_2^{\beta}y_3^{\alpha}t_4^{\alpha}} + \ket{y_1^{\alpha}o_2^{\beta}c_3^{\beta}t_4^{\alpha}} - \ket{y_1^{\alpha}c_2^{\beta}o_3^{\beta}t_4^{\alpha}} - \ket{c_1^{\beta}o_2^{\beta}y_3^{\alpha}t_4^{\alpha}} + \ket{c_1^{\beta}y_2^{\alpha}o_3^{\beta}t_4^{\alpha}}\big)\Bigr]
\end{align*}

\begin{align*}
    \bra{\Psi_{oc}^{ty}}S^2\ket{\Psi_{oc}^{ty}} =&\frac{1}{24}\cdot\Bigl[\big(\bra{t_1^{\alpha}o_2^{\beta}y_3^{\alpha}c_4^{\beta}} - \bra{t_1^{\alpha}o_2^{\beta}c_3^{\beta}y_4^{\alpha}} - \bra{t_1^{\alpha}y_2^{\alpha}o_3^{\beta}c_4^{\beta}} + \bra{t_1^{\alpha}y_2^{\alpha}c_3^{\beta}o_4^{\beta}} + \bra{t_1^{\alpha}c_2^{\beta}o_3^{\beta}y_4^{\alpha}} - \bra{t_1^{\alpha}c_2^{\beta}y_3^{\alpha}o_4^{\beta}} - \\
    &\bra{o_1^{\beta}t_2^{\alpha}y_3^{\alpha}c_4^{\beta}} + \bra{o_1^{\beta}t_2^{\alpha}c_3^{\beta}y_4^{\alpha}} + \bra{y_1^{\alpha}t_2^{\alpha}o_3^{\beta}c_4^{\beta}} - \bra{y_1^{\alpha}t_2^{\alpha}c_3^{\beta}o_4^{\beta}} - \bra{c_1^{\beta}t_2^{\alpha}o_3^{\beta}y_4^{\alpha}} + \bra{c_1^{\beta}t_2^{\alpha}y_3^{\alpha}o_4^{\beta}} + \\
    &\bra{o_1^{\beta}y_2^{\alpha}t_3^{\alpha}c_4^{\beta}} - \bra{o_1^{\beta}c_2^{\beta}t_3^{\alpha}y_4^{\alpha}} - \bra{y_1^{\alpha}o_2^{\beta}t_3^{\alpha}c_4^{\beta}} + \bra{y_1^{\alpha}c_2^{\beta}t_3^{\alpha}o_4^{\beta}} + \bra{c_1^{\beta}o_2^{\beta}t_3^{\alpha}y_4^{\alpha}} - \bra{c_1^{\beta}y_2^{\alpha}t_3^{\alpha}o_4^{\beta}} - \\
    &\bra{o_1^{\beta}y_2^{\alpha}c_3^{\beta}t_4^{\alpha}} + \bra{o_1^{\beta}c_2^{\beta}y_3^{\alpha}t_4^{\alpha}} + \bra{y_1^{\alpha}o_2^{\beta}c_3^{\beta}t_4^{\alpha}} - \bra{y_1^{\alpha}c_2^{\beta}o_3^{\beta}t_4^{\alpha}} - \bra{c_1^{\beta}o_2^{\beta}y_3^{\alpha}t_4^{\alpha}} + \bra{c_1^{\beta}y_2^{\alpha}o_3^{\beta}t_4^{\alpha}}\big)\times\\
    &(s_1^-+s_2^-+s_3^-+s_4^-)\times\\
    &\big(-\ket{o_1^{\alpha}t_2^{\alpha}y_3^{\alpha}c_4^{\beta}} + \ket{o_1^{\alpha}t_2^{\alpha}c_3^{\beta}y_4^{\alpha}} - \ket{c_1^{\alpha}t_2^{\alpha}o_3^{\beta}y_4^{\alpha}} + \ket{c_1^{\alpha}t_2^{\alpha}y_3^{\alpha}o_4^{\beta}} + \\
    &\ket{o_1^{\alpha}y_2^{\alpha}t_3^{\alpha}c_4^{\beta}} - \ket{o_1^{\alpha}c_2^{\beta}t_3^{\alpha}y_4^{\alpha}}+ \ket{c_1^{\alpha}o_2^{\beta}t_3^{\alpha}y_4^{\alpha}}-\ket{c_1^{\alpha}y_2^{\alpha}t_3^{\alpha}o_4^{\beta}} -\\
    &\ket{o_1^{\alpha}y_2^{\alpha}c_3^{\beta}t_4^{\alpha}} + \ket{o_1^{\alpha}c_2^{\beta}y_3^{\alpha}t_4^{\alpha}} -\ket{c_1^{\alpha}o_2^{\beta}y_3^{\alpha}t_4^{\alpha}} + \ket{c_1^{\alpha}y_2^{\alpha}o_3^{\beta}t_4^{\alpha}} + \\
    &\ket{t_1^{\alpha}o_2^{\alpha}y_3^{\alpha}c_4^{\beta}} -\ket{t_1^{\alpha}o_2^{\alpha}c_3^{\beta}y_4^{\alpha}} + \ket{t_1^{\alpha}c_2^{\alpha}o_3^{\beta}y_4^{\alpha}} -\ket{t_1^{\alpha}c_2^{\alpha}y_3^{\alpha}o_4^{\beta}} -\\
    &\ket{o_1^{\beta}c_2^{\alpha}t_3^{\alpha}y_4^{\alpha}} - \ket{y_1^{\alpha}o_2^{\alpha}t_3^{\alpha}c_4^{\beta}} + \ket{y_1^{\alpha}c_2^{\alpha}t_3^{\alpha}o_4^{\beta}} +\ket{c_1^{\beta}o_2^{\alpha}t_3^{\alpha}y_4^{\alpha}} + \\
    &\ket{o_1^{\beta}c_2^{\alpha}y_3^{\alpha}t_4^{\alpha}} + \ket{y_1^{\alpha}o_2^{\alpha}c_3^{\beta}t_4^{\alpha}} -\ket{y_1^{\alpha}c_2^{\alpha}o_3^{\beta}t_4^{\alpha}} -\ket{c_1^{\beta}o_2^{\alpha}y_3^{\alpha}t_4^{\alpha}} -\\
    &\ket{t_1^{\alpha}o_2^{\beta}c_3^{\alpha}y_4^{\alpha}} - \ket{t_1^[\alpha}y_2^{\alpha}o_3^{\alpha}c_4^{\beta}+
    \ket{t_1^{\alpha}y_2^{\alpha}c_3^{\alpha}o_4^{\beta}} + \ket{t_1^{\alpha}c_2^{\beta}o_3^{\alpha}y_4^{\alpha}}+\\
    &\ket{o_1^{\beta}t_2^{\alpha}c_3^{\alpha}y_4^{\alpha}} +\ket{y_1^{\alpha}t_2^{\alpha}o_3^{\alpha}c_4^{\beta}}-\ket{y_1^{\alpha}t_2^{\alpha}c_3^{\alpha}o_4^{\beta}} -\ket{c_1^{\beta}t_2^{\alpha}o_3^{\alpha}c_4^{\alpha}} -\\
    &\ket{o_1^{\beta}y_2^{\alpha}c_3^{\alpha}t_4^{\alpha}} + \ket{y_1^{\alpha}o_2^{\beta}c_3^{\alpha}t_4^{\alpha}} -\ket{y_1^{\alpha}c_2^{\beta}o_3^{\beta}t_4^{\alpha}} +\ket{c_1^{\beta}y_2^{\alpha}o_3^{\alpha}t_4^{\alpha}} +\\
    &\ket{t_1^{\alpha}o_2^{\beta}y_3^{\alpha}c_4^{\alpha}} - \ket{t_1^{\alpha}y_2^{\alpha}o_3^{\beta}c_4^{\alpha}} + \ket{t_1^{\alpha}y_2^{\alpha}c_3^{\beta}o_4^{\alpha}} -\ket{t_1^{\alpha}c_2^{\beta}y_3^{\alpha}o_4^{\alpha}} -\\
    &\ket{o_1^{\beta}t_2^{\alpha}y_3^{\alpha}c_4^{\alpha}} + \ket{y_1^{\alpha}t_2^{\alpha}o_3^{\beta}c_4^{\alpha}}-\ket{y_1^{\alpha}t_2^{\alpha}c_3^{\beta}o_4^{\alpha}} +\ket{c_1^{\beta}t_2^{\alpha}y_3^{\alpha}o_4^{\alpha}} +\\
    &\ket{o_1^{\beta}y_2^{\alpha}t_3^{\alpha}c_4^{\alpha}} -\ket{y_1^{\alpha}o_2^{\beta}t_3^{\alpha}c_4^{\alpha}} +\ket{y_1^{\alpha}c_2^{\beta}t_3^{\alpha}o_4^{\alpha}} -\ket{c_1^{\beta}y_2^{\alpha}t_3^{\alpha}o_4^{\alpha}}\big)\Bigr]
\end{align*}
{
\allowdisplaybreaks
\begin{align*}
     \bra{\Psi_{oc}^{ty}}S^2\ket{\Psi_{oc}^{ty}} =&\frac{1}{24}\cdot\Bigl[\big(...\big)\times
    \big(-\ket{o_1^{\beta}t_2^{\alpha}y_3^{\alpha}c_4^{\beta}} + \ket{o_1^{\beta}t_2^{\alpha}c_3^{\beta}y_4^{\alpha}} - \ket{c_1^{\beta}t_2^{\alpha}o_3^{\beta}y_4^{\alpha}} + \ket{c_1^{\beta}t_2^{\alpha}y_3^{\alpha}o_4^{\beta}} + \\
    &\ket{o_1^{\beta}y_2^{\alpha}t_3^{\alpha}c_4^{\beta}} - \ket{o_1^{\beta}c_2^{\beta}t_3^{\alpha}y_4^{\alpha}}+ \ket{c_1^{\beta}o_2^{\beta}t_3^{\alpha}y_4^{\alpha}}-\ket{c_1^{\beta}y_2^{\alpha}t_3^{\alpha}o_4^{\beta}} -\\
    &\ket{o_1^{\beta}y_2^{\alpha}c_3^{\beta}t_4^{\alpha}} + \ket{o_1^{\beta}c_2^{\beta}y_3^{\alpha}t_4^{\alpha}} -\ket{c_1^{\beta}o_2^{\beta}y_3^{\alpha}t_4^{\alpha}} + \ket{c_1^{\beta}y_2^{\alpha}o_3^{\beta}t_4^{\alpha}} + \\
    &\ket{t_1^{\beta}o_2^{\alpha}y_3^{\alpha}c_4^{\beta}} -\ket{t_1^{\beta}o_2^{\alpha}c_3^{\beta}y_4^{\alpha}} + \ket{t_1^{\beta}c_2^{\alpha}o_3^{\beta}y_4^{\alpha}} -\ket{t_1^{\beta}c_2^{\alpha}y_3^{\alpha}o_4^{\beta}} -\\
    &\ket{y_1^{\beta}o_2^{\alpha}t_3^{\alpha}c_4^{\beta}} + \ket{y_1^{\beta}c_2^{\alpha}t_3^{\alpha}o_4^{\beta}}+ \\
    &\ket{y_1^{\beta}o_2^{\alpha}c_3^{\beta}t_4^{\alpha}} -\ket{y_1^{\beta}c_2^{\alpha}o_3^{\beta}t_4^{\alpha}} -\\
    &\ket{t_1^{\beta}o_2^{\beta}c_3^{\alpha}y_4^{\alpha}} - \ket{t_1^{\beta}y_2^{\alpha}o_3^{\alpha}c_4^{\beta}}+
    \ket{t_1^{\beta}y_2^{\alpha}c_3^{\alpha}o_4^{\beta}} + \ket{t_1^{\beta}c_2^{\beta}o_3^{\alpha}y_4^{\alpha}}+\\
    &\ket{y_1^{\beta}t_2^{\alpha}o_3^{\alpha}c_4^{\beta}}-\ket{y_1^{\beta}t_2^{\alpha}c_3^{\alpha}o_4^{\beta}} +\\
    &\ket{y_1^{\beta}o_2^{\beta}c_3^{\alpha}t_4^{\alpha}} -\ket{y_1^{\beta}c_2^{\beta}o_3^{\alpha}t_4^{\alpha}} +\\
    &\ket{t_1^{\beta}o_2^{\beta}y_3^{\alpha}c_4^{\alpha}} - \ket{t_1^{\beta}y_2^{\alpha}o_3^{\beta}c_4^{\alpha}} + \ket{t_1^{\beta}y_2^{\alpha}c_3^{\beta}o_4^{\alpha}} -\ket{t_1^{\beta}c_2^{\beta}y_3^{\alpha}o_4^{\alpha}} +\\
    &\ket{y_1^{\beta}t_2^{\alpha}o_3^{\beta}c_4^{\alpha}}-\ket{y_1^{\beta}t_2^{\alpha}c_3^{\beta}o_4^{\alpha}} -\\
    &\ket{y_1^{\beta}o_2^{\beta}t_3^{\alpha}c_4^{\alpha}} +\ket{y_1^{\beta}c_2^{\beta}t_3^{\alpha}o_4^{\alpha}} -\\ 
    %s2
    &\ket{o_1^{\alpha}t_2^{\beta}y_3^{\alpha}c_4^{\beta}} + \ket{o_1^{\alpha}t_2^{\beta}c_3^{\beta}y_4^{\alpha}} - \ket{c_1^{\alpha}t_2^{\beta}o_3^{\beta}y_4^{\alpha}} + \ket{c_1^{\alpha}t_2^{\beta}y_3^{\alpha}o_4^{\beta}} + \\
    &\ket{o_1^{\alpha}y_2^{\beta}t_3^{\alpha}c_4^{\beta}} - \ket{c_1^{\alpha}y_2^{\beta}t_3^{\alpha}o_4^{\beta}} -\\
    &\ket{o_1^{\alpha}y_2^{\beta}c_3^{\beta}t_4^{\alpha}} + \ket{c_1^{\alpha}y_2^{\beta}o_3^{\beta}t_4^{\alpha}} + \\
    &\ket{t_1^{\alpha}o_2^{\beta}y_3^{\alpha}c_4^{\beta}} -\ket{t_1^{\alpha}o_2^{\beta}c_3^{\beta}y_4^{\alpha}} + \ket{t_1^{\alpha}c_2^{\beta}o_3^{\beta}y_4^{\alpha}} -\ket{t_1^{\alpha}c_2^{\beta}y_3^{\alpha}o_4^{\beta}} -\\
    &\ket{o_1^{\beta}c_2^{\beta}t_3^{\alpha}y_4^{\alpha}} - \ket{y_1^{\alpha}o_2^{\beta}t_3^{\alpha}c_4^{\beta}} + \ket{y_1^{\alpha}c_2^{\beta}t_3^{\alpha}o_4^{\beta}} +\ket{c_1^{\beta}o_2^{\beta}t_3^{\alpha}y_4^{\alpha}} + \\
    &\ket{o_1^{\beta}c_2^{\beta}y_3^{\alpha}t_4^{\alpha}} + \ket{y_1^{\alpha}o_2^{\beta}c_3^{\beta}t_4^{\alpha}} -\ket{y_1^{\alpha}c_2^{\beta}o_3^{\beta}t_4^{\alpha}} -\ket{c_1^{\beta}o_2^{\beta}y_3^{\alpha}t_4^{\alpha}} -\\
    &\ket{t_1^{\alpha}y_2^{\beta}o_3^{\alpha}c_4^{\beta}}+
    \ket{t_1^{\alpha}y_2^{\beta}c_3^{\alpha}o_4^{\beta}} + \\
    &\ket{o_1^{\beta}t_2^{\beta}c_3^{\alpha}y_4^{\alpha}} +\ket{y_1^{\alpha}t_2^{\beta}o_3^{\alpha}c_4^{\beta}}-\ket{y_1^{\alpha}t_2^{\beta}c_3^{\alpha}o_4^{\beta}} -\ket{c_1^{\beta}t_2^{\beta}o_3^{\alpha}c_4^{\alpha}} -\\
    &\ket{o_1^{\beta}y_2^{\beta}c_3^{\alpha}t_4^{\alpha}} + \ket{c_1^{\beta}y_2^{\beta}o_3^{\alpha}t_4^{\alpha}} -\\
    &\ket{t_1^{\alpha}y_2^{\beta}o_3^{\beta}c_4^{\alpha}} + \ket{t_1^{\alpha}y_2^{\beta}c_3^{\beta}o_4^{\alpha}} -\\
    &\ket{o_1^{\beta}t_2^{\beta}y_3^{\alpha}c_4^{\alpha}} + \ket{y_1^{\alpha}t_2^{\beta}o_3^{\beta}c_4^{\alpha}}-\ket{y_1^{\alpha}t_2^{\beta}c_3^{\beta}o_4^{\alpha}} +\ket{c_1^{\beta}t_2^{\beta}y_3^{\alpha}o_4^{\alpha}} +\\
    &\ket{o_1^{\beta}y_2^{\beta}t_3^{\alpha}c_4^{\alpha}} -\ket{c_1^{\beta}y_2^{\beta}t_3^{\alpha}o_4^{\alpha}} - \\
    %s3
    &\ket{o_1^{\alpha}t_2^{\alpha}y_3^{\beta}c_4^{\beta}} + \ket{c_1^{\alpha}t_2^{\alpha}y_3^{\beta}o_4^{\beta}} + \\
    &\ket{o_1^{\alpha}y_2^{\alpha}t_3^{\beta}c_4^{\beta}} - \ket{o_1^{\alpha}c_2^{\beta}t_3^{\beta}y_4^{\alpha}}+ \ket{c_1^{\alpha}o_2^{\beta}t_3^{\beta}y_4^{\alpha}}-\ket{c_1^{\alpha}y_2^{\alpha}t_3^{\beta}o_4^{\beta}} +\\
    &\ket{o_1^{\alpha}c_2^{\beta}y_3^{\beta}t_4^{\alpha}} -\ket{c_1^{\alpha}o_2^{\beta}y_3^{\beta}t_4^{\alpha}} +  \\
    &\ket{t_1^{\alpha}o_2^{\alpha}y_3^{\beta}c_4^{\beta}}  -\ket{t_1^{\alpha}c_2^{\alpha}y_3^{\beta}o_4^{\beta}} -\\
    &\ket{o_1^{\beta}c_2^{\alpha}t_3^{\beta}y_4^{\alpha}} - \ket{y_1^{\alpha}o_2^{\alpha}t_3^{\beta}c_4^{\beta}} + \ket{y_1^{\alpha}c_2^{\alpha}t_3^{\beta}o_4^{\beta}} +\ket{c_1^{\beta}o_2^{\alpha}t_3^{\beta}y_4^{\alpha}} + \\
    &\ket{o_1^{\beta}c_2^{\alpha}y_3^{\beta}t_4^{\alpha}} -\ket{c_1^{\beta}o_2^{\alpha}y_3^{\beta}t_4^{\alpha}} -\\
    &\ket{t_1^{\alpha}o_2^{\beta}c_3^{\beta}y_4^{\alpha}} - \ket{t_1^{\alpha}y_2^{\alpha}o_3^{\beta}c_4^{\beta}}+
    \ket{t_1^{\alpha}y_2^{\alpha}c_3^{\beta}o_4^{\beta}} + \ket{t_1^{\alpha}c_2^{\beta}o_3^{\beta}y_4^{\alpha}}+\\
    &\ket{o_1^{\beta}t_2^{\alpha}c_3^{\beta}y_4^{\alpha}} +\ket{y_1^{\alpha}t_2^{\alpha}o_3^{\beta}c_4^{\beta}}-\ket{y_1^{\alpha}t_2^{\alpha}c_3^{\beta}o_4^{\beta}} -\ket{c_1^{\beta}t_2^{\alpha}o_3^{\beta}c_4^{\alpha}} -\\
    &\ket{o_1^{\beta}y_2^{\alpha}c_3^{\beta}t_4^{\alpha}} + \ket{y_1^{\alpha}o_2^{\beta}c_3^{\beta}t_4^{\alpha}} -
    \ket{y_1^{\alpha}c_2^{\beta}o_3^{\beta}t_4^{\alpha}} +
    \ket{c_1^{\beta}y_2^{\alpha}o_3^{\beta}t_4^{\alpha}} +\\
    &\ket{t_1^{\alpha}o_2^{\beta}y_3^{\beta}c_4^{\alpha}} - \ket{t_1^{\alpha}c_2^{\beta}y_3^{\beta}o_4^{\alpha}} -\\
    &\ket{o_1^{\beta}t_2^{\alpha}y_3^{\beta}c_4^{\alpha}} +\ket{c_1^{\beta}t_2^{\alpha}y_3^{\beta}o_4^{\alpha}} +\\
    &\ket{o_1^{\beta}y_2^{\alpha}t_3^{\beta}c_4^{\alpha}} -\ket{y_1^{\alpha}o_2^{\beta}t_3^{\beta}c_4^{\alpha}} +\ket{y_1^{\alpha}c_2^{\beta}t_3^{\beta}o_4^{\alpha}} -\ket{c_1^{\beta}y_2^{\alpha}t_3^{\beta}o_4^{\alpha}} -\\
    %s4
    &\ket{o_1^{\alpha}t_2^{\alpha}c_3^{\beta}y_4^{\beta}} - \ket{c_1^{\alpha}t_2^{\alpha}o_3^{\beta}y_4^{\beta}} - \\
    &\ket{o_1^{\alpha}c_2^{\beta}t_3^{\alpha}y_4^{\beta}}+ \ket{c_1^{\alpha}o_2^{\beta}t_3^{\alpha}y_4^{\beta}}-\\
    &\ket{o_1^{\alpha}y_2^{\alpha}c_3^{\beta}t_4^{\beta}} + \ket{o_1^{\alpha}c_2^{\beta}y_3^{\alpha}t_4^{\beta}} -\ket{c_1^{\alpha}o_2^{\beta}y_3^{\alpha}t_4^{\beta}} + \ket{c_1^{\alpha}y_2^{\alpha}o_3^{\beta}t_4^{\beta}} - \\
    &\ket{t_1^{\alpha}o_2^{\alpha}c_3^{\beta}y_4^{\beta}} + \ket{t_1^{\alpha}c_2^{\alpha}o_3^{\beta}y_4^{\beta}} -\\
    &\ket{o_1^{\beta}c_2^{\alpha}t_3^{\alpha}y_4^{\beta}} +\ket{c_1^{\beta}o_2^{\alpha}t_3^{\alpha}y_4^{\beta}} + \\
    &\ket{o_1^{\beta}c_2^{\alpha}y_3^{\alpha}t_4^{\beta}} + \ket{y_1^{\alpha}o_2^{\alpha}c_3^{\beta}t_4^{\beta}} -\ket{y_1^{\alpha}c_2^{\alpha}o_3^{\beta}t_4^{\beta}} -\ket{c_1^{\beta}o_2^{\alpha}y_3^{\alpha}t_4^{\beta}} -\\
    &\ket{t_1^{\alpha}o_2^{\beta}c_3^{\alpha}y_4^{\beta}} + \ket{t_1^{\alpha}c_2^{\beta}o_3^{\alpha}y_4^{\beta}}+\\
    &\ket{o_1^{\beta}t_2^{\alpha}c_3^{\alpha}y_4^{\beta}} -\ket{c_1^{\beta}t_2^{\alpha}o_3^{\alpha}c_4^{\beta}} -\\
    &\ket{o_1^{\beta}y_2^{\alpha}c_3^{\alpha}t_4^{\beta}} + \ket{y_1^{\alpha}o_2^{\beta}c_3^{\alpha}t_4^{\beta}} -\ket{y_1^{\alpha}c_2^{\beta}o_3^{\alpha}t_4^{\beta}} +\ket{c_1^{\beta}y_2^{\alpha}o_3^{\alpha}t_4^{\beta}} +\\
    &\ket{t_1^{\alpha}o_2^{\beta}y_3^{\alpha}c_4^{\beta}} - \ket{t_1^{\alpha}y_2^{\alpha}o_3^{\beta}c_4^{\beta}} + \ket{t_1^{\alpha}y_2^{\alpha}c_3^{\beta}o_4^{\beta}} -\ket{t_1^{\alpha}c_2^{\beta}y_3^{\alpha}o_4^{\beta}} -\\
    &\ket{o_1^{\beta}t_2^{\alpha}y_3^{\alpha}c_4^{\beta}} + \ket{y_1^{\alpha}t_2^{\alpha}o_3^{\beta}c_4^{\beta}}-\ket{y_1^{\alpha}t_2^{\alpha}c_3^{\beta}o_4^{\beta}} +\ket{c_1^{\beta}t_2^{\alpha}y_3^{\alpha}o_4^{\beta}} +\\
    &\ket{o_1^{\beta}y_2^{\alpha}t_3^{\alpha}c_4^{\beta}} -\ket{y_1^{\alpha}o_2^{\beta}t_3^{\alpha}c_4^{\beta}} +\ket{y_1^{\alpha}c_2^{\beta}t_3^{\alpha}o_4^{\beta}} -\ket{c_1^{\beta}y_2^{\alpha}t_3^{\alpha}o_4^{\beta}}\\
    \bra{\Psi_{oc}^{ty}}S^2\ket{\Psi_{oc}^{ty}} =&\frac{1}{24}\cdot\Bigl[24*2\Bigr] \\
    =& 2
\end{align*}
}
Denoting the 6 $M_s =0$ determinants $\ckc,\ckd,\ckg,\cke,\ckf,\ckh$ as $\ket{1},\ket{2},\ket{3},\ket{4},\ket{5},\ket{6}$ respectively, the upper triangular matrix elements may be calculated by placing determinants in maximum coincidence, accounting for sign change corresponding to the number of swaps required to achieve max coincidence, and using the relevant matrix element derived above.
\begin{align*}
    \ckc =& \ket{t\bar{o}y\bar{c}}\\
    \ckd =& \ket{o\bar{t}c\bar{y}}\\
    \ckg =& \ket{o\bar{y}t\bar{c}} \\
    \cke =&  \ket{o\bar{t}y\bar{c}} \\
    \ckf =& \ket{t\bar{o}c\bar{y}} \\
    \ckh =&\ket{y\bar{o}c\bar{t}}
\end{align*}
Compiling results, the upper triangular matrix elements are evaluated as
\begin{align*}
    &\braket{1|2} = 0, \braket{1|3} = 1, \braket{1|4} = -1, \braket{1|5} = -1, \braket{1|6} = 1\\
    &\braket{2|3} = , \braket{2|4} = -1, \braket{2|5} = -1, \braket{2|6} =1\\
    &\braket{3|4} = -1, \braket{3|5} = -1, \braket{3|6} = 0\\
    &\braket{4|5} = 0, \braket{4|6} = -1\\
    &\braket{5|6} = -1
\end{align*}
Combined with the diagonal matrix elements determined above, the $S^2$ matrix in the basis $\ckc,\ckd,\ckg,\cke,\ckf,\ckh$ is determined to be 
\renewcommand{\kbldelim}{(}
\renewcommand{\kbrdelim}{)}
\[
    \kbordermatrix{
    & \ckc & \ckd & \ckg & \cke & \ckf & \ckh \\
    \ckc & 2 & 0 & 1 & -1 & -1 & 1 \\
    \ckd & 0 & 2 & 1 & -1 & -1 & 1\\
    \ckg & 1 & 1 & 2 & -1 & -1 & 0\\
    \cke & -1 & -1 & -1 & 2 & 0 & -1\\
    \ckf & -1 & -1 & -1 & 0 & 2 & -1\\
    \ckh & 1 & 1 & 0 & -1 & -1 & 2
    }
\]
Diagonalization yields the eigenvalue matrix, 
\[
\renewcommand{\arraystretch}{0.75}
    \begin{blockarray}{cccccc}
    \begin{block}{(cccccc)}
     0 & 0 & 0 & 0 & 0 & 0 \\
     0 & 0 & 0 & 0 & 0 & 0\\
     0 & 0 & 2 & 0 & 0 & 0\\
     0 & 0 & 0 & 2 & 0 & 0\\
     0 & 0 & 0 & 0 & 2 & 0\\
     0 & 0 & 0 & 0 & 0 & 6\\
     \end{block}
    \end{blockarray}
\]
representing the possible two linearly independent singlets, three linearly independent triplets, and quintet state possible for a 4 open-shell $M_s=0$ state, and the eigenvector matrix
\renewcommand{\kbldelim}{(}
\renewcommand{\kbrdelim}{)}
\[
    \kbordermatrix{
    & \ckc & \ckd & \ckg & \cke & \ckf & \ckh \\
    \ckc & -1 & 1 & 0 & 0 & -1 & 1 \\
    \ckd & -1 & 1 & 0 & 0 & 1 & 1\\
    \ckg & 1 & 0 & -1 & 0 & 0 & 1\\
    \cke & 0 & 1 & 0 & -1 & 0 & -1\\
    \ckf & 0 & 1 & 0 & 1 & 0 & -1\\
    \ckh & 1 & 0 & 1 & 0 & 0 & 1
    }
\]
Taking linear combinations of eigenvectors of degenerate eigenvalue when necessary to maintain orthogonality, the 2 distinct singlets, 3 triplets, and quintet are constructed as below, normalizing appropriately.
\begin{align*}
    \ket{^{1_A}\Phi_{oc}^{ty}} &= (12)^{-1/2}  \left(2\ckc + 2\ckd + \cke + \ckf - \ckg - \ckh \right) \\
    \ket{^{1_B}\Phi_{oc}^{ty}} &= (2)^{-1}   \left(\cke + \ckf + \ckg + \ckh \right) \\
    \ket{^{3_C}\Phi_{oc}^{ty}} &= (2)^{-1/2} \left(\ckc -\ckd \right) \\
    \ket{^{3_D}\Phi_{oc}^{ty}} &= (2)^{-1/2} \left(\cke - \ckf \right) \\
    \ket{^{3_E}\Phi_{oc}^{ty}} &= (2)^{-1/2} \left(\ckg-\ckh \right) \\
    \ket{^{5_F}\Phi_{oc}^{ty}} &= (6)^{-1/2} \left(\ckc + \ckd - \cke - \ckf + \ckg + \ckh \right)
\end{align*}

\subsection{Matrix elements for singlet CSFs}

There are two linearly-independent singlet CSFs
\begin{align*}
    \ket{^A \Psi_{oc}^{ty}} = & (12)^{-1/2} \big(2 \ckc + 2 \ckd + \cke + \ckf - \ckg - \ckh \big) \\ 
    = & (12)^{-1/2} 
    \big(2 \ket{C} + 2 \ket{D} + \ket{E} + \ket{F} - \ket{G} - \ket{H} \big) \\
    \ket{^B \Psi_{oc}^{ty}} = & (2)^{-1} \big(\cke + \ckf + \ckg + \ckh \big) \\
    = & (2)^{-1} 
    \big(\ket{E} + \ket{F} + \ket{G} + \ket{H} \big) \\
\end{align*}

Finding the Hamiltonian matrix elements of the CSFs consists of just adding the appropriate linear combination of Slater-determinant matrix elements. The contributions from each spin compliment will be doubled. 
\subsubsection{$\bra{^1 \Psi_{c}^{t}} H \ket{^1 \Psi_{c}^{t}}$}
\begin{align*}
    2 \bra{^1 \Psi_{c}^{t}} H \ket{^1 \Psi_{c}^{t}}  = &
    2 \cdot \big( H_{AA} + H_{AB} \big) \\ 
    \bra{^1 \Psi_{c}^{t}} H \ket{^1 \Psi_{c}^{t}} 
    = & \big( 
    f_{tt} - f_{cc} + 2\braket{tc | ct} - \braket{tc | tc} \big) 
\end{align*}

\subsubsection{$\bra{^1 \Psi_{c}^{t}} H \ket{^A \Psi_{oc}^{ty}}$}
\begin{align*}
    \sqrt{24} \bra{^1 \Psi_{c}^{t}} H \ket{^A \Psi_{oc}^{ty}}  = 
    2 \cdot \big( & 2H_{AC} + 2H_{AD} + H_{AE} + H_{AF} - H_{AG} - H_{AH} \big) \\ 
    \bra{^1 \Psi_{c}^{t}} H \ket{^A \Psi_{oc}^{ty}} 
    =  \frac{1}{\sqrt{6}} \cdot \Big[ &
    3 \delta_{ty} f_{ot} - 3 f_{oy} + \\ &
    3 \braket{to | yt} - 3 \braket{to | ty} + 3 \braket{oc | yc} - 3 \braket{oc | cy} + 
    \delta_{ty} \big( 
    3 \braket{oc | ct} - 3 \braket{oc | tc} \big) \Big] \\
    = \frac{\sqrt{6}}{2} \cdot \Big[ &
    \delta_{ty} f_{ot} - f_{oy} + \braket{to | yt} - \braket{to | ty} + \braket{oc | yc} - \braket{oc | cy} + 
    \delta_{ty} \big( 
    \braket{oc | ct} - \braket{oc | tc} \big) \Big]
\end{align*}

\begin{align*}
\text{Special case: } y \neq & \; t \\
    \bra{^1 \Psi_{c}^{t}} H \ket{^A \Psi_{oc}^{ty}} = \;
    = & \frac{\sqrt{6}}{2} \cdot \Big[ - f_{oy} + 
    \braket{to | yt} - \braket{to | ty} + \braket{oc | yc} - \braket{oc | cy} \Big] \\
\text{Special case: } y = & \; t   \\
    \bra{^1 \Psi_{c}^{t}} H \ket{^A \Psi_{oc}^{ty}} = \; & 0
\end{align*}

\subsubsection{$\bra{^1 \Psi_{c}^{t}} H \ket{^B \Psi_{oc}^{ty}}$}
\begin{align*}
    2\sqrt{2} \bra{^1 \Psi_{c}^{t}} H \ket{^B \Psi_{oc}^{ty}}  = 
    2 \cdot \big( & H_{AE} + H_{AF} + H_{AG} + H_{AH} \big) \\ 
    \bra{^1 \Psi_{c}^{t}} H \ket{^B \Psi_{oc}^{ty}} 
    = (2)^{-1/2} \cdot \Big[ &
    f_{oy} + \delta_{ty} f_{ot}  + \\ &
    \braket{to | yt} + \braket{to | ty} - \braket{oc | yc} - \braket{oc | cy} - 
    \delta_{ty} \big( 
    \braket{oc | tc} + \braket{oc | ct} \big) \Big]
\end{align*}

\begin{align*}
\text{Special case: } y \neq & \; t \\
    \bra{^1 \Psi_{c}^{t}} H \ket{^B \Psi_{oc}^{ty}} = \;
    & (2)^{-1/2} \cdot \Big[ 
    f_{oy} + 
    \braket{to | yt} + \braket{to | ty} - \braket{oc | yc} - \braket{oc | cy} \Big] \\
\text{Special case: } y = & \; t \text{. Additional factor of } (2)^{-1/2} \text{ to re-normalize} \ket{^B \Psi_{oc}^{ty}} \\
    \bra{^1 \Psi_{c}^{t}} H \ket{^B \Psi_{oc}^{ty}} = \;
    & (4)^{-1/2} \cdot \Big[
    2f_{ot} +
    \braket{to | tt} + \braket{to | tt} - 2 \braket{oc | tc} - 2 \braket{oc | ct}\Big] \\
    = \; & f_{ot} + \braket{to | tt} - \braket{oc | tc} - \braket{oc | ct}
\end{align*}

\subsubsection{$\bra{^A \Psi_{oc}^{ty}} H \ket{^A \Psi_{oc}^{ty}}$}

\begin{align*}
    12 \bra{^A \Psi_{oc}^{tx}} H \ket{^A \Psi_{oc}^{ty}} = 
    4 \cdot \big(& 2H_{CC} + 2H_{CD} + H_{CE} + H_{CF} - H_{CG} - H_{CH} \big) + \\
    2 \cdot \big(& 2H_{EC} + 2H_{ED} + H_{EE} + H_{EF} - H_{EG} - H_{EH} \big) + \\ 
    2 \cdot \big(& -2H_{GC} - 2H_{GD} - H_{GE} - H_{GF} + H_{GG} + H_{HG} \big) \\ = 
    4 \cdot \Big[& 2 \big( f_{xy} - \delta_{ty} f_{xt} - \delta_{xt} f_{ty} + \delta_{xy} (f_{tt} - f_{cc} - f_{oo}) + \delta_{xt}\delta_{ty} (f_{cc} + f_{oo}) \big) + \\ &
                 2\braket{tx | ty}  
               - 2\braket{tx | yt} 
               - 2\braket{xc | yc} 
               + 3\braket{xc | cy}  
               - 2\braket{xo | yo} 
               + 3\braket{xo | oy}  \\ 
               \delta_{xt} \big( &
                   2\braket{tc | yc} 
                 - 3\braket{tc | cy}  
                 + 2\braket{to | yo}  
                 - 3\braket{to | oy} \big) \\  
               \delta_{yt} \big( &
                   2\braket{xc | tc} 
                 - 3\braket{xc | ct}  
                 + 2\braket{xo | to}  
                 - 3\braket{xo | ot} \big) + \\ 
               \delta_{xy} \big( &
                   2\braket{oc | oc}  
                 - 2\braket{oc | co} 
                 - 2\braket{tc | tc} 
                 + 3\braket{tc | ct}  
                 - 2\braket{to | to}  
                 + 3\braket{to | ot} \big) - \\ 
               \delta_{xt} \delta_{yt} \big( &
                   2\braket{oc | oc}  
                 - 2\braket{oc | co} \big) \Big] + \\ 
    2 \cdot \Big[ & f_{xy} - \delta_{xt} f_{ty} - \delta_{ty} f_{xt} + \delta_{xy} (f_{tt} - f_{cc} - f_{oo}) + \delta_{xt}\delta_{ty} (f_{cc} + f_{oo}) + \\ &
                   \braket{tx | ty} 
                -  \braket{tx | yt}
                -  \braket{xc | yc}  
                + 3\braket{xc | cy}  
                -  \braket{xo | yo}  \\ 
                \delta_{xt} \big( &
                     \braket{tc | yc}
                  +  \braket{to | yo}
                  - 3\braket{to | oy}  
                \big) + \\ 
                \delta_{yt} \big( &
                     \braket{xo | to}
                  +  \braket{tc | xc}
                  - 3\braket{tc | cx}
                \big) + \\ 
                \delta_{xy} \big( &
                     \braket{oc | oc} 
                  -  \braket{oc | co}
                  -  \braket{tc | tc}  
                  -  \braket{to | to}  
                  + 3\braket{to | ot}
                \big) + \\ 
                \delta_{xt} \delta_{yt} \big( &
                  \braket{co | oc} - 
                  \braket{oc | oc}
                \big) \Big] + \\ 
    2 \cdot \Big[ & f_{xy} - \delta_{xt} f_{ty} - \delta_{ty} f_{xt} + \delta_{xy} (f_{tt} - f_{cc} - f_{oo}) + \delta_{xt}\delta_{ty} (f_{cc} + f_{oo}) + \\ &
                   \braket{xt | yt}
                -  \braket{xt | ty}
                -  \braket{xc | yc}  
                -  \braket{xo | yo}  
                + 3\braket{xo | oy} \\
                \delta_{xt} \big( &
                     \braket{to | yo}
                  +  \braket{tc | yc}
                  - 3\braket{tc | cy}
                \big) + \\
                \delta_{yt} \big( &
                     \braket{tc | xc}
                  +  \braket{to | xo}
                  - 3\braket{to | ox}
                \big) + \\
                \delta_{xy} \big( &
                    \braket{oc | oc} 
                  - \braket{oc | co}
                  - \braket{to | to} 
                  - \braket{tc | tc} 
                  + 3\braket{tc | ct}
                \big)  + \\
                \delta_{xt} \delta_{yt} \big( &
                  \braket{oc | co} - 
                  \braket{oc | oc}
                \big) \Big] 
\end{align*}

\begin{align*}
    \bra{^A \Psi_{oc}^{tx}} H \ket{^A \Psi_{oc}^{ty}} = 
    12 \cdot \Big[& f_{xy} - \delta_{xt} f_{ty} - \delta_{ty} f_{xt} + \delta_{xy} (f_{tt} - f_{cc} - f_{oo}) + \delta_{xt}\delta_{ty} (f_{cc} + f_{oo}) \Big] + \\ \Big[ &
                 12\braket{tx | ty}  
               - 12\braket{tx | yt} 
               - 12\braket{xc | yc} 
               + 18\braket{xc | cy}  
               - 12\braket{xo | yo} 
               + 18\braket{xo | oy}  \\ 
               \delta_{xt} \big( &
                   12\braket{tc | yc} 
                 - 18\braket{tc | cy} 
                 + 12\braket{to | yo}  
                 - 18\braket{to | oy} \big) \\  
               \delta_{yt} \big( &
                   12\braket{xc | tc} 
                 - 18\braket{xc | ct} 
                 + 12\braket{xo | to}  
                 - 18\braket{xo | ot} \big) + \\ 
               \delta_{xy} \big( &
                   12\braket{oc | oc}  
                 - 12\braket{oc | co} 
                 - 12\braket{tc | tc} 
                 + 18\braket{tc | ct}  
                 - 12\braket{to | to}  
                 + 18\braket{to | ot} \big) - \\ 
               \delta_{xt} \delta_{yt} \big( &
                   12\braket{oc | oc}  
                 - 12\braket{oc | co}
                 \big) \Big]  \\ = 
    \Big[& f_{xy} - \delta_{xt} f_{ty} - \delta_{ty} f_{xt} + \delta_{xy} (f_{tt} - f_{cc} - f_{oo}) + \delta_{xt}\delta_{ty} (f_{cc} + f_{oo})\Big] + \\ &
               \braket{tx | ty} - 
               \braket{tx | yt} -
               \braket{xc | yc} +
               \frac{3}{2}\braket{xc | cy} - 
               \braket{xo | yo} +
               \frac{3}{2}\braket{xo | oy} + \\ 
               \delta_{xt} \big( &
                 \braket{tc | yc} -
                 \frac{3}{2} \braket{tc | cy} + 
                 \braket{to | yo} - 
                 \frac{3}{2} \braket{to | oy} \big) \\  
               \delta_{yt} \big( &
                 \braket{xc | tc} -
                 \frac{3}{2} \braket{xc | ct} + 
                 \braket{xo | to} - 
                 \frac{3}{2}\braket{xo | ot} \big) + \\ 
               \delta_{xy} \big( &
                 \braket{oc | oc} - 
                 \braket{oc | co} -
                 \braket{tc | tc} +
                 \frac{3}{2}\braket{tc | ct} - 
                 \braket{to | to} + 
                 \frac{3}{2}\braket{to | ot} \big) - \\ 
               \delta_{xt} \delta_{yt} \big( &
                 \braket{oc | oc} - 
                 \braket{oc | co}
                 \big)              
\end{align*}

\newpage
\begin{align*}
\text{Special case: }x \neq y \text{ and } x, y \neq t \\
    \bra{^A \Psi_{oc}^{tx}} H \ket{^A \Psi_{oc}^{ty}} = \;
    & f_{xy} + \\ \Big[ &
               \braket{tx | ty} - 
               \braket{tx | yt} -
               \braket{xc | yc} +
               \frac{3}{2}\braket{xc | cy} - 
               \braket{xo | yo} +
               \frac{3}{2}\braket{xo | oy} \Big] \\
\text{Special case: }x = y \text{ and } x, y \neq t \\
    \bra{^A \Psi_{oc}^{tx}} H \ket{^A \Psi_{oc}^{ty}} = \;
    \Big[& f_{xy} + f_{tt} - f_{cc} - f_{oo} \Big] + \\ \Big[ &
               \braket{tx | tx} - 
               \braket{tx | xt} -
               \braket{xc | xc} +
               \frac{3}{2}\braket{xc | cx} - 
               \braket{xo | xo} +
               \frac{3}{2}\braket{xo | ox} + \\ 
               \big( &
                 \braket{oc | oc} - 
                 \braket{oc | co} -
                 \braket{tc | tc} +
                 \frac{3}{2}\braket{tc | ct} - 
                 \braket{to | to} + 
                 \frac{3}{2}\braket{to | ot} \big) \Big] \\
\text{Special case: }x \neq y \text{ and } x = t \\
    \bra{^A \Psi_{oc}^{ty}} H \ket{^A \Psi_{oc}^{ty}} = \;
    \Big[& f_{ty} - f_{ty} \Big] + \\ \Big[ &
               \braket{tt | ty} - 
               \braket{tt | yt} -
               \braket{tc | yc} +
               \frac{3}{2}\braket{tc | cy} - 
               \braket{to | yo} +
               \frac{3}{2}\braket{to | oy} + \\ 
               \big( &
                 \braket{tc | yc} -
                 \frac{3}{2} \braket{tc | cy} + 
                 \braket{to | yo} - 
                 \frac{3}{2} \braket{to | oy} \big) \Big] \\ = \; & 0 \\
\text{Special case: }x \neq y \text{ and } y = t \\
    \bra{^A \Psi_{oc}^{ty}} H \ket{^A \Psi_{oc}^{ty}} = \; & 0 \\
\text{Special case: }x = y \text{ and } x, y = t \\
    \bra{^A \Psi_{oc}^{ty}} H \ket{^A \Psi_{oc}^{ty}} = \; & 0
\end{align*}

\subsubsection{$\bra{^B \Psi_{oc}^{tx}} H \ket{^B \Psi_{oc}^{ty}}$}
\begin{align*}
    4 \bra{^B \Psi_{oc}^{tx}} H \ket{^B \Psi_{oc}^{ty}} = 
    2 \cdot \big(& H_{EE} + H_{EF} + H_{EG} + H_{EH} \big) + \\ 
    2 \cdot \big(& H_{GE} + H_{GF} + H_{GG} + H_{HG} \big) \\ = 
    2 \cdot \Big[ & f_{xy} + \delta_{xt} f_{ty} + \delta_{ty} f_{xt} + \delta_{xy} (f_{tt} - f_{cc} - f_{oo}) - \delta_{xt}\delta_{ty} (f_{cc} + f_{oo}) + \\ &
                   \braket{tx | ty}
                +  \braket{tx | yt}
                -  \braket{xc | yc}  
                +  \braket{xc | cy}  
                -  \braket{xo | yo}  \\ 
                \delta_{xt} \big( & 
                     \braket{to | oy}
                  -  \braket{to | yo}
                  -  \braket{tc | yc}
                \big) + \\ 
                \delta_{yt} \big( &
                     \braket{xc | ct}
                  -  \braket{xc | tc}
                  -  \braket{xo | to}
                \big) + \\ 
                \delta_{xy} \big( &
                     \braket{oc | oc}
                  +  \braket{oc | co}
                  -  \braket{tc | tc}  
                  -  \braket{to | to}  
                  +  \braket{to | ot}
                \big) + \\ 
                \delta_{xt} \delta_{yt} \big( &
                     \braket{co | oc} 
                  +  \braket{oc | oc}
                \big) \Big] + \\ 
    2 \cdot \Big[ & f_{xy} + \delta_{xt} f_{ty} + \delta_{ty} f_{xt} + \delta_{xy} (f_{tt} - f_{cc} - f_{oo}) - \delta_{xt}\delta_{ty} (f_{cc} + f_{oo}) + \\ &
                   \braket{xt | yt}
                +  \braket{xt | ty}
                -  \braket{xc | yc}  
                -  \braket{xo | yo}  
                +  \braket{xo | oy} \\
                \delta_{xt} \big( &
                     \braket{tc | cy}
                  -  \braket{tc | yc}
                  -  \braket{to | yo}
                \big) + \\
                \delta_{yt} \big( &
                     \braket{to | ox}
                  -  \braket{to | xo}
                  -  \braket{tc | xc}
                \big) + \\
                \delta_{xy} \big( &
                    \braket{oc | oc} 
                  + \braket{oc | co}
                  - \braket{to | to} 
                  - \braket{tc | tc} 
                  + \braket{tc | ct}
                \big)  + \\
                \delta_{xt} \delta_{yt} \big( &
                  \braket{oc | co} + 
                  \braket{oc | oc}
                \big) \Big] 
\end{align*}

\begin{align*}
    4 \bra{^B \Psi_{oc}^{tx}} H \ket{^B \Psi_{oc}^{ty}} = 
    4 \cdot \Big[ & f_{xy} + \delta_{xt} f_{ty} + \delta_{ty} f_{xt} + \delta_{xy} (f_{tt} - f_{cc} - f_{oo}) - \delta_{xt}\delta_{ty} (f_{cc} + f_{oo}) + \\ &
                   \braket{tx | ty}
                +  \braket{tx | yt}
                -  \braket{xc | yc}  
                +  \frac{1}{2} \braket{xc | cy}  
                -  \braket{xo | yo} 
                +  \frac{1}{2} \braket{xo | oy}\\ 
                \delta_{xt} \big( & 
                     \frac{1}{2} \braket{to | oy}
                  -  \braket{to | yo}
                  +  \frac{1}{2} \braket{tc | cy}
                  -  \braket{tc | yc}
                \big) + \\ 
                \delta_{yt} \big( &
                     \frac{1}{2} \braket{xc | ct}
                  -  \braket{xc | tc}
                  +  \frac{1}{2} \braket{xo | ot}
                  -  \braket{xo | to}
                \big) + \\ 
                \delta_{xy} \big( &
                     \braket{oc | oc}
                  +  \braket{oc | co}
                  +  \frac{1}{2} \braket{tc | ct}
                  -  \braket{tc | tc} 
                  +  \frac{1}{2} \braket{to | ot}
                  -  \braket{to | to}  
                \big) + \\ 
                \delta_{xt} \delta_{yt} \big( &
                     \braket{co | oc} 
                  +  \braket{oc | oc}
                \big) \Big] \\ 
    \bra{^B \Psi_{oc}^{tx}} H \ket{^B \Psi_{oc}^{ty}} = 
    \Big[ & f_{xy} + \delta_{xt} f_{ty} + \delta_{ty} f_{xt} + \delta_{xy} (f_{tt} - f_{cc} - f_{oo}) - \delta_{xt}\delta_{ty} (f_{cc} + f_{oo}) \Big] + \\ \Big[ &
                   \braket{tx | ty}
                +  \braket{tx | yt}
                -  \braket{xc | yc}  
                +  \frac{1}{2} \braket{xc | cy}  
                -  \braket{xo | yo} 
                +  \frac{1}{2} \braket{xo | oy}\\ 
                \delta_{xt} \big( & 
                     \frac{1}{2} \braket{to | oy}
                  -  \braket{to | yo}
                  +  \frac{1}{2} \braket{tc | cy}
                  -  \braket{tc | yc}
                \big) + \\ 
                \delta_{yt} \big( &
                     \frac{1}{2} \braket{xc | ct}
                  -  \braket{xc | tc}
                  +  \frac{1}{2} \braket{xo | ot}
                  -  \braket{xo | to}
                \big) + \\ 
                \delta_{xy} \big( &
                     \braket{oc | oc}
                  +  \braket{oc | co}
                  +  \frac{1}{2} \braket{tc | ct}
                  -  \braket{tc | tc} 
                  +  \frac{1}{2} \braket{to | ot}
                  -  \braket{to | to}  
                \big) + \\ 
                \delta_{xt} \delta_{yt} \big( &
                     \braket{co | oc} 
                  +  \braket{oc | oc}
                \big) \Big] \\     
\end{align*}

\begin{align*}
\text{Special case: }x \neq y \text{ and } x, y & \neq t \\
    \bra{^B \Psi_{oc}^{tx}} H \ket{^B \Psi_{oc}^{ty}} = \; & 
    f_{xy} + \\ &
                   \braket{tx | ty}
                +  \braket{tx | yt}
                -  \braket{xc | yc}  
                +  \frac{1}{2} \braket{xc | cy}  
                -  \braket{xo | yo} 
                +  \frac{1}{2} \braket{xo | oy}  \\
\text{Special case: }x = y \text{ and } x, y & \neq t \\
    \bra{^B \Psi_{oc}^{tx}} H \ket{^B \Psi_{oc}^{ty}} = \; & 
    (f_{xy} + f_{tt} - f_{cc} - f_{oo})  + \\ &
                   \braket{tx | tx}
                +  \braket{tx | xt}
                -  \braket{xc | xc}  
                +  \frac{1}{2} \braket{xc | cx}  
                -  \braket{xo | xo} 
                +  \frac{1}{2} \braket{xo | ox} + \\ 
                \big( &
                     \braket{oc | oc}
                  +  \braket{oc | co}
                  +  \frac{1}{2} \braket{tc | ct}
                  -  \braket{tc | tc} 
                  +  \frac{1}{2} \braket{to | ot}
                  -  \braket{to | to}  
                \big) 
\end{align*}
\begin{align*}
\text{Special case: }x \neq y \text{ and } x & = t \text{. Additional factor of } (2)^{-1/2} \text{ to re-normalize} \bra{^B \Psi_{oc}^{tx}} \\
    \bra{^B \Psi_{oc}^{tx}} H \ket{^B \Psi_{oc}^{ty}} = \; 
    \frac{\sqrt{2}}{2} \cdot \Big[ &
    (f_{ty} + f_{ty}) + \\ &
                   \braket{tt | ty}
                +  \braket{tt | yt}
                -  \braket{tc | yc}  
                +  \frac{1}{2} \braket{tc | cy}  
                -  \braket{to | yo} 
                +  \frac{1}{2} \braket{to | oy} + \\ 
                \big( & 
                     \frac{1}{2} \braket{to | oy}
                  -  \braket{to | yo}
                  +  \frac{1}{2} \braket{tc | cy}
                  -  \braket{tc | yc}
                \big) \Big] \\ = \; \frac{\sqrt{2}}{2} \cdot \Big[ &
    2 f_{ty} +
                  2\braket{tt | ty}
                - 2\braket{tc | yc}  
                +  \braket{tc | cy}  
                - 2\braket{to | yo} 
                +  \braket{to | oy} \Big] \\ 
\text{Special case: }x \neq y \text{ and } y & = t \text{. Additional factor of } (2)^{-1/2} \text{ to re-normalize} \ket{^B \Psi_{oc}^{ty}} \\
    \bra{^B \Psi_{oc}^{ty}} H \ket{^B \Psi_{oc}^{ty}} = \; 
    \frac{\sqrt{2}}{2} \cdot \Big[ &
    f_{xt} + f_{xt}  + \\ &
                   \braket{tx | tt}
                +  \braket{tx | tt}
                -  \braket{xc | tc}  
                +  \frac{1}{2} \braket{xc | ct}  
                -  \braket{xo | yo} 
                +  \frac{1}{2} \braket{xo | ot} + \\  
                \big( &
                     \frac{1}{2} \braket{xc | ct}
                  -  \braket{xc | tc}
                  +  \frac{1}{2} \braket{xo | ot}
                  -  \braket{xo | to}
                \big) \Big] \\ = \;
    \frac{\sqrt{2}}{2} \cdot \Big[ &
    2 f_{xt} +
                  2\braket{tx | tt}
                -  \braket{xc | tc}  
                +  \braket{xc | ct}  
                - 2\braket{xo | yo} 
                +  \braket{xo | ot} 
                \big) \Big] \\
\text{Special case: }x = y \text{ and } x, y & = t \text{. Additional factor of } (2)^{-1} \text{ to re-normalize...} \\
    \bra{^B \Psi_{oc}^{ty}} H \ket{^B \Psi_{oc}^{ty}} = \; \frac{1}{2} \cdot
    \Big[ & f_{tt} + f_{tt} + f_{tt} + (f_{tt} - f_{cc} - f_{oo}) - 
    (f_{cc} + f_{oo}) \Big] + \\ \frac{1}{2} \cdot \Big[ &
                   \braket{tt | tt}
                +  \braket{tt | tt}
                -  \braket{tc | tc}  
                +  \frac{1}{2} \braket{tc | ct}  
                -  \braket{to | to} 
                +  \frac{1}{2} \braket{to | ot}\\ 
                \big( & 
                     \frac{1}{2} \braket{to | ot}
                  -  \braket{to | to}
                  +  \frac{1}{2} \braket{tc | ct}
                  -  \braket{tc | tc}
                \big) + \\ 
                \big( &
                     \frac{1}{2} \braket{tc | ct}
                  -  \braket{tc | tc}
                  +  \frac{1}{2} \braket{to | ot}
                  -  \braket{to | to}
                \big) + \\ 
                \big( &
                     \braket{oc | oc}
                  +  \braket{oc | co}
                  +  \frac{1}{2} \braket{tc | ct}
                  -  \braket{tc | tc} 
                  +  \frac{1}{2} \braket{to | ot}
                  -  \braket{to | to}  
                \big) + \\ 
                \big( &
                     \braket{co | oc} 
                  +  \braket{oc | oc}
                \big) \Big] \\ = \; \frac{1}{2} \cdot
    \Big[ & 4 f_{tt} - 2 f_{cc} - 2 f_{oo} \Big] + \\ \frac{1}{2} \cdot \Big[ &
                  2\braket{tt | tt}
                - 4\braket{tc | tc}  
                + 2\braket{tc | ct}  
                - 4\braket{to | to} 
                + 2\braket{to | ot}
                + 2\braket{oc | oc}
                + 2\braket{oc | co} \Big] \\ = \; 
    \big(& 2 f_{tt} - f_{cc} - f_{oo} \big) + \\ \Big[ & 
                   \braket{tt | tt}
                - 2\braket{tc | tc}  
                +  \braket{tc | ct}  
                - 2\braket{to | to} 
                +  \braket{to | ot}
                +  \braket{oc | oc}
                +  \braket{oc | co} \Big]    
\end{align*}

\subsubsection{$\bra{^A \Psi_{oc}^{tx}} H \ket{^B \Psi_{oc}^{ty}}$}

\begin{align*}
    2 \sqrt{12} \bra{^A \Psi_{oc}^{tx}} H \ket{^B \Psi_{oc}^{ty}} = 
    4 \cdot \big(& H_{CE} + H_{CF} + H_{CG} + H_{CH} \big) + \\
    2 \cdot \big(& H_{EE} + H_{EF} + H_{EG} + H_{EH} \big) - \\ 
    2 \cdot \big(& H_{GE} + H_{GF} + H_{GG} + H_{HG} \big) \\ = 
    4 \cdot \Big[&
                  \braket{xc | cy} 
                - \braket{xo | oy}\\ 
               \delta_{xt} \big( &
                    \braket{to | oy} 
                 -  \braket{tc | cy} \big) \\  
               \delta_{yt} \big( &
                    \braket{xc | tc} 
                 -  \braket{xo | ot} \big) + \\ 
               \delta_{xy} \big( & 
                    \braket{to | ot}  
                 -  \braket{tc | ct} \big) \Big] + \\ 
    2 \cdot \Big[ & f_{xy} + \delta_{xt} f_{ty} + \delta_{ty} f_{xt} + \delta_{xy} (f_{tt} - f_{cc} - f_{oo}) - \delta_{xt}\delta_{ty} (f_{cc} + f_{oo}) + \\ &
                   \braket{tx | ty}
                +  \braket{tx | yt}
                -  \braket{xc | yc}  
                +  \braket{xc | cy}  
                -  \braket{xo | yo}  \\ 
                \delta_{xt} \big( & 
                     \braket{to | oy}
                  -  \braket{to | yo}
                  -  \braket{tc | yc}
                \big) + \\ 
                \delta_{yt} \big( &
                     \braket{xc | ct}
                  -  \braket{xc | tc}
                  -  \braket{xo | to}
                \big) + \\ 
                \delta_{xy} \big( &
                     \braket{oc | oc}
                  +  \braket{oc | co}
                  -  \braket{tc | tc}  
                  -  \braket{to | to}  
                  +  \braket{to | ot}
                \big) + \\ 
                \delta_{xt} \delta_{yt} \big( &
                     \braket{co | oc} 
                  +  \braket{oc | oc}
                \big) \Big] - \\ 
    2 \cdot \Big[ & f_{xy} + \delta_{xt} f_{ty} + \delta_{ty} f_{xt} + \delta_{xy} (f_{tt} - f_{cc} - f_{oo}) - \delta_{xt}\delta_{ty} (f_{cc} + f_{oo}) + \\ &
                   \braket{xt | yt}
                +  \braket{xt | ty}
                -  \braket{xc | yc}  
                -  \braket{xo | yo}  
                +  \braket{xo | oy} \\
                \delta_{xt} \big( &
                     \braket{tc | cy}
                  -  \braket{tc | yc}
                  -  \braket{to | yo}
                \big) + \\
                \delta_{yt} \big( &
                     \braket{to | ox}
                  -  \braket{to | xo}
                  -  \braket{tc | xc}
                \big) + \\
                \delta_{xy} \big( &
                    \braket{oc | oc} 
                  + \braket{oc | co}
                  - \braket{to | to} 
                  - \braket{tc | tc} 
                  + \braket{tc | ct}
                \big)  + \\
                \delta_{xt} \delta_{yt} \big( &
                  \braket{oc | co} + 
                  \braket{oc | oc}
                \big) \Big] 
\end{align*}

\begin{align*}
    2 \sqrt{12} \bra{^A \Psi_{oc}^{tx}} H \ket{^B \Psi_{oc}^{ty}} = 
    4 \cdot \Big[&
                  \braket{xc | cy} 
                - \braket{xo | oy} + \\ 
               \delta_{xt} \big( &
                    \braket{to | oy} 
                 -  \braket{tc | cy} \big) + \\  
               \delta_{yt} \big( &
                    \braket{xc | ct} 
                 -  \braket{xo | ot} \big) + \\ 
               \delta_{xy} \big( & 
                    \braket{to | ot}  
                 -  \braket{tc | ct} \big) \Big] + \\ 
    2 \cdot \Big[ &  
                   \braket{xc | cy}  
                -  \braket{xo | oy} + \\ 
                \delta_{xt} \big( & 
                     \braket{to | oy}
                  -  \braket{tc | cy}
                \big) + \\ 
                \delta_{yt} \big( &
                     \braket{xc | ct}
                  -  \braket{xo | ot}
                \big) + \\ 
                \delta_{xy} \big( &
                     \braket{to | ot}  
                  -  \braket{tc | ct}  
                \big) \Big] + \\ 
    \bra{^A \Psi_{oc}^{tx}} H \ket{^B \Psi_{oc}^{ty}} = 
    \sqrt{\frac{3}{4}} \cdot \Big[&
                  \braket{xc | cy} 
                - \braket{xo | oy} + \\ 
               \delta_{xt} \big( &
                    \braket{to | oy} 
                 -  \braket{tc | cy} \big) + \\  
               \delta_{yt} \big( &
                    \braket{xc | ct} 
                 -  \braket{xo | ot} \big) + \\ 
               \delta_{xy} \big( & 
                    \braket{to | ot}  
                 -  \braket{tc | ct} \big) \Big]
\end{align*}

\begin{align*}
\text{Special case: }x \neq y \text{ and } x, y & \neq t \\
    \bra{^A \Psi_{oc}^{tx}} H \ket{^B \Psi_{oc}^{ty}} = \; 
    \sqrt{\frac{3}{4}} \cdot \big( &
                  \braket{xc | cy} 
                - \braket{xo | oy} \big) \\ 
\text{Special case: }x = y \text{ and } x, y & \neq t \\
    \bra{^A \Psi_{oc}^{tx}} H \ket{^B \Psi_{oc}^{ty}} = \; 
    \sqrt{\frac{3}{4}} \cdot \big( &
                  \braket{xc | cx} 
                - \braket{xo | ox}  +
                  \braket{to | ot} 
                - \braket{tc | ct} \big) \\ 
\text{Special case: }x \neq y \text{ and } x & = t \\ 
    \bra{^A \Psi_{oc}^{tx}} H \ket{^B \Psi_{oc}^{ty}} = \; 
    \sqrt{\frac{3}{4}} \cdot \big( &
                  \braket{tc | cy} 
                - \braket{to | oy}  +
                  \braket{to | oy} 
                - \braket{tc | cy} \big) \\ = \; & 0 \\ 
\text{Special case: }x \neq y \text{ and } y & = t 
\text{. Additional factor of } (2)^{-1/2} \text{ to re-normalize} \ket{^B \Psi_{oc}^{ty}} \\
    \bra{^A \Psi_{oc}^{tx}} H \ket{^B \Psi_{oc}^{ty}} = \; 
    \sqrt{\frac{3}{8}} \cdot \big( &
                  \braket{xc | ct} 
                - \braket{xo | ot}  +
                  \braket{xc | ct} 
                - \braket{xo | ot} \big) \\ = \; 
    \sqrt{\frac{3}{2}} \cdot \big( &
                  \braket{xc | ct} 
                - \braket{xo | ot} \big) \\
\text{Special case: }x = y \text{ and } x, y & = t 
\text{. Additional factor of } (2)^{-1/2} \text{ to re-normalize...} \\
    \bra{^A \Psi_{oc}^{tx}} H \ket{^B \Psi_{oc}^{ty}} = \; 
    \sqrt{\frac{3}{8}} \cdot \big( &
                  \braket{tc | ct} 
                - \braket{to | ot}   
                + \braket{to | ot} 
                - \braket{tc | ct} + \\ &
                  \braket{tc | ct} 
                - \braket{to | ot} +  
                  \braket{to | ot}  
                - \braket{tc | ct} \big) \\ = & \; 0
\end{align*}

\subsection{Matrix elements for Doublet CSFs}
\subsubsection{$\bra{^{2_G} \Phi_{oc}^t}H\ket{^{2_G} \Phi_{oc}^t}$}
\begin{align*}
    \bra{^{2_G} \Phi_{o\bar{c}}^t}H\ket{^{2_G} \Phi_{o\bar{c}}^t} = & \frac{1}{6}\cdot\big(4H_{II} - 2H_{IJ} - 2H_{IK} - 2H_{JI} + H_{JJ} + H_{JK} -2H_{KI} + H_{KJ} + H_{KK}\big) \\
    =& \frac{1}{6}\cdot\big(4H_{II} - 4H_{IJ} - 4H_{IK} +2H_{JK} + H_{JJ} + H_{KK}\big) \\
    =& \frac{1}{6}\cdot\big(-4f_{cc} - 4f_{oo} + 4f_{tt} + 4\braket{to|ot} - 4\braket{to|to} + 4\braket{tc|ct} - \\
    &4\braket{tc|tc} + 4\braket{oc|oc} - 4\braket{oc|co} + 4\braket{to|ot} + 4\braket{tc|ct} - 2 \braket{oc|co}  \\
    &-f_{cc} - f_{oo} + f_{tt} + \braket{to|ot} - \braket{to|to} - \braket{tc|tc} + \braket{oc|oc} \\
    &-f_{cc} - f_{oo} + f_{tt} + \braket{tc|ct} - \braket{tc|tc} - \braket{to|to} + \braket{oc|oc}\big) \\
    &= f_{tt} - f_{cc} - f_{oo} + \frac{3}{2}\braket{to|ot} - \braket{to|to} + \frac{3}{2} \braket{tc|ct} - \braket{tc|tc} + \braket{oc|oc} - \braket{oc|co}
\end{align*} 

\subsubsection{$\bra{^{2_H} \Phi_{oc}^t}H\ket{^{2_H} \Phi_{oc}^t}$}

\begin{align*}
    \bra{^{2_H}\Phi_{o\bar{c}}^t}H\ket{^{2_H} \Phi_{o\bar{c}}^t} =& \frac{1}{2} \cdot\big(
    H_{JJ} - H_{JK} - H_{KJ} + H_{KK} \big)\\
    =& \frac{1}{2}\cdot\big(H_{JJ} - 2H_{JK} + H_{KK}\big)\\
    =&\frac{1}{2}\cdot\big(-f_{cc} - f_{oo} + f_{tt} + \braket{to|ot} - \braket{to|to} - \braket{tc|tc} + \braket{oc|oc} + \\
    &-f_{cc} + f_{tt} - f_{oo} + \braket{tc|ct} - \braket{tc|tc} - \braket{to|to} + \braket{oc|oc} + 2\braket{oc|co}\big)\\
    =&f_{tt} -  f_{cc} - f_{oo} + \frac{1}{2}\braket{to|ot} - \braket{to|to} - \braket{tc|tc} + \braket{oc|oc} + \braket{oc|co} + \frac{1}{2}\braket{tc|ct}
\end{align*}

\subsubsection{$\bra{^{2_G} \Phi_{oc}^t}H\ket{^{2_H} \Phi_{oc}^t}$}
\begin{align*}
    \bra{^{2_G} \Phi_{o\bar{c}}^t}H\ket{^{2_H} \Phi_{o\bar{c}}^t} =& \frac{1}{\sqrt{12}}\cdot\big(
    2H_{IJ} - 2H_{IK} - H_{JJ} - H_{JK} + H_{KJ} + H_{KK}\big) \\
    =& \frac{\sqrt{3}}{6}\cdot\big(-2\braket{to|ot} + 2\braket{tc|ct} + \\
    &f_{cc} + f_{oo} - f_{tt} - \braket{to|ot} + \braket{to|to} +\braket{tc|tc} - \braket{oc|oc} -\\
    &f_{cc} - f_{oo} + f_{tt} + \braket{tc|ct} - \braket{tc|tc} - \braket{to|to} + \braket{oc|oc}\big)\\
    =& \frac{\sqrt{3}}{2}\cdot\big(\braket{tc|ct} - \braket{to|ot}\big)
\end{align*}

\subsection{Matrix elements for triplet CSFs}
\subsubsection{$ \bra{^3 \Psi_{c}^{t}} H \ket{^3 \Psi_{c}^{t}}$}
\begin{align*}
    \bra{^3 \Psi_{c}^{t}} H \ket{^3 \Psi_{c}^{t}} =& \frac{1}{2}\cdot\big(H_{AA} - H_{AB} - H_{BA} + H_{BB}\big) \\
    =&f_{tt}-f_{cc}-\braket{tc|tc}
\end{align*}
\subsubsection{$ \bra{^3 \Psi_{c}^{t}} H \ket{^C \Psi_{oc}^{ty}}$}
\begin{align*}
    \bra{^3 \Psi_{c}^{t}} H \ket{^C \Psi_{oc}^{ty}} =&\frac{1}{2}\cdot\big(H_{AC} - H_{AD} - H_{BC} + H_{BD}\big) \\
    =& \delta_{ty}f_{ot} - f_{oy} + \braket{oc|yc} - \braket{oc|cy} - \braket{to|ty} + \braket{to|yt} + \delta_{yt}(\braket{oc|ct}-\braket{oc|tc})
\end{align*}
\begin{align*}
\text{Special case: } y\neq &t \\
\bra{^3 \Psi_{c}^{t}} H \ket{^C \Psi_{oc}^{ty}} =&-f_{oy} + \braket{oc|yc} - \braket{oc|cy} - \braket{to|ty} + \braket{to|yt} \\
\text{Special case: }y = &t\\
\bra{^3 \Psi_{c}^{t}} H \ket{^C \Psi_{oc}^{ty}} =&0
\end{align*}
\subsubsection{$ \bra{^3 \Psi_{c}^{t}} H \ket{^D \Psi_{oc}^{ty}}$}
\begin{align*}
    \bra{^3 \Psi_{c}^{t}} H \ket{^D \Psi_{oc}^{ty}} =& \frac{1}{2}\big(H_{AE} - H_{AF} - H_{BE} + H_{BF}\big)\\
    =& \delta_{ty}f_{ot} + \braket{to|yt} - \delta_{yt}\braket{oc|tc} + \braket{oc|cy}
\end{align*}
\begin{align*}
\text{Special case: } y\neq &t \\
\bra{^3 \Psi_{c}^{t}} H \ket{^D \Psi_{oc}^{ty}} =&\braket{to|yt} + \braket{oc|cy} \\
\text{Special case: }y = &t\\
\bra{^3 \Psi_{c}^{t}} H \ket{^D \Psi_{oc}^{ty}} =&f_{ot} + \braket{to|tt} -\braket{oc|tc} + \braket{oc|ct}
\end{align*}
\subsubsection{$ \bra{^3 \Psi_{c}^{t}} H \ket{^E \Psi_{oc}^{ty}}$}
\begin{align*}
    \bra{^3 \Psi_{c}^{t}} H \ket{^E \Psi_{oc}^{ty}} =& \frac{1}{2}\cdot\big(H_{AG} - H_{AH} - H_{BG} - H_{BH}\big) \\
    =&f_{oy} + \braket{to|ty} - \braket{oc|yc} + \delta_{ty}\braket{oc|ct}
\end{align*}
\begin{align*}
\text{Special case: } y\neq &t \\
\bra{^3 \Psi_{c}^{t}} H \ket{^E \Psi_{oc}^{ty}} =&f_{oy} + \braket{to|ty} - \braket{oc|yc} \\
\text{Special case: }y = &t\\
\bra{^3 \Psi_{c}^{t}} H \ket{^E \Psi_{oc}^{ty}} =&f_{ot} +\braket{to|tt} - \braket{oc|tc} + \braket{oc|ct}
\end{align*}
\subsubsection{$ \bra{^C \Psi_{oc}^{tx}} H \ket{^C \Psi_{oc}^{ty}}$}
\begin{align*}
    \bra{^C \Psi_{oc}^{tx}} H \ket{^C \Psi_{oc}^{ty}} = & \frac{1}{2}\cdot\big(H_{CC} - H_{CD} - H_{DC} + H_{DD}\big) \\ =&
    f_{xy} - \delta_{ty} f_{xt} - \delta_{xt} f_{ty} + \delta_{xy} (f_{tt} - f_{cc} - f_{oo}) + \delta_{xt}\delta_{ty} (f_{cc} + f_{oo}) + \\
             & \braket{tx | ty} - 
               \braket{tx | yt} -
               \braket{xc | yc} +
               \braket{xc | cy} - 
               \braket{xo | yo} +
               \braket{xo | oy} + \\ & 
               \delta_{xt} \big( 
                 \braket{tc | yc} -
                 \braket{tc | cy} + 
                 \braket{to | yo} - 
                 \braket{to | oy} \big)+ \\ & 
               \delta_{yt} \big(
                 \braket{xc | tc} -
                 \braket{xc | ct} + 
                 \braket{xo | to} - 
                 \braket{xo | ot} \big) + \\ & 
               \delta_{xy} \big( 
                 \braket{oc | oc} - 
                 \braket{oc | co} -
                 \braket{tc | tc} +
                 \braket{tc | ct} - 
                 \braket{to | to} + 
                 \braket{to | ot} \big) - \\ &
               \delta_{xt} \delta_{yt} \big(
                 \braket{oc | oc} - 
                 \braket{oc | co} \big)
\end{align*}
\begin{align*}
\text{Special case: }x \neq y \text{ and } x, y & \neq t \\
\bra{^C \Psi_{oc}^{tx}} H \ket{^C \Psi_{oc}^{ty}} =& f_{xy} + \braket{tx|ty} - \braket{tx|yt} - \braket{xc|yc} + \braket{xc|cy} - \braket{xo|yo} + \braket{xo|oy} \\
\text{Special case: }x = y \text{ and } x, y & \neq t  \\
\bra{^C \Psi_{oc}^{tx}} H \ket{^C \Psi_{oc}^{tx}} =& f_{xx} +f_{tt}-f_{cc}-f_{oo} + \\ &\braket{tx|tx} - \braket{tx|xt} - \braket{xc|xc} + \braket{xc|cx} - \braket{xo|xo} + \braket{xo|ox} + \braket{oc|oc} - \\ &\braket{oc|co} - \braket{tc|tc} + \braket{tc|ct} - \braket{to|to} + \braket{to|ot}\\
\text{Special case: }x \neq y \text{ and } x & = t \\
\bra{^C \Psi_{oc}^{tt}} H \ket{^C \Psi_{oc}^{ty}} =& 0\\
\text{Special case: }x \neq y \text{ and } y & = t \\
\bra{^C \Psi_{oc}^{tx}} H \ket{^C \Psi_{oc}^{tt}} =& 0\\
\text{Special case: }x = y \text{ and } x, y & = t \\
\bra{^C \Psi_{oc}^{tt}} H \ket{^C \Psi_{oc}^{tt}} = & 0
\end{align*}

\subsubsection{$ \bra{^D \Psi_{oc}^{tx}} H \ket{^D \Psi_{oc}^{ty}}$}
\begin{align*}
    \bra{^D \Psi_{oc}^{tx}} H \ket{^D \Psi_{oc}^{ty}} =&
    \frac{1}{2}\cdot\big(H_{EE} - H_{EF} - H_{FE} + H_{FF}\big) \\=&
    f_{xy} + \delta_{xy}(f_{tt} - f_{cc} - f_{oo}) + \braket{tx|ty} - \braket{xc|yc} + \braket{xc|cy} - \braket{xo|yo} + \\
    &\delta_{xy}(\braket{oc|oc} - \braket{tc|tc} - \braket{to|to} + \braket{to|ot}) - \delta_{xt}\delta_{yt}\braket{co|oc} 
\end{align*}
\begin{align*}
\text{Special case: }x \neq y \text{ and } x, y & \neq t \\
\bra{^D \Psi_{oc}^{tx}} H \ket{^D \Psi_{oc}^{ty}} =& f_{xy} + \braket{tx|ty} - \braket{xc|yc} + \braket{xc|cy} - \braket{xo|yo} \\
\text{Special case: }x = y \text{ and } x, y & \neq t  \\
\bra{^D \Psi_{oc}^{tx}} H \ket{^D \Psi_{oc}^{tx}} =& f_{xx} + f_{tt} - f_{cc} - f_{oo} + \braket{tx|tx} - \braket{xc|xc} +\\
&\braket{xc|cx} - \braket{xo|xo} + \braket{oc|oc} - \braket{tc|tc} - \braket{to|to} + \braket{to|ot}\\
\text{Special case: }x \neq y \text{ and } x & = t \\
\bra{^D \Psi_{oc}^{tt}} H \ket{^D \Psi_{oc}^{ty}} =& f_{ty} + \braket{tt|ty} - \braket{tc|yc}+\braket{tc|cy}-\braket{to|yo}\\
\text{Special case: }x \neq y \text{ and } y & = t \\
\bra{^D \Psi_{oc}^{tx}} H \ket{^D \Psi_{oc}^{tt}} =& f_{xt} +\braket{tx|tt} -\braket{xc|tc}+\braket{xc|ct} - \braket{xo|to}\\
\text{Special case: }x = y \text{ and } x, y & = t \\
\bra{^D \Psi_{oc}^{tt}} H \ket{^D \Psi_{oc}^{tt}} = & 2f_{tt} - f_{cc}-f_{oo} +\braket{tt|tt} - 
2\braket{tc|tc} - 2\braket{to|to} + \braket{tc|ct}  + \braket{to|ot} + \braket{oc|oc} - \braket{oc|co}
\end{align*}
\subsubsection{$ \bra{^E \Psi_{oc}^{tx}} H \ket{^E \Psi_{oc}^{ty}}$}
\begin{align*}
    \bra{^E \Psi_{oc}^{tx}} H \ket{^E \Psi_{oc}^{ty}} =&
    \frac{1}{2}\cdot\big(H_{GG}-H_{GH}-H_{HG}+H_{HH}\big) \\ =&f_{xy} + \delta_{xy} (f_{tt} - f_{cc} - f_{oo}) +  \\
             & \braket{xt | yt} - 
               \braket{xc | yc} - 
               \braket{xo | yo} + 
               \braket{xo | oy} +\\ 
               %There may be a missing + here%
               & \delta_{xy} \big(
               \braket{oc | oc} - 
               \braket{to | to} -
               \braket{tc | tc} +
               \braket{tc | ct} \big) -\delta_{xt}\delta_{yt}\braket{oc|co} 
\end{align*}
\begin{align*}
\text{Special case: }x \neq y \text{ and } x, y & \neq t \\
\bra{^E \Psi_{oc}^{tx}} H \ket{^E \Psi_{oc}^{ty}} =& f_{xy} + \braket{tx|ty} - \braket{xc|yc} + \braket{xo|oy} - \braket{xo|yo}  \\
\text{Special case: }x = y \text{ and } x, y & \neq t  \\
\bra{^E \Psi_{oc}^{tx}} H \ket{^E \Psi_{oc}^{tx}} =& f_{xx} + f_{tt} - f_{cc} - f_{oo} + \braket{tx|tx} - \braket{xc|xc} + \\
&\braket{xo|ox} - \braket{xo|xo}  + 
\braket{oc|oc}-\braket{tc|tc}-\braket{to|to}+\braket{tc|ct}\\
\text{Special case: }x \neq y \text{ and } x & = t \\
\bra{^E \Psi_{oc}^{tt}} H \ket{^E \Psi_{oc}^{ty}} =& f_{ty} +\braket{tt|yt}-\braket{tc|yc} + \braket{to|oy} - \braket{to|yo} \\
\text{Special case: }x \neq y \text{ and } y & = t \\
\bra{^E \Psi_{oc}^{tx}} H \ket{^E \Psi_{oc}^{tt}} =& f_{xt} +\braket{xt|tt} -\braket{xc|tc} + \braket{xo|ot}  - \braket{xo|to}\\
\text{Special case: }x = y \text{ and } x, y & = t \\ =&
2f_{tt}-f_{cc}-f_{oo} + \braket{tt|tt} - 2\braket{tc|tc} - 2\braket{to|to} + \braket{tc|ct} + \braket{to|ot} + \braket{oc|oc} 
- \braket{oc|co}
\end{align*}
\subsubsection{$ \bra{^C \Psi_{oc}^{tx}} H \ket{^D \Psi_{oc}^{ty}}$}
\begin{align*}
    \bra{^C \Psi_{oc}^{tx}} H \ket{^D \Psi_{oc}^{ty}} =& \frac{1}{2}\cdot\big( H_{CE} - H_{CF} - H_{DE} + H_{DF}\big) \\
    =& \frac{1}{2}\cdot\Bigl[2\delta_{xy}\braket{to|ot} - 2 \delta_{yt}\braket{xo|ot} - \\
    &2\braket{xc|cy} + 2\delta_{xt}\braket{tc|cy}\Bigr]\\
    =&\delta_{xy}\braket{to|ot} + \delta_{xt}\braket{tc|cy} - \delta_{yt}\braket{xo|ot} - \braket{xc|cy}
\end{align*}
\begin{align*}
\text{Special case: }x \neq y \text{ and } x, y & \neq t \\
\bra{^C \Psi_{oc}^{tx}} H \ket{^D \Psi_{oc}^{ty}} =& -\braket{xc|cy}\\
\text{Special case: }x = y \text{ and } x, y & \neq t  \\
\bra{^C \Psi_{oc}^{tx}} H \ket{^D \Psi_{oc}^{tx}} =& \braket{to|ot} - \braket{xc|cx}\\
\text{Special case: }x \neq y \text{ and } x & = t \\
\bra{^C \Psi_{oc}^{tt}} H \ket{^D \Psi_{oc}^{ty}} =& 0\\
\text{Special case: }x \neq y \text{ and } y & = t \\
\bra{^C \Psi_{oc}^{tx}} H \ket{^D \Psi_{oc}^{tt}} =& -\braket{xo|ot}-\braket{xc|ct}\\
\text{Special case: }x = y \text{ and } x, y & = t \\ 
\bra{^C \Psi_{oc}^{tt}} H \ket{^D \Psi_{oc}^{tt}} =&0
\end{align*}
\subsubsection{$ \bra{^C \Psi_{oc}^{tx}} H \ket{^E \Psi_{oc}^{ty}}$}
\begin{align*}
    \bra{^C \Psi_{oc}^{tx}} H \ket{^E \Psi_{oc}^{ty}} =&
    \frac{1}{2}\cdot\big(H_{CG} - H_{CH} - H_{DG} + H_{DH}\big) \\
    =&\frac{1}{2}\cdot\Bigl[2(\delta_{xt}\braket{to|oy}-\braket{xo|oy}) - 2(\delta_{yt}\braket{xc|ct}-\delta_{xy}\braket{tc|ct})\Bigr] \\
    =& \delta_{xy}\braket{tc|ct} + \delta_{xt}\braket{to|oy} - \delta_{yt}\braket{xc|ct} - \braket{xo|oy}
\end{align*}
\begin{align*}
\text{Special case: }x \neq y \text{ and } x, y & \neq t \\
\bra{^C \Psi_{oc}^{tx}} H \ket{^E \Psi_{oc}^{ty}} =& -\braket{xo|oy} \\
\text{Special case: }x = y \text{ and } x, y & \neq t  \\
\bra{^C \Psi_{oc}^{tx}} H \ket{^E \Psi_{oc}^{tx}} =&\braket{tc|ct} - \braket{xo|ox} \\
\text{Special case: }x \neq y \text{ and } x & = t \\
\bra{^C \Psi_{oc}^{tt}} H \ket{^E \Psi_{oc}^{ty}} =& 0\\
\text{Special case: }x \neq y \text{ and } y & = t \\
\bra{^C \Psi_{oc}^{tx}} H \ket{^E \Psi_{oc}^{tt}} =& -\braket{xc|ct}-\braket{xo|ot}\\
\text{Special case: }x = y \text{ and } x, y & = t \\ 
\bra{^C \Psi_{oc}^{tt}} H \ket{^E \Psi_{oc}^{tt}} =&0
\end{align*}

\subsubsection{$ \bra{^D \Psi_{oc}^{tx}} H \ket{^E \Psi_{oc}^{ty}}$}
\begin{align*}
    \bra{^D \Psi_{oc}^{tx}} H \ket{^E \Psi_{oc}^{ty}} =&
    \frac{1}{2}\cdot\big(H_{EG} - H_{EH} - H_{FG} + H_{FH}\big) \\
    =&\frac{1}{2}\cdot\Bigl[2\big(\delta_{xt}f_{ty} + \delta_{ty}f_{xt} - \delta_{xt}\delta_{ty}(f_{cc}+f_{oo}) + \\
    &\braket{tx|yt} + \\
    &\delta_{xt}(\braket{to|oy} - \braket{to|yo} - \braket{tc|yc}) + \\
    &\delta_{yt}(\braket{xc|ct} - \braket{xc|tc} - \braket{xo|to}) + \\
    &\delta_{xt}\delta_{ty}(\braket{oc|oc})\big) - \\
    &2\delta_{xy}\braket{oc|co}\Bigr] \\
    =& \delta_{xt}f_{ty} + \delta_{ty}f_{xt} - \delta_{xt}\delta_{ty}(f_{cc}+f_{oo}) + \\
    &\braket{tx|yt} + \\
    &\delta_{xt}(\braket{to|oy} - \braket{to|yo} - \braket{tc|yc}) + \\
    &\delta_{yt}(\braket{xc|ct} - \braket{xc|tc} - \braket{xo|to}) + \\
    &\delta_{xt}\delta_{ty}(\braket{oc|oc}) - \\
    &\delta_{xy}\braket{oc|co} 
\end{align*}
\begin{align*}
\text{Special case: }x \neq y \text{ and } x, y & \neq t \\
\bra{^D \Psi_{oc}^{tx}} H \ket{^E \Psi_{oc}^{ty}} =& \braket{tx|yt}\\
\text{Special case: }x = y \text{ and } x, y & \neq t  \\
\bra{^D \Psi_{oc}^{tx}} H \ket{^E \Psi_{oc}^{tx}} =& \braket{tx|xt} - \braket{oc|co} \\
\text{Special case: }x \neq y \text{ and } x & = t \\
\bra{^D \Psi_{oc}^{tt}} H \ket{^E \Psi_{oc}^{ty}} =& f_{ty} + \braket{tt|yt} - \braket{tc|yc} + \braket{to|oy} - \braket{to|yo} \\
\text{Special case: }x \neq y \text{ and } y & = t \\
\bra{^D \Psi_{oc}^{tx}} H \ket{^E \Psi_{oc}^{tt}} =& f_{xt} + \braket{tx|tt} - \braket{xc|tc} + \braket{xc|ct}  - \braket{xo|to}\\
\text{Special case: }x = y \text{ and } x, y & = t \\ 
\bra{^D \Psi_{oc}^{tt}} H \ket{^E \Psi_{oc}^{tt}} =&f_{tt} + f_{tt} - f_{cc} - f_{oo} + \braket{tt|tt} \\
&\braket{to|ot} - \braket{to|to} - \braket{tc|tc} + \\
&\braket{tc|ct} - \braket{tc|tc} - \braket{to|to} + \\
&\braket{oc|oc} - \braket{oc|co} \\
    =&  2f_{tt} - f_{cc}-f_{oo} +\braket{tt|tt} - 2\braket{tc|tc} - 2\braket{to|to} + \braket{tc|ct}+ \braket{to|ot}  + \braket{oc|oc} - \braket{oc|co}
\end{align*}

\subsection{Matrix element for Quartet}
\subsubsection{$\bra{^{2_I} \Phi_{oc}^t}H\ket{^{2_I} \Phi_{oc}^t}$}

\begin{align*}
    \bra{^{2_I} \Phi_{oc}^t}H\ket{^{2_I} \Phi_{oc}^t} =& \frac{1}{3}\cdot\big(H_{II} + H_{IJ} + H_{IK} + H_{JI} + H_{JJ} + H_{JK} + H_{KI} + H_{KJ} + H_{KK}\big) \\
    &=\frac{1}{3}\cdot\big(H_{II} + H_{JJ} + H_{KK} +2H_{IJ} + 2H_{IK} +2H_{JK} \big) \\
    &= \frac{1}{3}\cdot\big(f_{tt} - f_{cc} - f_{oo} + \braket{to|ot} - \braket{to|to} + \braket{tc|ct} -\braket{tc|tc} + \braket{oc|oc} - \braket{oc|co} + \\
    &f_{tt} - f_{cc} - f_{oo} + \braket{to|ot} - \braket{to|to} - \braket{tc|tc} + \braket{oc|oc} + \\
    &f_{tt} - f_{cc} - f_{oo} + \braket{tc|ct} - \braket{tc|tc} - \braket{to|to} + \braket{oc|oc} - \\
    &2\braket{to|ot} - 2\braket{tc|ct} - 2\braket{oc|co}\big) \\
    =&f_{tt} - f_{oo} - f_{cc} - \braket{to|to} - \braket{tc|tc} + \braket{oc|oc} - \braket{oc|co}
\end{align*}

\subsection{Matrix elements for Quintet}
\subsubsection{$ \bra{^F \Psi_{oc}^{tx}} H \ket{^F \Psi_{oc}^{ty}}$}
\begin{align*}
    \bra{^F \Psi_{oc}^{tx}} H \ket{^F \Psi_{oc}^{ty}} =&\frac{1}{6}\cdot\Bigl[H_{CC} + H_{CD} - H_{CE} - H_{CF} + H_{CG} + H_{CH} + \\
    &H_{DC} + H_{DD} - H_{DE} - H_{DF} + H_{DG} + H_{DH}  \\
    &-H_{EC}-H_{ED} + H_{EE} + H_{EF} - H_{EG} - H_{EH} \\
    &-H_{FC} - H_{FD} + H_{FE} + H_{FF} - H_{FG} - H_{FH} + \\
    &H_{GC} + H_{GD} - H_{GE} - H_{GF} + H_{GG} + H_{GH} + \\
    &H_{HC} + H_{HD} - H_{HE} - H_{HF} + H_{HG} + H_{HH}\Bigr] \\
    = & \frac{1}{3}\cdot\Bigl[H_{CC} + H_{CD} - H_{CE} - H_{CF} + H_{CG} + H_{CH} \\
    &-H_{EC}-H_{ED} + H_{EE} + H_{EF} - H_{EG} - H_{EH} +\\
     &H_{GC} + H_{GD} - H_{GE} - H_{GF} + H_{GG} + H_{GH}\Bigr]
\end{align*}
\begin{align*}
  \bra{^F \Psi_{oc}^{tx}} H \ket{^F \Psi_{oc}^{ty}} =&
     \frac{1}{3}\cdot\Bigl[f_{xy} - \delta_{ty}f_{xt} - \delta_{xt}f_{ty} + \delta_{xy}(f_{tt}-f_{cc}-f_{oo}) + \delta_{xt}\delta_{yt}(f_{cc}+f_{oo}) + \\
     &\braket{tx|ty} - \braket{tx|yt} - \braket{xc|yc} + \braket{xc|cy} - \braket{xo|yo} + \braket{xo|oy} + \\
     &\delta_{xt}(\braket{tc|yc} - \braket{tc|cy} + \braket{to|yo} - \braket{to|oy}) + \\
     &\delta_{yt}(\braket{xc|tc} - \braket{xc|ct} + \braket{xo|to} - \braket{xo|ot}) + \\
     &\delta_{xy}(\braket{oc|oc} - \braket{oc|co} - \braket{tc|tc} + \braket{tc|ct} - \braket{to|to} + \braket{to|ot})- \\
     &\delta_{xt}\delta_{yt}(\braket{oc|oc}-\braket{oc|co})-\delta_{xy}\braket{to|ot} + \delta_{yt}\braket{xo|ot} - \\
     &\braket{xc|cy} + \delta_{xt}\braket{tc|cy} + \delta_{xt}\braket{to|oy} - \braket{xo|oy} + \\
     &\delta_{yt}\braket{xc|ct}-\delta_{xy}\braket{tc|ct} - \delta_{xy}\braket{to|ot} + \delta_{xt}\braket{to|oy}-\\
     &\braket{yc|cx} + \delta_{yt}\braket{tc|cx} + \\
     &f_{xy} + \delta_{xy}(f_{tt}-f_{cc} - f_{oo}) + \\
     &\braket{tx|ty} - \braket{xc|yc} + \braket{xc|cy} - \braket{xo|yo} + \\
     &\delta_{xy}(\braket{oc|oc} - \braket{tc|tc}-\braket{to|to}+\braket{to|ot})+\\
     &\delta_{xt}\delta_{yt}\braket{co|oc} -\\
     &\delta_{xt}f_{ty}-\delta_{yt}f_{xt}+\delta_{xt}\delta_{ty}(f_{cc}+f_{oo}) -\\
     &\braket{tx|yt} - \delta_{xt}(\braket{to|oy}-\braket{to|yo}-\braket{tc|yc})-\delta_{yt}(\braket{xc|ct}-\braket{xc|tc}-\braket{xo|to})-\\ &\delta_{xt}\delta_{yt}\braket{oc|oc}-
     \delta_{xy}\braket{oc|co}+\\
     &\delta_{yt}\braket{to|ox}-\braket{yo|ox} + \\
     &\delta_{xt}\braket{yc|ct}-\delta_{xy}\braket{tc|ct} - \\
     &\delta_{xt}f_{ty}-\delta_{ty}f_{xt} + \delta_{xt}\delta_{ty}(f_{cc}+f_{oo})-\\
     &\braket{ty|xt} -\delta_{ty}(\braket{to|ox}-\braket{to|xo} - \braket{tc|xc}) - \delta_{xt}(\braket{yc|ct} - \braket{yc|tc} - \braket{yo|to})-\\ &\delta_{xt}\delta_{yt}\braket{oc|oc}-\delta_{xy}\braket{oc|co} + \\
     &f_{xy} + \delta_{xy}(f_{tt}-f_{cc}-f_{oo}) + \braket{xt|yt} - \braket{xc|yc} -\braket{xo|yo} + \braket{xo|oy} + \\
     &\delta_{xy}(\braket{oc|oc} - \braket{to|to} - \braket{tc|tc} + \braket{tc|ct}) + \\
     &\delta_{xt}\delta_{yt}\braket{oc|co}\Bigr] 
\end{align*}
\begin{align*}
    \bra{^F \Psi_{oc}^{tx}} H \ket{^F \Psi_{oc}^{ty}} =&\frac{1}{3}\cdot\Bigl[3f_{xy}+3\delta_{xy}(f_{tt}-f_{cc}-f_{oo}) - 3\delta_{ty}f_{xt} - 3\delta_{xt}f_{ty} + 3\delta_{xt}\delta_{yt}(f_{cc}+f_{oo}) + \\
    &3\braket{tx|ty}-3\braket{tx|yt} - 3\braket{xc|yc} - 3\braket{xo|yo} + \\
    &\delta_{xt}(3\braket{tc|yc} + 3\braket{to|yo}) +\\
    &\delta_{yt}(3\braket{xc|tc}+3\braket{xo|to})  +\\
    &\delta_{xt}\delta_{yt}(3\braket{oc|co}-3\braket{oc|oc}) + \\
    &\delta_{xy}(3\braket{oc|oc} - 3\braket{tc|tc} - 3\braket{to|to} - 3\braket{oc|co})
    \Bigr] \\
    =& f_{xy}+\delta_{xy}(f_{tt}-f_{cc}-f_{oo}) - \delta_{ty}f_{xt} - \delta_{xt}f_{ty} + \delta_{xt}\delta_{yt}(f_{cc}+f_{oo}) + \\
    &\braket{tx|ty}-\braket{tx|yt} - \braket{xc|yc} - \braket{xo|yo} + \\
    &\delta_{xt}(\braket{tc|yc} + \braket{to|yo}) +\\
    &\delta_{yt}(\braket{xc|tc}+\braket{xo|to})  +\\
    &\delta_{xt}\delta_{yt}(\braket{oc|co}-\braket{oc|oc}) + \\
    &\delta_{xy}(\braket{oc|oc} - \braket{tc|tc} - \braket{to|to} - \braket{oc|co})
\end{align*}
\begin{align*}
\text{Special case: }x \neq y \text{ and } x, y & \neq t \\
\bra{^F \Psi_{oc}^{tx}} H \ket{^F \Psi_{oc}^{ty}} =& f_{xy} + \braket{tx|ty} - \braket{tx|yt} - \braket{xc|yc} - \braket{xo|yo}\\
\text{Special case: }x = y \text{ and } x, y & \neq t  \\
\bra{^F \Psi_{oc}^{tx}} H \ket{^F \Psi_{oc}^{tx}} =& f_{xx} + f_{tt} - f_{cc}-f_{oo} + \braket{tx|tx} - \braket{tx|xt} - \braket{xc|xc} - \braket{xo|xo} + \\
&\braket{oc|oc} - \braket{tc|tc} - \braket{to|to} - \braket{oc|co}\\
\text{Special case: }x \neq y \text{ and } x & = t \\
\bra{^F \Psi_{oc}^{tt}} H \ket{^F \Psi_{oc}^{ty}} =& 0\\
\text{Special case: }x \neq y \text{ and } y & = t \\
\bra{^F \Psi_{oc}^{tx}} H \ket{^F \Psi_{oc}^{tt}} =&0 \\
\text{Special case: }x = y \text{ and } x, y & = t \\ 
&= 0
\end{align*}

\section{Simplification to non-orthogonal matrix elements due to spin-adaptation}\label{sec:non_orthogonal}

In this section we will derive a general result for non-orthogonal (NO) matrix elements for an arbitrary operator $\hat{O}$ - which does not depend on spin - between CSFs. This holds for matrix elements for three operators we are concerned about: the identity operator $I$ for overlaps, the vector dipole operator $\hat{\mu}$ for the computation of transition dipoles, and the Hamiltonian operator $\hat{H}$ for constructing a CI matrix that 
ensures properly orthogonal initial pump states and final pump-probe states.

Two notes are that a) since we are dealing with Hermitian matrices,
\begin{align*}
    \bra{^1\Phi_o^t} \hat{O} \ket{^M\Phi_{oc}^{tx}} = 
    \bra{^M\Phi_{oc}^{tx}} \hat{O} \ket{^1\Phi_o^t}
\end{align*}
and b) that we'll only need NO matrix elements between 2eOS CSFs and 4eOS CSFs. No NO matrix element between two 4eOS CSFs is needed in this work.

\subsection{Singlet CSFs}
The relevant singlet CSFs are 
\begin{align}
    \ket{^{1}\Phi_{o}^{t}} &= (2)^{-1/2} \left(\ket{\Phi_{o}^{t}} + 
                                               \ket{\Phi_{\bar{o}}^{\bar{t}}} \right)
\end{align}
for the valence excited state and 
\begin{align}
    \ket{^{1}\Phi_{c}^{t}} &= (2)^{-1/2} \left(\ket{\Phi_{c}^{t}} +
                                               \ket{\Phi_{\bar{c}}^{\bar{t}}} \right) \\
    \ket{^{1_A}\Phi_{oc}^{ty}} &= (12)^{-1/2}  \left(2\ckc + 2\ckd + \cke + \ckf - \ckg - \ckh \right) \\
    \ket{^{1_B}\Phi_{oc}^{ty}} &= (2)^{-1}   \left(\cke + \ckf + \ckg + \ckh \right) 
\end{align}
for the final pump-probe excited states. Note that due to the symmetry of singlets, only the contribution due to one of the configurations of $\ket{^{1}\Phi_{o}^{t}}$ to the matrix elements needs to be explicitly calculated, with the contribution due to the second configuration being completely equal.
\subsubsection{$\bra{^1\Phi_o^t} \hat{O} \ket{^1\Phi_{c}^{t}}$}
\begin{align*}
    \bra{^1\Phi_o^t} \hat{O} \ket{^1\Phi_{c}^{t}} = &
    (2)^{-1} \cdot 2 \cdot \bigg( 
    \bra{\Phi_{o}^{t}} \hat{O} \ket{\Phi_{c}^{t}} + 
    \bra{\Phi_{o}^{t}} \hat{O} \ket{\Phi_{\bar{c}}^{\bar{t}}} \bigg) \\ = &
    \bra{\Phi_{o}^{t}} \hat{O} \ket{\Phi_{c}^{t}} + 
    \bra{\Phi_{o}^{t}} \hat{O} \ket{\Phi_{\bar{c}}^{\bar{t}}}
\end{align*}

\subsubsection{$\bra{^1\Phi_o^t} \hat{O} \ket{^{1_A}\Phi_{oc}^{ty}}$}
\begin{align*}
    \bra{^1\Phi_o^t} \hat{O} \ket{^{1_A}\Phi_{oc}^{ty}} = &
    (24)^{-1/2} \cdot 2 \cdot \bigg[ 2 \bigg( 
    \bra{\Phi_{o}^{t}} \hat{O} \ket{\Phi_{oc}^{ty}} + 
    \bra{\Phi_{o}^{t}} \hat{O} \ket{\Phi_{\bar{o}\bar{c}}^{\bar{t}\bar{y}}} \bigg) + 
    \bra{\Phi_{o}^{t}} \hat{O} \ket{\Phi_{\bar{o}c}^{\bar{t}y}} +
    \bra{\Phi_{o}^{t}} \hat{O} \ket{\Phi_{o\bar{c}}^{t\bar{y}}} -
    \bra{\Phi_{o}^{t}} \hat{O} \ket{\Phi_{\bar{o}c}^{\bar{y}t}} -
    \bra{\Phi_{o}^{t}} \hat{O} \ket{\Phi_{o\bar{c}}^{y\bar{t}}}
    \bigg] \\ = & 
    (6)^{-1/2} \bigg[ 2 \bigg( 
    \bra{\Phi_{o}^{t}} \hat{O} \ket{\Phi_{oc}^{ty}} + 
    \bra{\Phi_{o}^{t}} \hat{O} \ket{\Phi_{\bar{o}\bar{c}}^{\bar{t}\bar{y}}} \bigg) + 
    \bra{\Phi_{o}^{t}} \hat{O} \ket{\Phi_{\bar{o}c}^{\bar{t}y}} +
    \bra{\Phi_{o}^{t}} \hat{O} \ket{\Phi_{o\bar{c}}^{t\bar{y}}} -
    \bra{\Phi_{o}^{t}} \hat{O} \ket{\Phi_{\bar{o}c}^{\bar{y}t}} -
    \bra{\Phi_{o}^{t}} \hat{O} \ket{\Phi_{o\bar{c}}^{y\bar{t}}}
    \bigg]
\end{align*}

\subsubsection{$\bra{^1\Phi_o^t} \hat{O} \ket{^{1_B}\Phi_{oc}^{ty}}$}
\begin{align*}
    \bra{^1\Phi_o^t} \hat{O} \ket{^{1_B}\Phi_{oc}^{ty}} = &
    (8)^{-1/2} \cdot 2 \cdot \bigg( 
    \bra{\Phi_{o}^{t}} \hat{O} \ket{\Phi_{\bar{o}c}^{\bar{t}y}} +
    \bra{\Phi_{o}^{t}} \hat{O} \ket{\Phi_{o\bar{c}}^{t\bar{y}}} +
    \bra{\Phi_{o}^{t}} \hat{O} \ket{\Phi_{\bar{o}c}^{\bar{y}t}} +
    \bra{\Phi_{o}^{t}} \hat{O} \ket{\Phi_{o\bar{c}}^{y\bar{t}}}
    \bigg) \\ = &
    (2)^{-1/2} \bigg( 
    \bra{\Phi_{o}^{t}} \hat{O} \ket{\Phi_{\bar{o}c}^{\bar{t}y}} +
    \bra{\Phi_{o}^{t}} \hat{O} \ket{\Phi_{o\bar{c}}^{t\bar{y}}} +
    \bra{\Phi_{o}^{t}} \hat{O} \ket{\Phi_{\bar{o}c}^{\bar{y}t}} +
    \bra{\Phi_{o}^{t}} \hat{O} \ket{\Phi_{o\bar{c}}^{y\bar{t}}}
    \bigg)
\end{align*}
If re-normalization has been taken care of, this also holds for the matrix elements for the special case when y = t. 

\subsection{Triplet CSFs}
The relevant triplet CSFs are 
\begin{align}
    \ket{^{3}\Phi_{o}^{t}} &= (2)^{-1/2} \left(\ket{\Phi_{o}^{t}} - 
                                               \ket{\Phi_{\bar{o}}^{\bar{t}}} \right)
\end{align}
for the valence excited state and 
\begin{align}
    \ket{^{3}\Phi_{c}^{t}} &= (2)^{-1/2} \left(\ket{\Phi_{c}^{t}} -
                                               \ket{\Phi_{\bar{c}}^{\bar{t}}} \right) \\
    \ket{^{3_C}\Phi_{oc}^{ty}} &= (2)^{-1/2} \left(\ckc - \ckd \right) \\
    \ket{^{3_D}\Phi_{oc}^{ty}} &= (2)^{-1/2} \left(\cke - \ckf \right) \\
    \ket{^{3_E}\Phi_{oc}^{ty}} &= (2)^{-1/2} \left(\ckg - \ckh \right) 
\end{align}
for the final pump-probe excited states. As with the singlets, the symmetry of triplets, the Hermiticity of the matrix elements, and the ascence of SOC in the operators results in the need of only the contribution due to one of the configurations of $\ket{^{3}\Phi_{o}^{t}}$ to be calculated explicitly. The contribution due to the second one is equivalent, and is accounted for by a factor of 2.

\subsubsection{$\bra{^3\Phi_o^t} \hat{O} \ket{^3\Phi_{c}^{t}}$}
\begin{align*}
    \bra{^3\Phi_o^t} \hat{O} \ket{^3\Phi_{c}^{t}} = &
    (2)^{-1} \cdot 2 \cdot \bigg( 
    \bra{\Phi_{o}^{t}} \hat{O} \ket{\Phi_{c}^{t}} - 
    \bra{\Phi_{o}^{t}} \hat{O} \ket{\Phi_{\bar{c}}^{\bar{t}}} \bigg) \\ = &
    \bra{\Phi_{o}^{t}} \hat{O} \ket{\Phi_{c}^{t}} - 
    \bra{\Phi_{o}^{t}} \hat{O} \ket{\Phi_{\bar{c}}^{\bar{t}}}
\end{align*}

\subsubsection{$\bra{^3\Phi_o^t} \hat{O} \ket{^{3_C}\Phi_{oc}^{ty}}$}
\begin{align*}
    \bra{^3\Phi_o^t} \hat{O} \ket{^{3_C}\Phi_{oc}^{ty}} = &
    (2)^{-1} \cdot 2 \cdot \bigg( 
    \bra{\Phi_{o}^{t}} \hat{O} \ket{\Phi_{oc}^{ty}} -
    \bra{\Phi_{o}^{t}} \hat{O} \ket{\Phi_{\bar{o}\bar{c}}^{\bar{t}\bar{y}}}
    \bigg) \\ = &
    \bra{\Phi_{o}^{t}} \hat{O} \ket{\Phi_{oc}^{ty}} -
    \bra{\Phi_{o}^{t}} \hat{O} \ket{\Phi_{\bar{o}\bar{c}}^{\bar{t}\bar{y}}}
\end{align*}

\subsubsection{$\bra{^3\Phi_o^t} \hat{O} \ket{^{3_D}\Phi_{oc}^{ty}}$}
\begin{align*}
    \bra{^3\Phi_o^t} \hat{O} \ket{^{3_D}\Phi_{oc}^{ty}} = &
    (2)^{-1} \cdot 2 \cdot \bigg( 
    \bra{\Phi_{o}^{t}} \hat{O} \ket{\Phi_{\bar{o}c}^{\bar{t}y}} -
    \bra{\Phi_{o}^{t}} \hat{O} \ket{\Phi_{o\bar{c}}^{t\bar{y}}}
    \bigg) \\ = &
    \bra{\Phi_{o}^{t}} \hat{O} \ket{\Phi_{\bar{o}c}^{\bar{t}y}} -
    \bra{\Phi_{o}^{t}} \hat{O} \ket{\Phi_{o\bar{c}}^{t\bar{y}}}
\end{align*}

\subsubsection{$\bra{^3\Phi_o^t} \hat{O} \ket{^{3_E}\Phi_{oc}^{ty}}$}
\begin{align*}
    \bra{^3\Phi_o^t} \hat{O} \ket{^{3_E}\Phi_{oc}^{ty}} = &
    (2)^{-1} \cdot 2 \cdot \bigg( 
    \bra{\Phi_{o}^{t}} \hat{O} \ket{\Phi_{\bar{o}c}^{\bar{y}t}} -
    \bra{\Phi_{o}^{t}} \hat{O} \ket{\Phi_{o\bar{c}}^{y\bar{t}}}
    \bigg) \\ = &
    \bra{\Phi_{o}^{t}} \hat{O} \ket{\Phi_{\bar{o}c}^{\bar{y}t}} -
    \bra{\Phi_{o}^{t}} \hat{O} \ket{\Phi_{o\bar{c}}^{y\bar{t}}}
\end{align*}

\newpage
\section{Supporting information for calculations on water}\label{sec:H2O}
\subsection{Ill-behavior of STEX states beyond the ionization threshold}

\begin{figure}[h!]
    \includegraphics[width=1.00\linewidth]{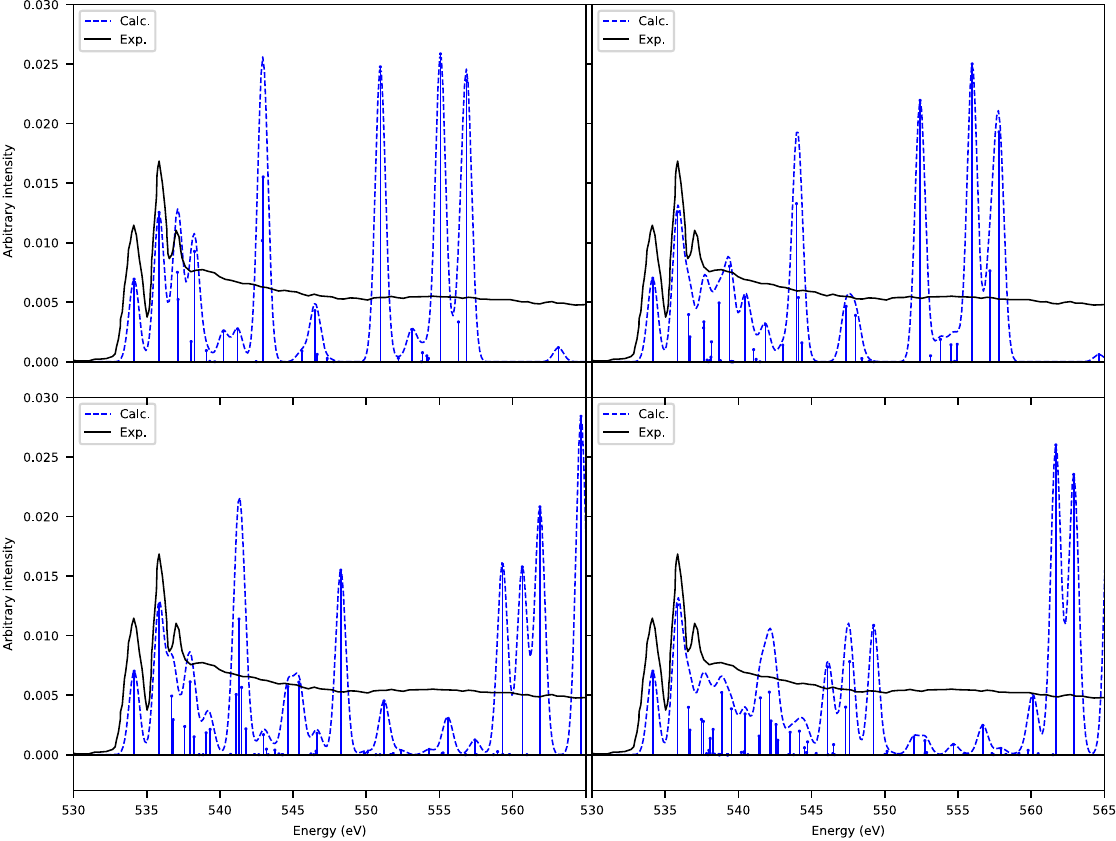}
    \caption{STEX calculation of the ground state of water at the the O K-edge spectra (both NEXAS and EXAFS) employing different basis. Top-left: aug-pcX-2 (O) / aug-pcseg-1 (H), bottom left: aug-pcX-3 (O) / aug-pcseg-2 (H), top right: d-aug-pcX-2 (O) / aug-pcseg-1 (H), bottom right: d-aug-pcX-3 (O) / aug-pcseg-2 (H). The experimental results are those of Ishii and coworkers.\cite{Ishii1987}}
    \label{fgr:H2O_EXAFS}
\end{figure}

Unless specialized techniques are employed, core-excited states beyond the core ionization are ill-described by square-integrable, L$^2$ basis functions, such as the gaussian-type orbitals (GTOs) employed in the present work.\cite{Gokhberg2009} The poor description of these states may manifest in erroneously bright transitions, as observed in the STEX spectra for water (IP of 540 eV) in Figure \ref{fgr:H2O_EXAFS}. Interestingly and despite the former, as large, diffuse basis functions are included the bright transitions immediatley past the ionization threshold begin to tame off.

\newpage
\subsection{On the intense NEXAS oscillator strengths when employing the ROKS-optimized $\ket{^1\Phi_c^t}$ orbitals}

Employing the $\ket{^1\Phi_c^t}$ orbitals to produce the NEXAS resulted in overly intense transitions in the high-energy region when the two open-shells, \textbf{c} and \textbf{t} corresponded to the O$_{1s}$ and O$_{3s}$ orbitals. As has been noted in the literature, strong mixing between the open-shells during the ROKS procedure enhances the non-orthogonality to the ground state, and delivers energies that are closer approximations to the corresponding triplet.\cite{Kowalczyk2013} By employing a constant energy-shift between the two open-shells, as prescribed by Kowalczyk and coworkers, we managed to find alternative ROKS solutions whose overlaps with the ground state were inversely commensurate with the energy shift. To our surprise, while the predicted transition energies where relatively unaffected by the seemingly small non-orthogonality between the ground state and the $\ket{^1\Phi_c^t}$ configuration, the transition dipole is extremely sensitive to such quantity. Ensuring that the final pump-probe excited-states are orthogonal to the initial valence excited state via projection operators seems unable to avert this phenomena and ensuring orthogonality within the SCF optimization via level shifts is required. We suspect this may be related to overly-intense ROKS N$_{1s} \xrightarrow[]{}$ N$_{3s}$ and C$_{1s} \xrightarrow[]{}$ C$_{3s}$ transitions in ammonia and trans-butadiene reported in the literature.\cite{Carter-Fenk2022a} 

\begin{figure}[h!]
    \includegraphics[width=1.00\linewidth]{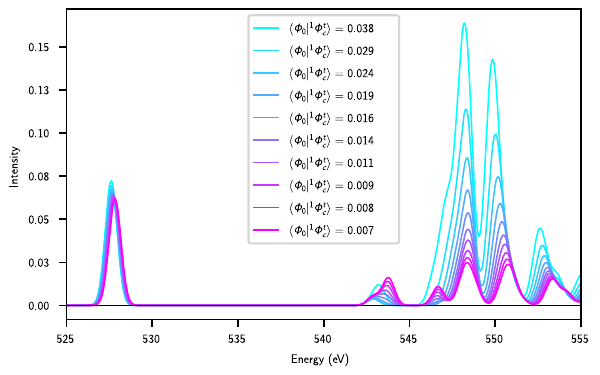}
    \caption{1C-NOCIS 2eOS / aug-pcX-2 (O) / aug-pcseg-1 (H) spectra for the 2p$_x$ (b$_1$) $\xrightarrow[]{}$ 3s (a$_1$) excited state of water, employing the ROKS-optimized $\ket{^1\Phi_c^t}$ orbitals and enforcing increasing energy shifts between the open-shells to ensure lower overlaps with the ground state.}
    \label{fgr:H2O-ROKS-overlap}
\end{figure}

For the three calculations carried out in this work that require ROKS-optimized O$_{1s}\xrightarrow[]{}$ O$_{3s}$ $\ket{^1\Phi_c^t}$ orbitals, namely those for the 1C-NOCIS 2eOS spectra of the 2p$_x$ (b$_1$) $\xrightarrow[]{}$ 3s (a$_1$), 2p$_z$ (a$_1$) $\xrightarrow[]{}$ 3s (a$_1$), and 2p$_y$ (b$_2$) $\xrightarrow[]{}$ 3s (a$_1$) employing the aforementioned choice of orbitals, we apply an energy shift in the ROKS calculation to ensure the overlap between the ground state and the $\ket{^1\Phi_c^t}$ ROKS configuration remains below 0.007.

\newpage
\subsection{Choice of orbitals for the construction of the CSFs}

\begin{figure}[h!]
    \includegraphics[width=1.00\linewidth]{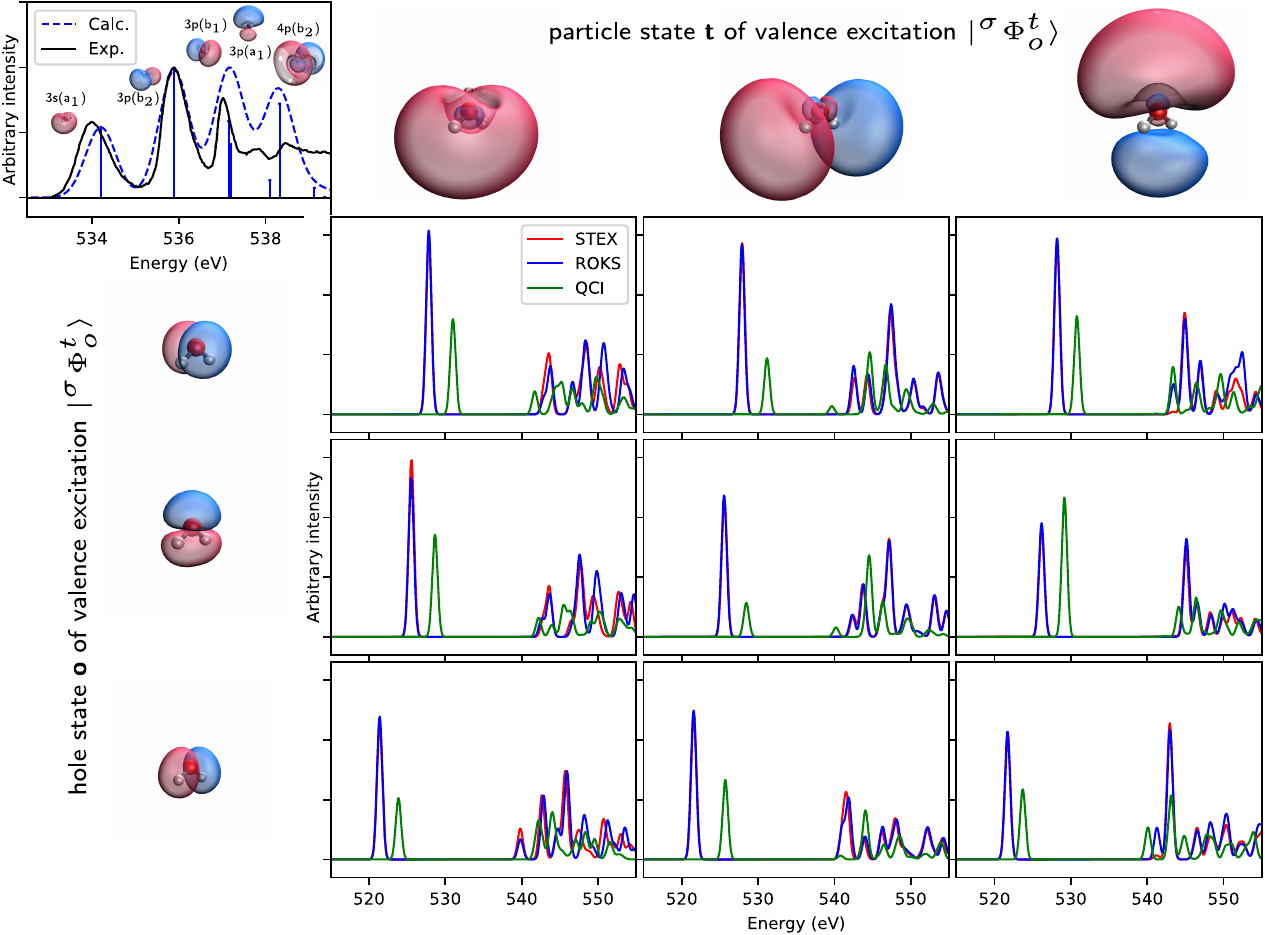}
    \caption{1C-NOCIS 2eOS / aug-pcX-2 (O) / aug-pcseg-1 (H) spectra for different valence excited states of water, employing different orbitals for the construction of the CSFs.}
    \label{fgr:orbital_choice}
\end{figure}

As plotted in Figure \ref{fgr:orbital_choice}, the 1C-NOCIS NEXAS spectra for the different excited states of water generated with the 1C-NOCIS NTO orbitals as a reference are nearly indistinguishable to those generated with the orbitals from the ROKS-optimized $\ket{^1\Phi_c^t}$ configuration (when ensuring a small overlap between the ground state and the $\ket{^1\Phi_c^t}$ ROKS-optimized orbitals, as specified in the previous section). The most salient different is a small discrepancy in the intensity of the high-energy transitions in the 549 - 552 eV region for the 2p$_x$ (b$_1$) $\xrightarrow[]{}$ 3p$_z$ (a$_1$) state. This implies that the relaxation of the orbitals due to an electron in the particle state \textbf{t} is negligible compared to the relaxation in the orbitals due to the presence of a core hole \textbf{c} for these valence excited states. This makes intuitive sense since the valence excited states of water are of full-fledged Rydberg character.

The spectra generated with the $\ket{^4\Phi_{oc}^t}$ quartet core ion orbitals differs drastically from the other two choices of orbitals. The energy of the c $\xrightarrow[]{}$ SOMO(o) transition is blue-shifted by roughly 2 eVs, and there are visible differences in the high-energy region of the spectrum. A physical reason to prefer either the 1C-NOCIS NTO or $\ket{^1\Phi_c^t}$ as reference orbitals for the singlet pump-probe core excited states comes from the fact that the spatial description of quartet-optimized orbitals probably differs significantly to those of a related singlet state. On the other hand, the quartet core ion orbitals are likely a better choice to describe triplet pump-probe core excited states.

\newpage
\subsection{Visualizing the different contributions to the excited state spectra}
\begin{figure}[h!]
    \includegraphics[width=0.85\linewidth]{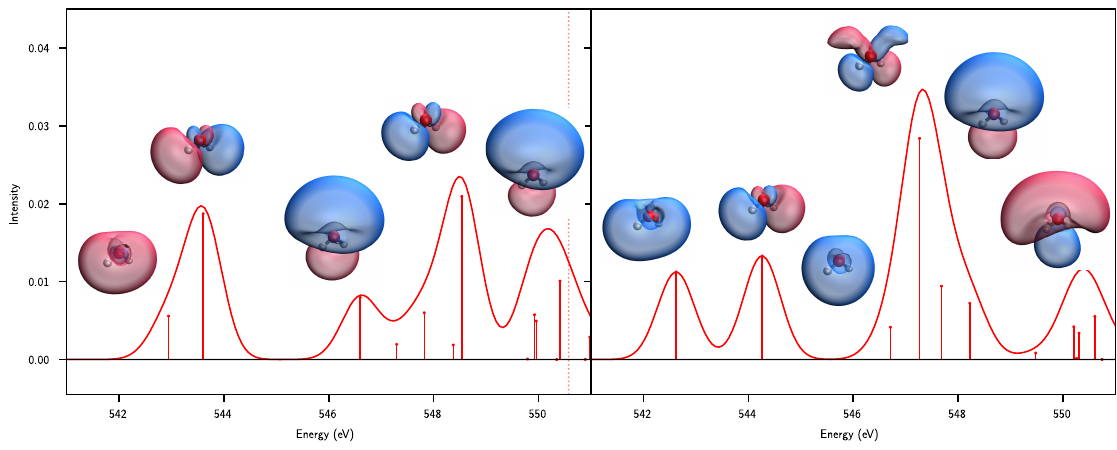}
    \caption{High-energy region of the 1C-NOCIS 2eOS (STEX NTOs) / aug-pcX-2 (O) / aug-pcseg-1 (H) spectra for the 2p$_x$ (b$_1$) $\xrightarrow[]{}$ 3s (a$_1$) (left) and 2p$_x$ (b$_1$) $\xrightarrow[]{}$ 3p$_y$ (b$_2$) (right). The NTO-like particle pump-probe excited states of bright / interesting transitions are plotted right next to them.}
    \label{fgr:H2O_NTOs}
\end{figure}

To visualize the character of the pump-probe excited states, we take each of the eigenvectors of the 1C-NOCIS 2eOS Hamiltonian. 

\begin{equation}
\begin{blockarray}{ccc}
\begin{block}{c[c]c}
          & ^1b_c^t & \ket{^1\Phi_c^t} \\
          & ^Ab_{oc}^{ty} & \ket{^{1_A}\Phi_{oc}^{ty}} \\
\mathbf{C_f} = & \vdots & \vdots \\
          & ^Bb_{oc}^{ty} & \ket{^{1_B}\Phi_{oc}^{ty}} \\
          & \vdots & \vdots \notag\\
\end{block}
\end{blockarray} 
\end{equation}
and singular-value decompose (SVD) them 
\begin{align*}
    \mathbf{\sigma} = \mathbf{L} \cdot \mathbf{C_f} \cdot \mathbf{R}
\end{align*}
This allows to rotate the final pump-probe state into a compact representation, akin to a natural transition orbital (NTO).\cite{Mayer2007}
\begin{equation}
\begin{blockarray}{ccc}
\begin{block}{c[c]c}
          & \vdots & \\
          & 0 & \\
          & \vdots & \\
\mathbf{C^{'}_{f}} = \mathbf{C_f} \cdot \mathbf{R} = & 1 & \\
          & \vdots & \\
          & 0 & \\
          & \vdots  \notag\\
\end{block}
\end{blockarray} 
\end{equation}
Employing the aforementioned procedure, we plot the NTO-like quantities for the two lowest valence excited states of water in Figure \ref{fgr:H2O_NTOs}. Beyond the c $\xrightarrow[]{}$ SOMO(o) transition, a transition into the 3s (a$_1$) particle state (c $\xrightarrow[]{}$ SOMO(t)) features as the lowest-energy transition for the 2p$_x$ (b$_1$) $\xrightarrow[]{}$ 3s (a$_1$) excited state. Since the 3s (a$_1$) particle state is fully-vacant for the 2p$_x$ (b$_1$) $\xrightarrow[]{}$ 3p$_y$ (b$_2$) valence excited state, there are two individual transitions associated with said spatial orbital, arising from the two linearly-independent singlets that can be constructed for a 4eOS system. The 2p$_x$ (b$_1$) $\xrightarrow[]{}$ 3s (a$_1$) valence excited state features prominent spin-split transitions into the 3p$_y$ (b$_2$) and 3p$_z$ (a$_1$) particle states.

\newpage
\section{Ionization thresholds for thymine}\label{sec:thymine}
\begin{figure}[h!]
  \includegraphics[width=0.50\linewidth]{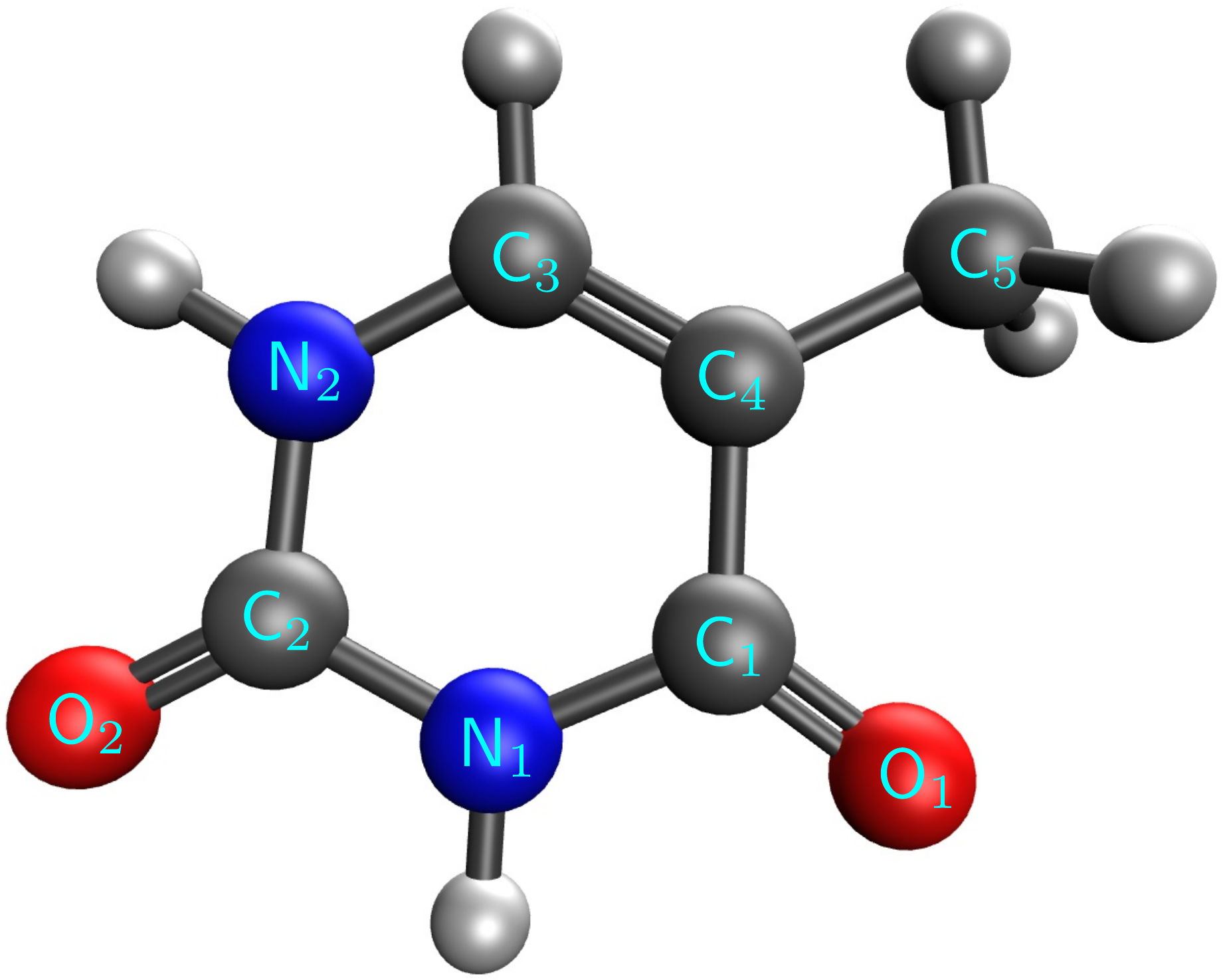}
  \caption{Atom labeling for thymine.}
  \label{fgr:thymine}
\end{figure}
For all the calculations on thymine, the aug-pcX-2 basis is employed for the element of interest, and aug-pcseg-1 for the others. The tables below list the carbon, nitrogen, and oxygen core ionization potentials (IPs) for the ground state, as well as those for the $\pi \xrightarrow[]{} \pi^*$ and O$_{LP}\xrightarrow[]{} \pi^*$ excited states. For the excited states, there are two doublet ($\ket{^{2_G} \Phi_{oc}^{t }}$, $\ket{^{2_H} \Phi_{oc}^{t}}$) and one quartet IPs ($\ket{^{4_I} \Phi_{oc}^t}$); refer to the main text for a definition of the CSFs. Note that since the energies of these CSFs are evaluated with ROKS-optimized $\ket{^{1}\Phi_{c}^{t}}$ configuration and, as a result, the quartet IPs are likely overestimated. The coupling between the two doublet states is also provided for the excited states. All energies are reported in eV, without any shifts applied.

\begin{table}[h!]
  \caption{$\Delta$SCF(HF) IPs for the ground state. }
  \label{tbl:IP-thymine-GS}
  \begin{tabular}{cc|cc|cc}
    \hline
    carbon & IP & nitrogen & IP & oxygen & IP \\
    \hline
    1 & 295.27 & 1 & 406.66 & 1 & 536.68 \\
    2 & 296.50 & 2 & 406.86 & 2 & 537.00 \\
    3 & 292.91 & & & & \\
    4 & 291.03 & & & & \\
    5 & 291.35 & & & & \\
    \hline
  \end{tabular}
\end{table}
Carbon atom 1 features a red-shift on the order of 2 - 3 eVs on its core IP relative to the ground state, when the molecule is on its $\pi \xrightarrow[]{} \pi^*$ excited state. All the other carbon atoms, as well as the nitrogen and oxygen atoms, feature IPs more or less close to those of the ground state. It's interesting to note that while both nitrogens have similar IPs on the ground state, their IPs differ by more than 1 eV for this excited state.
\begin{table}[h!]
  \caption{IPs, and the coupling between the two doublet ions, in eV for the $\pi \xrightarrow[]{} \pi^*$ excited state of thymine.}
  \label{tbl:IP-thymine-pi-pistar}
  \begin{tabular}{ccccc}
    \hline
    carbon & $\ket{^{2_G} \Phi_{oc}^{t }}$ IP & $\ket{^{2_H} \Phi_{oc}^{t}}$ IP & $\ket{^{4_I} \Phi_{oc}^t}$ IP & $\bra{^{2_G} \Phi_{oc}^{t }} H \ket{^{2_H} \Phi_{oc}^{t}}$ \\
    \hline
    1 & 293.38 & 292.07 & 291.41 & -0.49 \\
    2 & 297.91 & 296.23 & 295.25 & -1.41 \\
    3 & 292.32 & 291.72 & 290.86 & -0.63 \\
    4 & 291.73 & 291.59 & 290.98 & -0.34 \\
    5 & 292.40 & 291.15 & 290.50 & -1.09 \\
    \hline
    nitrogen & $\ket{^{2_G} \Phi_{oc}^{t }}$ IP & $\ket{^{2_H} \Phi_{oc}^{t}}$ IP & $\ket{^{4_I} \Phi_{oc}^t}$ IP & $\bra{^{2_G} \Phi_{oc}^{t }} H \ket{^{2_H} \Phi_{oc}^{t}}$ \\
    \hline
    1 & 407.18 & 405.75 & 405.03 & -1.17 \\
    2 & 408.99 & 407.18 & 406.09 & -1.64 \\
    \hline
    oxygen & $\ket{^{2_G} \Phi_{oc}^{t }}$ IP & $\ket{^{2_H} \Phi_{oc}^{t}}$ IP & $\ket{^{4_I} \Phi_{oc}^t}$ IP & $\bra{^{2_G} \Phi_{oc}^{t }} H \ket{^{2_H} \Phi_{oc}^{t}}$ \\
    \hline
    1 & 538.66 & 537.54 & 536.89 & -0.97 \\
    2 & 538.76 & 537.03 & 536.13 & -1.50 \\
    \hline
  \end{tabular}
\end{table}

For the O$_{LP} \xrightarrow[]{} \pi^*$ excited state, all carbon atoms have IPs relatively close to the ground state, including carbon 1. The IP of nitrogen 1 seems to be blue-shifted from the ground by about 2 eVs and, like in the $\pi \xrightarrow[]{} \pi^*$ excited state, the difference in IPs between the two nitrogens is on the order on 1 eV. Finally, the IP of oxygen 1 features the largest IP blue-shift of all relative to the ground state - around 10 eV. Since this is the oxygen contributing the lone pair for the O$_{LP} \xrightarrow[]{} \pi^*$ valence excited state, a large amount of charge migrates out of this atom's vicinity, effectively behaving as an ionized system for the subsequent core ionization and thus incurring in large energetic costs for the removal of an electron electron from an atom that has already been ``effectively'' ionized. Finally, an interesting note is that both the difference in IP between the two linearly-independent doublets, as well as the coupling between them, seem to be smaller for the O$_{LP} \xrightarrow[]{} \pi^*$ excited state compared to the $\pi \xrightarrow[]{} \pi^*$ excited state.
\begin{table}[h!]
  \caption{IPs, and the coupling between the two doublet ions, in eV for the O$_{LP} \xrightarrow[]{} \pi^*$ excited state of thymine.}
  \label{tbl:IP-thymine-O_LP-pistar}
  \begin{tabular}{ccccc}
    \hline
    carbon & $\ket{^{2_G} \Phi_{oc}^{t }}$ IP & $\ket{^{2_H} \Phi_{oc}^{t}}$ IP & $\ket{^{4_I} \Phi_{oc}^t}$ IP & $\bra{^{2_G} \Phi_{oc}^{t }} H \ket{^{2_H} \Phi_{oc}^{t}}$ \\
    \hline
    1 & 296.05 & 295.51 & 295.11 & 0.11 \\
    2 & 298.10 & 297.92 & 297.79 & -0.05 \\
    3 & 292.64 & 292.37 & 292.21 & -0.04 \\
    4 & 294.19 & 294.10 & 293.93 & -0.09 \\
    5 & 292.77 & 292.69 & 292.64 & -0.07 \\
    \hline
    nitrogen & $\ket{^{2_G} \Phi_{oc}^{t }}$ IP & $\ket{^{2_H} \Phi_{oc}^{t}}$ IP & $\ket{^{4_I} \Phi_{oc}^t}$ IP & $\bra{^{2_G} \Phi_{oc}^{t }} H \ket{^{2_H} \Phi_{oc}^{t}}$ \\
    \hline
    1 & 408.94 & 408.94 & 408.63 & -0.11 \\
    2 & 408.19 & 407.97 & 407.85 & -0.16 \\
    \hline
    oxygen & $\ket{^{2_G} \Phi_{oc}^{t }}$ IP & $\ket{^{2_H} \Phi_{oc}^{t}}$ IP & $\ket{^{4_I} \Phi_{oc}^t}$ IP & $\bra{^{2_G} \Phi_{oc}^{t }} H \ket{^{2_H} \Phi_{oc}^{t}}$ \\
    \hline
    1 & 546.36 & 546.62 & 546.05 & -0.12 \\
    2 & 538.70 & 538.49 & 538.36 & -0.18 \\
    \hline
  \end{tabular}
\end{table}

\newpage
\section{TR-NEXAS simulation of acetylacetone}\label{sec:acetylacetone}
\subsection{Ultra-fast non-adiabatic dynamics between the $\pi \xrightarrow{} \pi^*$ and O$_{n.b.} \xrightarrow{} \pi^*$ }

\begin{figure}[h!]
  \includegraphics[width=1.00\linewidth]{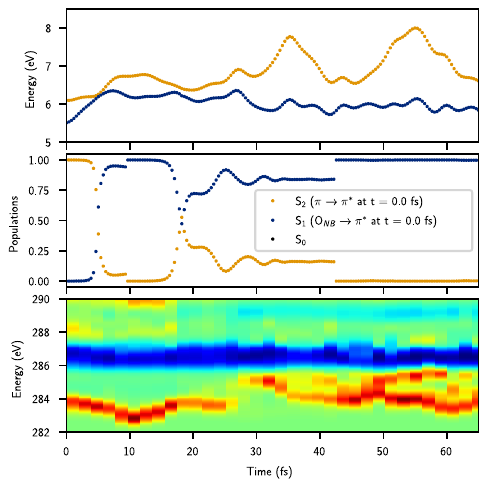}
  \caption{0 - 65 fs zoom-in of the TR-NEXAS simulation displayed in Figure 5 of the manuscript. Top) Energies of the (O$_{n.b.} \xrightarrow{} \pi^*$) and ($\pi \xrightarrow{} \pi^*$) excited states during the AFSSH NAMD simulation. Middle) Their populations Bottom) Simulated TR-NEXAS}
  \label{fgr:acetylacetone-NAMD}
\end{figure}

\newpage 
\subsection{An additional TR-NEXAS simulated from a different trajectory}

\begin{figure}[h!]
  \includegraphics[width=1.00\linewidth]{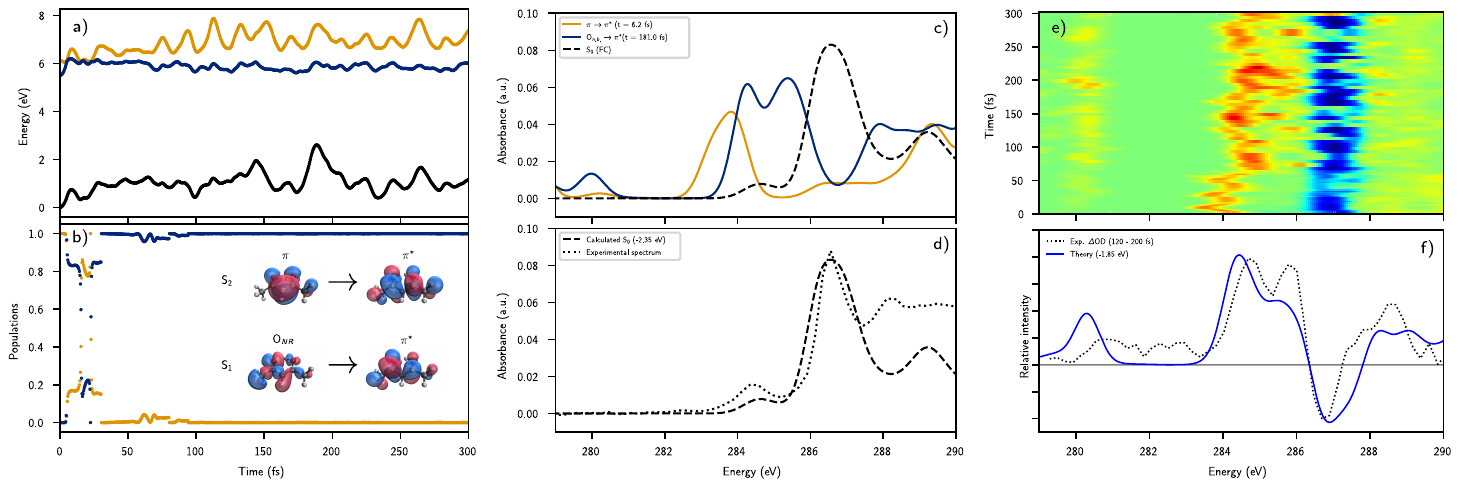}
  \caption{a) Energies of the three lowest singlet states (the ground state and the (O$_{n.b.} \xrightarrow{} \pi^*$) and ($\pi \xrightarrow{} \pi^*$) excited states) during an AFSSH NAMD simulation. b) The populations of the two excited states during the NAMD trajectory. c) Comparison of the ground state NEXAS spectrum (Wigner-broadened with 150 structures) and the excited-state NEXAS spectra at relevant times (i.e. when they are populated in the dynamics.) d) Comparison of the ground state NEXAS spectrum calculated with 1C-NOCIS / aug-pcX-1 and the experimental NEXAS of Bhattacherjee and coworkers.\cite{Bhattacherjee2017} A shift of -2.35 eV is required to match experiment and theory. e) Simulated TR-NEXAS, shifted by -1.85 eV; see text for details f) Comparison of experimental spectrum, averaged over the 120 - 200 fs time bins compared to the simulated spectrum averaged over the same time range.\cite{Bhattacherjee2017}}
  \label{fgr:acetylacetone-transient}
\end{figure}
Note that in this alternative trajectory, the doubled-peak feature in the simulated time-slice of the TR-NEXAS (Figure \ref{fgr:acetylacetone-transient}f), while still evident, is less pronounced than in the experiment and in the simulated transient in the main text. This provides evidence for the relevance of employing a representative sampling of the nuclear configurations present in the evolution of the system.

\newpage
\subsection{Visualization of the contributions to the NEXAS for the O$_{LP} \xrightarrow{} \pi^*$ excited state}
\begin{figure}[h!]
    \includegraphics[width=1.00\linewidth]{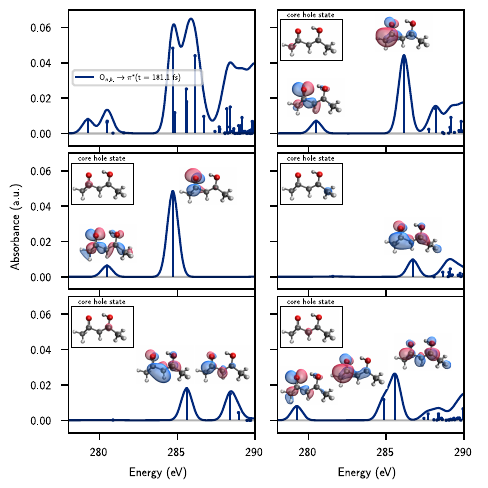}
    \caption{Top right: Sample 1C-NOCIS 2eOS / aug-pcX-1 (C) / aug-pcseg-1 (O, H) spectra for the O$_{LP} \xrightarrow{} \pi^*$ excited state, at a particular point in the trajectory presented in Section S6.1. Rest of the panels: contributions of excitations out of each of the five carbon atoms, plotted within the insets, to the spectra. The particle state associated with the brightest transitions, obtained by rotation into a compact basis per the previously explained procedure, is plotted as well.}
    \label{fgr:NTOs}
\end{figure}

Figure \ref{fgr:NTOs} visualizes the NTO-like pump-probe particle states for the O$_{LP} \xrightarrow{} \pi^*$ excited state of acetylacetone at a particular point in a NAMD trajectory employing the same approach explained in S5.2. The low-energy features, at around 280 - 281 eV come from C$_{1s} \xrightarrow{}$SOMO (O$_{LP}$) transitions. Since the O$_{LP}$ orbital is strongly localized in the keto oxygen, the transition intensity for these states is small and coming exclusively from the carbons in the vicinity of said oxygen. The main feature at 284 - 287 eV, displaying a double-peaked profile, arises predominantly from C$_{1s} \xrightarrow{}$SOMO ($\pi^*$) transitions. A major contribution from a 4eOS C$_{1s} \xrightarrow{}$ ($\pi^*$) transition out of the central carbon, where the $\pi^*$ orbital is the higher-lying one not involved in the valence excitation, partially accounts for the high energy peak at 286.5 eV.

\bibliography{achemso-demo.bib}